\documentclass[a4paper,twocolumn,11pt,accepted=2026-07-27]{quantumarticle}

\pdfoutput=1
\usepackage[utf8]{inputenc}
\usepackage{graphicx,cite}
\graphicspath{ {Images/} }
\usepackage[margin=10pt,font=small,labelfont=bf]{caption}
\usepackage{subcaption}
\usepackage{setspace}
\usepackage{mciteplus}
\usepackage{enumitem}
\usepackage[english]{babel}
\usepackage{physics}
\usepackage{tensor}
\usepackage{amsthm}
\usepackage{amsmath}
\usepackage{amsfonts}
\usepackage{amssymb}
\usepackage{dsfont, mathrsfs, tikz-cd}
\usepackage{multicol}
\usepackage{xcolor}
\usepackage{hyperref}
\usepackage{url}
\usepackage{soul}
\usepackage{CJKutf8}

\newcommand{\eps}{\varepsilon}
\newcommand{\dirac}[3]{\ensuremath{\hat{O}_{#2}^{#3}(#1)}}

\DeclareMathOperator*{\SumInt}{
\mathchoice
{\ooalign{$\displaystyle\sum$\cr\hidewidth$\displaystyle\int$\hidewidth\cr}}
  {\ooalign{\raisebox{.14\height}{\scalebox{.7}{$\textstyle\sum$}}\cr\hidewidth$\textstyle\int$\hidewidth\cr}}
  {\ooalign{\raisebox{.2\height}{\scalebox{.6}{$\scriptstyle\sum$}}\cr$\scriptstyle\int$\cr}}
  {\ooalign{\raisebox{.2\height}{\scalebox{.6}{$\scriptstyle\sum$}}\cr$\scriptstyle\int$\cr}}
}

\title{On the relation between perspective-neutral, algebraic, and effective quantum reference frames}

\author{Julian De Vuyst}
\email{julian.devuyst@oist.jp}
\orcid{0000-0002-3832-1112}

\author{Philipp A.\ H\"ohn}
\email{philipp.hoehn@oist.jp}
\orcid{}

\affiliation{Qubits and Spacetime Unit, Okinawa Institute of Science and Technology (\begin{CJK}{UTF8}{min}沖縄科学技術大学院大学\end{CJK}), 1919-1 Tancha, Onna-son, Kunigami-gun, Okinawa, Japan 904-0495}

\author{Artur Tsobanjan}
\email{tsobanjana@hollins.edu}
\orcid{}
\affiliation{Theoretical Sciences Visiting Program (TSVP), Okinawa Institute of Science and Technology (\begin{CJK}{UTF8}{min}沖縄科学技術大学院大学\end{CJK}), 1919-1 Tancha, Onna-son, Kunigami-gun, Okinawa, Japan 904-0495}
\affiliation{Hollins University, 8003 Fishburn Dr., Roanoke, VA 24020, USA}
\affiliation{King's College, 133 N River St, Wilkes-Barre, PA 18711, USA}

\begin{document}

\maketitle

\begin{abstract}
    The framework of internal quantum reference frames (QRFs) constitutes a universal toolset for dealing with symmetries in quantum theory and has led to new revelations in quantum gravity, gauge theories and foundational physics.
    Multiple approaches have emerged, sometimes differing in scope and the way symmetries are implemented, raising the question as to their relation. Here, we investigate the relation between three approaches to QRFs for gauge symmetries, namely the \emph{effective} semiclassical,  \emph{algebraic}, and \emph{perspective-neutral} (PN) approaches. Rather than constructing Hilbert spaces, as the PN approach, the effective approach is based on a quantum phase space parametrized by expectation values and fluctuations, while the emphasis of the algebraic approach is on the state space of complex linear functionals on a kinematical algebra. Nevertheless, external frame information is treated as gauge in all three formalisms, manifested in constraints on  states and algebra. We  show that these three approaches are, in fact, equivalent for ideal QRFs,  distinguished by sharp orientations, which is the previous setting of the first two approaches. Our demonstration pertains to single constraints, including relativistic ones, and encompasses QRF changes. In particular, the QRF transformations of the PN framework agree semiclassically with those of the older effective approach, by which it was inspired. As a physical application, we explore the QRF covariance of uncertainties and fluctuations, which turn out to be frame dependent. This is particularly well-suited for the effective and algebraic approaches, for which these quantities form a natural basis. Finally, we pave the way towards extending these two approaches to non-ideal QRFs by studying the projection and gauge-fixing operations of the Page-Wootters formalism, built into the PN framework, on algebraic states.
\end{abstract}

\tableofcontents

\hrulefill
\vspace{0.5cm}

\section{Introduction } 

Reference frames usually serve no  purpose other than defining a variety of descriptions of a given physical system of interest to which they are then assumed to be \emph{external} and with which they do not interact.
However, the advent of general relativity has brought with it a profound lesson: generally covariant theories are background independent, meaning that their description should not depend on the coordinatization of the underlying spacetime manifold. This condition manifests itself as diffeomorphism invariance and means that physics is independent of standard external notions of reference frames. A crucial consequence of this is that any reference frame with respect to which we describe the system should be \emph{internal}, comprised of \emph{a priori} dynamical degrees of freedom that become quantum `rods and clocks' once we subject the entire system to the laws of quantum theory.

Such quantum `rods and clocks' have become known as internal \emph{quantum reference frames} (QRFs) and are important far beyond gravity 
\cite{Aharonov:1967zza,Aharonov:1967zz,Aharonov:1984,angeloPhysicsQuantumReference2011a,rovelli_2004,Rovelli:1990pi,CRovelli_1991a,PhysRevD.43.442,Dittrich_2006,Dittrich:2004cb,Giddings:2005id,Gary:2006mw,Giddings:2025xym,Thiemann_2006,Bartlett:2006tzx,Bartlett_2006,Bartlett_2009,Palmer:2013zza,Smithcomm2019,Krumm:2020fws,Castro-Ruiz:2021vnq,PhysRevA.94.012333,Gour_2008,PhysRevD.77.104012,Susskind:2023rxm,Miyadera_2016,Loveridge_2017,Loveridge:2017pcv,Carette:2023wpz,Fewster:2024pur,Riera:2024ehk,EffectiveConstr, EffectiveConstrRel, Bojowald:2009jk, EffectivePoT1, EffectivePoT2, EffectivePoTChaos, SemiclassicalLie,QDSR,AlgebraicPoT,AlgebraicClocks,delaHamette:2021oex,Hoehn:2023ehz,Trinity,TrinityRel,Chataignier:2024eil,Hohn:2018toe,Hohn:2018iwn,AliAhmad:2021adn,DeVuyst:2024pop,DeVuyst:2024uvd,Araujo-Regado:2025ejs,Hoehn:2021flk,Carrozza:2024smc,Giacomini:2021gei,Castro-Ruiz:2019nnl,delaHamette:2021piz,Vanrietvelde:2018dit,Vanrietvelde:2018pgb,Suleymanov:2023wio,Hoehn:2023axh,Suleymanov:2025nrr,Giacomini:2017zju,PhysRevLett.123.090404,delaHamette:2020dyi,Streiter2021,Ballesteros:2020lgl,Cepollaro:2024rss,Kabel:2024lzr,delaHamette:2021iwx,Kabel:2023jve,Mikusch:2021kro,AliAhmad:2024wja,AliAhmad:2024vdw}. They embody the idea that all reference frames are associated with physical systems and ultimately subjected to the universality of quantum theory. They  are the tool for describing quantum physics ``from the inside'', i.e.\ in purely relational terms. Like external frames, QRFs must thus transform non-trivially under some pertinent symmetry group, the key difference being that they are now some suitable subsystem of a given composite physical system of interest. 

In the gravitational case, this means that they must transform under the group of bulk diffeomorphisms \cite{Goeller:2022rsx,Carrozza:2022xut,Kabel:2024lzr}, while in gauge theories, they are some suitable kinematical degrees of freedom that transform under the pertinent gauge group \cite{Carrozza:2021gju,Araujo-Regado:2024dpr,Araujo-Regado:2025ejs}. In both gauge theory and gravity, QRFs associated with gauge symmetries thus naturally arise. However, this idea applies more generally, including to symmetries that are not gauge: two agents in separate laboratories who wish to communicate information, yet have never met and thus do not share an external lab frame, may wish to resort to either encoding reference frame information in a communicated subsystem, or to encode information relationally in an external lab frame-independent manner \cite{Bartlett:2006tzx,Bartlett_2009,Smithcomm2019}. Similarly, there are thought experiments in the foundations of quantum theory where an external lab frame may not be accessible or shared between agents \cite{Giacomini:2017zju,angeloPhysicsQuantumReference2011a,Krumm:2020fws,Castro-Ruiz:2021vnq,Baumann:2019fbd}.

QRFs thus have wide-ranging applicability and may be viewed as a universal tool set for dealing with symmetries in quantum systems. As a consequence of applications in differing contexts with different meanings of symmetries, multiple approaches to QRFs have emerged over time. These differ mostly in how symmetries are implemented. On the somewhat more operational side, where symmetries are not treated as gauge, there are the \emph{quantum information approach} \cite{Bartlett:2006tzx,Bartlett_2006,Bartlett_2009,Palmer:2013zza,Smithcomm2019,Krumm:2020fws,Castro-Ruiz:2021vnq,PhysRevA.94.012333,Gour_2008,PhysRevD.77.104012} (going back to \cite{Aharonov:1967zza,Aharonov:1967zz,Aharonov:1984}) and the \emph{operational approach} \cite{Miyadera_2016,Loveridge_2017,Loveridge:2017pcv,Carette:2023wpz,Fewster:2024pur,Riera:2024ehk}, which in general are inequivalent. On the side that treats symmetries as gauge, there are the \emph{effective semiclassical approach} \cite{EffectiveEq1, EffectiveEq2,EffectiveConstr, EffectiveConstrRel, Bojowald:2009jk, EffectivePoT1, EffectivePoT2, EffectivePoTChaos, SemiclassicalLie}, the  \emph{algebraic approach} \cite{QDSR,AlgebraicPoT,AlgebraicClocks}, and the \emph{perspective-neutral} framework \cite{delaHamette:2021oex,Hoehn:2023ehz,Trinity,TrinityRel,Chataignier:2024eil,Hohn:2018toe,Hohn:2018iwn,AliAhmad:2021adn,Hoehn:2021flk,Carrozza:2024smc,Giacomini:2021gei,Castro-Ruiz:2019nnl,delaHamette:2021piz,Vanrietvelde:2018dit,Vanrietvelde:2018pgb,Suleymanov:2023wio,DeVuyst:2024pop,DeVuyst:2024uvd,Hoehn:2023axh,Araujo-Regado:2025ejs,Suleymanov:2025nrr}, which will be the focus of this manuscript. Roughly somewhere in-between the two sides one may place the \emph{perspectival approach} \cite{Giacomini:2017zju,PhysRevLett.123.090404,delaHamette:2020dyi,Streiter2021,Ballesteros:2020lgl,Cepollaro:2024rss,Kabel:2024lzr,delaHamette:2021iwx,Mikusch:2021kro,Kabel:2023jve} from quantum foundations, which does not in general treat symmetries as gauge, as well as the recent approach based on crossed product von Neumann algebras \cite{AliAhmad:2024wja,AliAhmad:2024vdw}, which  constrains observable algebras but not states.

A core feature of using external reference frames in standard textbook treatments is \emph{covariance}: we imagine that there are many distinct choices of external frame between which we can transform by applying a suitable unitary transformation to the quantum system of interest, such as a Lorentz transformation in quantum field theory in Minkowski space, and this should not change the physics but merely the description of it. Covariance, however, is not a distinguishing feature of external frames. In any sufficiently complex quantum system, there will always be many choices of internal subsystems that justifiably may serve the role of an internal QRF. Naturally, the question arises as to how one may change internal perspective by going from one internal QRF perspective on the composite system to that of another, i.e.\ what are the transformations between QRFs? 

\emph{A priori} this appears like a rather complicated question, given that QRFs are physical subsystems that may interact and entangle with other systems and that may be in complicated superpositions of their orientations. However, the last 15 years have seen several waves of work deriving QRF transformations in various circumstances and exploring the notion of covariance in a genuinely quantum regime, where the frame itself is subject to fluctuations. Despite the ideas underlying QRFs dating back to at least the 1960s \cite{Bergmann:1960wb,Bergmann:1961wa,Bergmann:1961zz,Aharonov:1967zza,Aharonov:1967zz}, the first QRF transformations  only appeared in 2010 at semiclassical level in the effective approach (for clocks as temporal QRFs) \cite{EffectivePoT1,EffectivePoT2,EffectivePoTChaos}, while the first QRF transformations in the quantum information approach were formulated in 2013 \cite{Palmer:2013zza} (and extended later in \cite{Castro-Ruiz:2021vnq}). The seminal work that first formulated QRF transformations in the full quantum theory along with the question of the covariance of the laws of physics is the one by Giacomini et al.~within the perspectival approach~\cite{Giacomini:2017zju}, which ignited the subsequent wave of works on QRF covariance. The perspective-neutral framework, which was inspired by the older effective approach and the insights of \cite{Giacomini:2017zju}, followed soon after and has QRF transformations built in systematically from the start. QRF transformations within the operational approach were then formulated in \cite{Carette:2023wpz}.

The study of QRF covariance is thus relatively young but has already led to some striking discoveries over the past years. On the foundational side, studying the covariance has led to many indications that the physics of a subsystem depends on the QRF one chooses to describe it. A key insight is  \emph{subsystem relativity} \cite{AliAhmad:2021adn,Hoehn:2023ehz,delaHamette:2021oex,Castro-Ruiz:2021vnq,DeVuyst:2024pop,DeVuyst:2024uvd,Araujo-Regado:2025ejs}: different choices of QRF decompose the total system in different ways into gauge-invariant subsystems (associated with the same kinematical subsystems). That is, the definition of a subsystem is QRF dependent, and so properties such as the purity of a subsystem state, thermal equilibrium, temperature, entropy, variances, etc., turn out to be QRF dependent as well \cite{Giacomini:2017zju,AliAhmad:2021adn,delaHamette:2021oex,Castro-Ruiz:2021vnq,Hoehn:2023ehz,DeVuyst:2024pop,DeVuyst:2024uvd,Suleymanov:2025nrr,Cepollaro:2024rss,Araujo-Regado:2025ejs}.  

In particular, these insights carry over to high-energy physics. Recent work on subregion observable algebras in perturbative quantum gravity \cite{Witten:2021unn,Chandrasekaran:2022cip,Jensen:2023yxy,Kudler-Flam:2023qfl}, showed that, remarkably, explicitly taking an observer into account, leads to an intrinsic regularization of the entanglement entropy associated to a subregion containing quantized matter fields and graviton degrees of freedom (see \cite{Faulkner:2024gst,AliAhmad:2023etg,KirklinGSL,Witten:2023xze,Klinger:2023auu,Aguilar-Gutierrez:2023odp,Gomez:2023wrq}
for a flurry of follow-up work). Crucially, it turns out that this explicit inclusion of an observer is equivalent to including a QRF \cite{DeVuyst:2024pop,DeVuyst:2024uvd,Ali:2024jkx,Fewster:2024pur}, further leading to the aforementioned phenomenon of a QRF- or observer-dependent entanglement entropy \cite{DeVuyst:2024pop,DeVuyst:2024uvd} as a consequence of subsystem relativity. The same QRF-dependence of entanglement entropies has also been exhibited in lattice gauge theories \cite{Araujo-Regado:2025ejs}.

Let us now turn to the key topic that we will be concerned with in this article. With seven \emph{a priori} distinct QRF approaches, the need for some order among them arises. What are the relations and the precise differences between them? And what are the precise scope and range of applicability of each approach? Given the different meanings ascribed to symmetries and QRFs in different physical contexts, it is clear that there can be no `one-fits-all' approach to QRFs and the inequivalence between some approaches is physically appropriate. For example, some key consequences of the different ways of implementing symmetries between the quantum information, operational and perspective-neutral approaches have been discussed in \cite[Sect.~II]{Hoehn:2023ehz} (see also \cite{Krumm:2020fws}) and will be further illustrated in \cite{Castro-Ruiz:2025yvi}. The perspectival and perspective-neutral approaches have been shown to be equivalent for the special case of \emph{ideal} QRFs \cite{Vanrietvelde:2018pgb,Krumm:2020fws,Hoehn:2021flk,delaHamette:2021oex}, characterized by featuring perfectly distinguishable orientation states, but are inequivalent for non-ideal QRFs with fuzzy orientation states \cite{delaHamette:2021oex,Trinity}. This is because the perspectival approach works in the case of fuzzy frames with Hilbert spaces that are bigger than permitted by gauge invariance \cite{delaHamette:2021oex,delaHamette:2020dyi}. However, one would hope that approaches treating symmetries in the same way (yet use different formalisms) ultimately agree on the description of physics. As noted, the effective, algebraic and perspective-neutral approaches share the feature of treating the external frame transformations as gauge and their relation is precisely the question we will set out to answer in this work. 

In canonical formulations, treating symmetries as gauge means eliminating unobservable degrees of freedom by imposing quantum constraints \emph{\`a la} Dirac. Gauge symmetries manifest themselves as one or more constraints $C_i = 0$ at the classical level, which then turn into Wheeler--DeWitt type equations at the quantum level: $\hat{C}_i \ket{\psi} = 0$. In addition, observables living on the space of solutions to these constraints are invariant and must commute with the constraints. The three approaches we will henceforth be interested in implement this program in different ways. 
\begin{enumerate}
    \item The \emph{perspective-neutral (PN) framework} \cite{delaHamette:2021oex,Hoehn:2023ehz,Trinity,TrinityRel,Chataignier:2024eil,Hohn:2018toe,Hohn:2018iwn,AliAhmad:2021adn,Hoehn:2021flk,Carrozza:2024smc,Giacomini:2021gei,Castro-Ruiz:2019nnl,delaHamette:2021piz,Vanrietvelde:2018dit,Vanrietvelde:2018pgb,Suleymanov:2023wio,DeVuyst:2024pop,DeVuyst:2024uvd,Hoehn:2023axh,Araujo-Regado:2025ejs,Suleymanov:2025nrr} directly implements constraint quantization at the Hilbert space level. The external, frame-independent, internally accessible physics is encoded in the physical Hilbert space built using solutions to the quantum constraints. This constitutes an internal QRF perspective-neutral structure that encodes and links all the internal QRF perspectives, much like a manifold links coordinate descriptions. Internal QRF perspectives are then created either by constructing relational observables or reduced Hilbert spaces, with explicit reduction maps connecting them to the perspective-neutral view.\footnote{The technique used to impose the constraint is that of \emph{Refined Algebraic Quantization }\cite{Giulini:1998kf,Giulini:1998rk,Landsman:1993xe,AHiguchi_1991,Marolf:2000iq}, but the PN approach goes further by also linking it to the various perspectival spaces.}
    \item The \emph{effective semiclassical approach} \cite{EffectiveEq1, EffectiveEq2,EffectiveConstr, EffectiveConstrRel, Bojowald:2009jk, EffectivePoT1, EffectivePoT2, EffectivePoTChaos, SemiclassicalLie}
is based on canonical quantum phase space methods, where quantum states are replaced by the expectation values, variances, and higher-order moments that they assign to some 
complete collection of kinematical observables. This quantum phase space inherits a Poisson structure from expectation values of the commutator. One then employs the semiclassical approximation to truncate the degrees of freedom at some order in $\hbar$, and applies essentially classical methods for treating gauge freedom associated with frame transformations.
    \item The  \emph{algebraic approach} \cite{QDSR,AlgebraicPoT,AlgebraicClocks} is essentially an extension of the effective approach beyond the semiclassical approximation.\footnote{The distinct operational approach \cite{Miyadera_2016,Loveridge_2017,Loveridge:2017pcv,Carette:2023wpz,Fewster:2024pur,Riera:2024ehk}, as well as the von Neumann algebra based approach in \cite{AliAhmad:2024wja,AliAhmad:2024vdw} may likewise be called algebraic approaches with their starting point also being observable algebras. The key difference is that these approaches do not implement constraints on states in contrast to here. When we refer to the algebraic approach, we strictly mean the constrained-based one formulated in \cite{QDSR,AlgebraicPoT,AlgebraicClocks}, as it was previously called that way.} As in the effective approach, one focuses on the algebra generated by a complete set of kinematical observables, with quantum states now provided by linear functionals on this algebra, characterized by the values they assign to kinematical algebra elements. Constraints are solved by placing restrictions on these values, while the freedom to choose a QRF perspective takes the form of gauge orbits on the space of quantum states.
 \end{enumerate}

Our key result is the equivalence of these three approaches in the
setting in which they can currently be jointly applied. This is essentially the case of \emph{ideal} QRFs associated with a one-dimensional gauge group, corresponding to a single constraint of the form
\begin{equation}\label{eq:con1}
    \hat{C} = \sum_i \hat{p}_i + \hat{G}_S \,, 
\end{equation}
consisting of mutually commuting generators of the observed system $\hat{G}_S$ and one or more reference systems $\hat{p}_i$, describing their transformation under external frame transformations. In line with their idealness, the spectrum of the $\hat{p}_i$'s is the full real line $\sigma_i := \text{Spec}(\hat{p}_i) = \mathds{R}$. While our explicit argument is thus presented for a reference frame with a non-degenerate spectrum, we argue in detail that the equivalence still holds for reference frames with degenerate generators, degeneracy-sector-by-degeneracy-sector. The PN framework applies to general non-ideal reference frames (and general unimodular groups \cite{delaHamette:2021oex}), the configuration measurements of which need not be projective. In this case, the frame configurations correspond to a positive operator-valued measure  rather than a self-adjoint orientation operator. Because the algebraic and the effective semiclassical approaches in their current formulation explicitly rely on self-adjoint orientation observables, they require a non-trivial extension in order to cover non-ideal reference frames.  On the other hand, an advantage of the algebraic and effective approaches is that their current formulation naturally encompasses the case when the and the observed system $S$ and QRFs $i$  couple, leading to `interaction' terms in~\eqref{eq:con1}. A direct comparison of the three approaches for non-ideal frames and `interactions' thus requires additional work that we will comment on towards the end of this article.

A powerful feature of the PN framework is the availability of explicit transformations of both states and observables from one reference frame into another. Semiclassically, this is also true for expectation values of observables in the effective approach, where frame changes are formulated using gauge transformations \cite{EffectivePoT1,EffectivePoT2,EffectivePoTChaos}. Like its effective counterpart, the algebraic approach possesses a well-defined notion of frame-choice-related gauge transformations. However, up until now, the focus within this approach has been on fixing those freedoms to extract physically meaningful relations rather than on exploiting gauge flows to transform between frames. Here we use the equivalence between the three approaches and the QRF transformation machinery from the PN framework to further study frame transformations in algebraic and effective settings, finding that, at semiclassical order, the older effective frame changes \cite{EffectivePoT1,EffectivePoT2,EffectivePoTChaos} encode the same information as those of the PN framework. Since we formulate this for ideal QRFs, this also means that, at semiclassical order, the effective QRF transformations agree with those of the perspectival approach \cite{Giacomini:2017zju,delaHamette:2020dyi}, which, as noted above, agrees with the PN framework in that case. For non-ideal QRFs, however, we anticipate this agreement to break down due to the insistence of the effective approach on constraint implementation like the PN framework.

Altogether, our analysis therefore brings some clarity into the relation between four of the seven approaches to QRFs. The equivalence of the three constraint-based approaches and the fact that they have advantages in different settings can also be viewed as expanding the arsenal of mathematical tools for that case. The PN framework previously linked three \emph{a priori} distinct formulations of relational physics, namely that of relational observables on the gauge-invariant Hilbert space, the Page-Wootters formalism \cite{1984IJTP...23..701W,PhysRevD.27.2885} and a certain quantum deparametrization, altogether dubbed the `trinity' \cite{Trinity,TrinityRel,Chataignier:2024eil,delaHamette:2021oex}. Adding the effective and algebraic formulations to this, the trinity thus turns into a `quinternity', at least for ideal QRFs.

As an immediate physical application of the QRF transformations, we further explore the frame dependence of uncertainties and covariances. This effect was recently studied within the PN framework in~\cite{Suleymanov:2025nrr} in the example of a translationally invariant system, where it was found that  under a QRF transformation, the uncertainty of a system observable as computed in one perspective gains additional terms proportional to the covariance with other observables within the original frame.\footnote{Uncertainties (but not covariances) in the setting of the operational approach to QRFs  were also recently explored in \cite{Riera:2024ehk}.} We note that the effective semiclassical approach is especially well-suited for looking at frame dependence of uncertainties and covariances, which are the basic quantities of interest within this approach. In fact, some frame covariance effects of uncertainties and fluctuations already appeared in the earlier work \cite{EffectivePoT1,EffectivePoT2,EffectivePoTChaos}, though not in quite as general a form.  Using a new formulation of QRF transformations derived for the effective approach, we obtain general frame transformations for uncertainties and covariances, which reproduce the findings of~\cite{Suleymanov:2025nrr} when applied to the same translation-invariant model.

Lastly, we reconcile one further aspect that \emph{a priori} appears quite differently in the three approaches: broadly speaking, within the PN framework, the relational nature of physically meaningful information manifests itself through the existence of multiple maps from physical states to reduced Hilbert spaces complemented by maps from observed system's degrees of freedom to invariant observables---these maps are labeled by the choice of a particular reference system and of its configuration. Within the algebraic and the effective semiclassical approaches to QRFs one focuses entirely on the values assigned by quantum states to \emph{a priori} non-invariant observables and relationality appears in the form of continuous gauge flows on those values that need to be fixed by specifying the values assigned by states to a reference observable. We explicitly reconcile these two manifestations of relationality by constructing a direct analogue of algebraic states within the PN framework.\\
\\
\noindent Let us summarize the structure of the manuscript: 

Sect.~\ref{sect:approaches} reviews the main features of the three approaches that are important to our discussion, emphasizing the close relation between the algebraic approach and the effective semiclassical treatment. As an illustration, we apply all three formalisms to a one-dimensional translationally invariant system.  Sect.~\ref{sect:equivalence} demonstrates the equivalence of the three approaches for ideal reference frames that are non-degenerate or have fixed degeneracy sectors. We do this by explicitly linking the Page--Wootters portion of the PN framework with the algebraic approach, which then also extends to the semiclassical effective construction. Sect.~\ref{sect:switchingIdeal} shows the relation between ideal QRF transformations in the PN framework and the older effective approach and uses this to derive them for the algebraic approach and extend them for the effective formalism.
The laws for transforming variances and moments of the effective treatment are then applied to the discussion of frame dependence of uncertainties and variances in Sect.~\ref{sect:rel}. In Sect.~\ref{sect:generalGaugeFreedom} we describe the way in which reference-frame-related gauge freedom within the PN framework can be shifted onto the generalized matrix elements of non-invariant operators in a way that parallels the gauge dependence of quantum states in the algebraic approach.
We summarize our findings and offer an outlook for future work in Sect.~\ref{sect:discussion}. The interested reader will find details of some of our calculations as well as additional examples in the appendix.

\section{Overview of the three approaches}
\label{sect:approaches}

The approaches discussed in this section will be used to treat quantum systems with a single constraint. We assume that the kinematical degrees of freedom of the system are represented on the kinematical Hilbert space $\mathcal{H}$, and that the constraint that generates the frame transformation has been promoted to a self-adjoint operator $\hat{C}$ on $\mathcal{H}$. All of the approaches attempt to implement the Dirac constraint quantization condition
\cite{dirac1967lectures,Henneaux:1992ig}
\begin{align} \label{eq:DiracQConstraint}
\hat{C} |\psi \rangle =0 \ .
\end{align}
Algebraically, the kinematics of the system is captured by an algebra of kinematical operators $\mathcal{A}$. We note that the collection of all operators on $\mathcal{H}$\ does not in general form an algebra, since there may not be a dense subset of $\mathcal{H}$\ on which all linear operators are defined. Therefore, $\mathcal{A}$\ will typically be some subset of kinematical operators, sufficiently large to resolve the classical degrees of freedom, containing the identity $\mathds{1}$\ and closed with respect to the $*$-involution that sends an operator to its adjoint. For example, the set of \emph{bounded} operators $\mathcal{B}(\mathcal{H})$ is of course an algebra. However, we may in some cases be interested in unbounded kinematical operators, such as momentum and position observables. In this case, one may also consider the algebra of polynomials formed by Heisenberg pairs and the identity.

The Hilbert space-based reference frame constructions \cite{Trinity,TrinityRel,Chataignier:2024eil,Hohn:2018iwn,Hohn:2018toe,Vanrietvelde:2018pgb,DeVuyst:2024uvd} explicitly assume that the kinematical Hilbert space factorizes into the states associated with a chosen reference subsystem $R$ and those associated with ``the rest of the system'' $S$, $\mathcal{H} = \mathcal{H}_R \otimes \mathcal{H}_S$.\footnote{See references \cite{Smith2019quantizingtime,EffectivePoTChaos,Hoehn:2023axh,Castro-Ruiz:2019nnl} for when there is also interaction between the system and clock.} On the algebraic side, we assume $\mathcal{A}$\ is algebraically generated by a pair of commuting $*$--subalgebras $\mathcal{A}_R$ (containing operators of the form $\hat{a}_R = \hat{\tilde{a}}_R\otimes \mathds{1}_S$) and $\mathcal{A}_S$ (containing operators of the form $\hat{a}_S = \mathds{1}_R\otimes\hat{\tilde{a}}_S $) with \cite{QDSR,AlgebraicPoT,AlgebraicClocks} 
\begin{itemize}
\item $[\mathcal{A}_R, \mathcal{A}_S] = 0$; 
\item $\mathcal{A}_R\cap \mathcal{A}_S = \mathbb{C} \mathds{1} $;
\item $\hat{a}\in \mathcal{A}$\ $\Rightarrow$ $\hat{a}=\sum_{i=1}^N \hat{a}_{R\,i} \hat{a}_{S\, i}$\ for some  $\hat{a}_{R\, i } \in \mathcal{A}_R$,  $\hat{a}_{S\, i } \in \mathcal{A}_S$\ and $N<\infty$.
\end{itemize}
So that $\mathcal{A}=\left\langle \mathcal{A}_R \cup\mathcal{A}_S \right\rangle \cong \mathcal{A}_R \otimes_\mathbb{C} \mathcal{A}_S$, where ``$\langle \text{subset} \rangle$'' denotes the (complex) algebra generated by an algebraic subset (subject to some closure, where appropriate).
Throughout the rest of the manuscript, we will explicitly write out the tensor product structure of Hilbert space states. However, for the purpose of compactness, we will use the same symbol to identify an operator acting on a factor space $\hat{\tilde{a}}_S$\ and its extension to the full kinematical Hilbert space $\hat{a}_S=\mathds{1}_R \otimes \hat{\tilde{a}}_S$, writing $\hat{a}_S$\ to denote both. Similarly, where this does not create ambiguities, we will interchangeably treat $\mathcal{A}_S$\ and $\mathcal{A}_R$\ as subalgebras and as tensor product factor algebras of $\mathcal{A}$.

Currently, a detailed comparison between the three approaches to QRFs is only possible when the reference frame is provided by a one-component canonical system, with extension of the algebraic approach to other scenarios still to be worked out. Therefore, much of our discussion in subsequent sections will focus on the case where $\mathcal{A}_R$\ is the algebra generated by complex polynomials of a canonical pair of observables, while the only assumption we will make about $\mathcal{A}_S$\ is that it contains the generator of system transformations $\hat{G}_S$ (see equation~\eqref{eq:Constraint} for the form of the constraint).

\subsection{The perspective--neutral framework}
\label{sect:trinity}

The basic idea of the perspective-neutral (PN) framework is to implement the constraint(s) on both the Hilbert space and observable algebra level to construct a gauge-invariant stand-alone quantum theory from the original kinematical data provided at the outset. The group generated by the constraint(s) is interpreted as one corresponding to \emph{external} reference frame transformations, while the kinematical data is interpreted as the degrees of freedom distinguishable relative to such an external (possibly) fictitious frame \cite[Sect.~II]{Hoehn:2023ehz}. For example, when $\hat C$ generates translations in space, then these can be viewed as translations of a background position frame. When $\hat C$ generates time translations, then these constitute evolution relative to some external time (e.g., in gravity some unphysical coordinate time). Imposing the constraints thereby means wiping out any such external (possibly fictitious) reference structures. 

The hurdle one must then jump over is to identify a suitable \emph{internal} reference frame relative to which the remaining degrees of freedom may be described. Such internal reference frames, transforming under the same group generated by the constraint(s), are what have become known as quantum reference frames (QRFs). They are the tool to describe the composite quantum system of interest from ``the inside'' in a purely relational manner. Note that a QRF need not necessarily have a spatiotemporal interpretation. When it does and $\hat C$ generates a spatial symmetry, it is a spatial QRF, e.g.\ a reference position for other subsystems; if it generates a temporal symmetry, such as reparametrizations in Hamiltonian constrained systems, then it is a temporal QRF, usually simply called a (quantum) clock, which keeps track of the evolution of the remaining degrees of freedom. The gauge-invariant theory obtained upon implementing the constraint(s) then corresponds to the data that is internally distinguishable, and gauge invariance is now another term for ``external reference frame independence''. 

In general, there is no preferred choice for an internal QRF and many different choices of subsystem may assume that role. For example, in Hamiltonian constrained systems, the reduced theory does not possess a preferred internal dynamical flow and this leads to a ``multiple choice problem'' in choosing a suitable internal time \cite{doi:10.1142/S0218271811019347,Isham:1992ms,Anderson2017,EffectivePoT1,EffectivePoT2,EffectivePoTChaos}. The point is that this choice of internal QRF is not dictated by the gauge-invariant stand-alone theory; it actually encodes all such choices. This is why the gauge-invariant theory assumes the role of an internal QRF \emph{perspective-neutral} structure that encodes and links all the different QRF choices, akin to how a spacetime manifold links all the possible coordinate frame choices, providing the framework with its name.

In what follows, we will keep the discussion mostly agnostic as to what the precise nature of the QRFs is, except that they will be associated with a one-dimensional translation group. To that end, we will follow the presentation of the PN framework in \cite{Trinity}, which was formulated for Hamiltonian constrained systems and unified three previously distinct approaches to relational dynamics, but applies equally to any translation group QRFs (see \cite{TrinityRel,Chataignier:2024eil,delaHamette:2021oex} for generalizations). Let us now summarize the constructions relevant to our discussion.

\subsubsection{The perspective-neutral Hilbert space}\label{sec:PN_HilbertSpace}

Most approaches to implementing a quantum constraint start by finding the set of solutions to~(\ref{eq:DiracQConstraint}), and we will do this along the lines of \emph{Refined Algebraic Quantization} \cite{Giulini:1998kf,Giulini:1998rk,Landsman:1993xe,AHiguchi_1991,Marolf:2000iq} (or, equivalently, the co-invariant method \cite{Chandrasekaran:2022cip,Held:2024rmg,DeVuyst:2024uvd}), to which the PN framework adds reduced perspectival spaces and explicit maps between them and the space of solutions. Constructing a consistent reduced system along this path requires one to overcome several technical and conceptual issues. For most constraints of interest ``0'' is in the continuous part of their spectrum, which means that solutions to~(\ref{eq:DiracQConstraint}) are typically not normalizable and are properly elements of the dual of the kinematical Hilbert space $\mathcal{H}^*$. One needs to endow the space of solutions with a meaningful \emph{physical inner product} and come up with a way of projecting a sufficiently large number of physically interesting operators to the resulting \emph{physical Hilbert space} \cite{AHiguchi_1991,Landsman:1993xe,Giulini:1998kf,Giulini:1998rk,Marolf:2000iq}.

The methods of \cite{Trinity} apply to the situation where the system is subject to a single  constraint of the form
\begin{align}\label{eq:Constraint}
    \hat{C}=\hat{G}_R+\hat{G}_S \ ,
\end{align}
where $\hat{G}_R=\hat{G}_R^* \in \mathcal{A}_R$\ and $\hat{G}_S=\hat{G}_S^* \in \mathcal{A}_S$, so that the translation group is represented via a unitary tensor product representation on $\mathcal{H}$ (which for Hamiltonian constraints means $R$ and $S$ do not interact). We will assume that $\sigma_R$, the spectrum of $\hat{G}_R$, is purely continuous and non-degenerate.\footnote{The degenerate case is treated in \cite{TrinityRel,DeVuyst:2024uvd}, whilst the discrete case is considered in \cite{Chataignier:2024eil} for periodic clocks.} 

Let us denote the eigenvalues of $\hat{G}_R$ by $\varepsilon$ and those of $\hat{G}_S$ by $\lambda$ (which may be continuous and/or discrete), possibly augmented by another degeneracy label $s_\lambda$ to complete a basis for $\mathcal{H}_S$.
Using these bases on $\mathcal{H}_S$\ and $\mathcal{H}_R$, an arbitrary kinematical state, written as 
\begin{equation}
    \label{eq:kinematicalState}
    \ket{\psi_\text{kin}} = \int_{\sigma_R} \dd{\eps} \SumInt_{\lambda \in\sigma_S} \sum_{s_\lambda}\psi_\text{kin}(\eps,\lambda,s_\lambda) \ket{\eps}_R \ket{\lambda,s_\lambda}_S
\end{equation}
can be formally mapped to a generalized solution state as\footnote{In  our case, where zero is in the continuous part of the spectrum of the constraint operator, $\Pi$\ is not a projection operator, since $\Pi^2$\ is undefined.}
\begin{align}
    \label{eq:projector}
    &|\psi_{\rm phys} \rangle = \Pi |\psi_{\rm kin} \rangle :=\frac{1}{2\pi\hbar} \int_{\mathbb{R}} ds\, e^{is\hat{C}/\hbar} |\psi_{\rm kin} \rangle \\ &=\SumInt_{\lambda\in\sigma_S\cap (-\sigma_R)}\sum_{s_\lambda} \psi_{\rm kin} (-\lambda, \lambda,s_\lambda) |-\lambda \rangle_R | \lambda,s_\lambda \rangle_S \ . \nonumber
\end{align}
These solution states are not normalizable with respect to the kinematical inner product. We use the  coherent group averaging operator to define the physical inner product on the space of solutions as
\begin{align}
\label{eq:PhysicalIP}
&\left( \psi_{\rm phys} | \phi_{\rm phys}\right) := \langle \psi_{\rm kin}| \Pi | \phi_{\rm kin} \rangle \\
&=  \SumInt_{\lambda\in\sigma_S\cap (-\sigma_R)}\sum_{s_\lambda} \bar{\psi}_{\rm kin} (-\lambda, \lambda,s_\lambda)\, \phi_{\rm kin} (-\lambda, \lambda,s_\lambda) \ . \nonumber
\end{align}
Here, $\langle\cdot|\cdot\rangle$ denotes the kinematical inner product on $\mathcal{H}$. In what follows, we assume the inner product defined by~\eqref{eq:PhysicalIP} to be positive semi-definite and to converge (see \cite{Higuchi:1991tk,Higuchi:1991tm,Giulini:1998kf,Giulini:1998rk,Marolf:2008hg,Kaplan:2024xyk} for some discussion on this). The space of generalized solutions to the constraint, completed in the norm defined by this inner product, then constitutes the gauge-invariant physical Hilbert space, henceforth denoted as $\mathcal{H}_{\rm phys}$. We will also call it the \emph{perspective-neutral} Hilbert space, given that it will encode the different QRF choices (e.g., $S$ may contain other subsystems that may serve as an internal reference).

\subsubsection{QRF orientations}\label{sssec_orientations}

Let us now turn to modelling the QRF $R$, especially its orientations. A key building block in \cite{Trinity} is the construction of a generalized orientation measurement on its Hilbert space $\mathcal{H}_R$, via a one-parameter family of generalized orientation states
\begin{align}
    |\rho \rangle = \int_{\sigma_R} \dd{\eps} e^{i g(\eps)} e^{-i\rho\eps/\hbar} |\eps \rangle \ ,
    \label{eq:clock states}
\end{align}
where $g(\eps)$\ is an arbitrary real-valued function which we will set to zero from here onward. $\rho$ labels the orientations of $R$. For a spatial QRF, this may be a position and for a clock, it will be its reading. These generalized states resolve the identity
\begin{align}
    \frac{1}{2 \pi \hbar} \int_\mathbb{R} \dd{\rho} \dyad{\rho} = \mathds{1}_R \ ,
    \label{eq:idResolution}
\end{align}
and are covariant with respect to reorientations (time evolution for clocks) of the reference
\begin{align}
    e^{-i\rho'\hat{G}_R/\hbar} |\rho \rangle =|\rho+\rho' \rangle \ .
    \label{eq:covariance}
\end{align}
Such a collection of states constitutes a covariant positive operator-valued measure (POVM) \cite{Holevo2011,Busch1995,Busch2016,Trinity}. For each measurable subset $X\subset \mathbb{R}$, these states define a self-adjoint positive effect operator $\hat{E}(X) = \frac{1}{2 \pi\hbar} \int_X \dd{\rho} |\rho \rangle \langle \rho |$, whose expectation values are the probabilities to find the QRF's orientation in $X$.

Unless $\sigma_R=\mathbb{R}$, the above generalized states are not orthogonal $\langle \rho | \rho' \rangle \neq 2\pi \delta (\rho-\rho')$\ and the orientation operator (the POVM's first moment) 
\begin{equation}
    \hat{R} := \frac{1}{2 \pi \hbar} \int_\mathbb{R} \dd{\rho} \rho \dyad{\rho} 
    \label{eq:time operator}
\end{equation}
is not self-adjoint. Nevertheless, the effect operators $\hat{E}(X)$ still provide a well-defined, albeit not perfectly distinguishable probability distribution over the QRF's orientations. The case $\sigma_R=\mathbb{R}$ corresponds to an \emph{ideal} QRF, characterized by perfectly distinguishable orientation states; only in this case are the orientation states also eigenstates of the orientation operator, $\hat{R}\ket{\rho}=\rho\ket{\rho}$. Most of this work will be concerned with such QRFs.

\subsubsection{Relational observables}

The orientation states allow us to construct relational Dirac observables---operators that commute with the constraint, and encode the physics described with respect to the QRF $R$. For every $\hat{f}_S \in \mathcal{A}_S$\ and for every $\rho \in \mathbb{R}$ we define an operator on $\mathcal{H}_\text{kin}$
via an incoherent group average ($G$-twirl) over the group generated by $\hat{C}$
\begin{align}
    \hat{O}^\rho_R(\hat{f}_S) &:=  \frac{1}{2\pi\hbar} \int_{\mathbb{R}} \dd{\rho'} e^{-i\rho'\hat{C}/\hbar}\left(|\rho \rangle \langle \rho|\otimes \mathds{1}_S \right)\, \hat{f}_S  e^{i\rho' \hat{C}/\hbar} \ .
    \label{eq:diracObservable}
\end{align}
When the reference frame is ideal, these take a simpler form owing to the fact that the orientation states are eigenstates of $\hat{R}$\footnote{For compactness, we omit the identity operator where it acts on the entire kinematical Hilbert space, such as writing $(\hat{R} - \rho)$\ in place of $(\hat{R} - \rho\mathds{1})$\ in~\eqref{eq:idealDirac}.}
\begin{equation}
    \hat{O}^\rho_R(\hat{f}_S) = e^{-i (\hat{R} - \rho) \hat{G}_S /\hbar} \hat{f}_S e^{i (\hat{R} - \rho) \hat{G}_S /\hbar} \ .
    \label{eq:idealDirac}
\end{equation}
According to \cite[Thm.~1]{Trinity}, $\hat{O}^\rho_R(\hat{f}_S)$\ is a Dirac observable $[\hat{O}^\rho_R(\hat{f}_S), \hat{C}]=\hat{0}$, interpreted as the relational observable measuring $f_S$, conditional on the QRF $R$ being in orientation $\rho$. 
Note that, at this stage, these are still operators on the kinematical Hilbert space. The full gauge-invariant algebra on the kinematical Hilbert space can be written as \cite[App.~A.2]{DeVuyst:2024uvd}
\begin{equation}
    \mathcal{A}_\text{inv} := \mathcal{A}^{\hat{C}} =  \Big\langle\hat{O}^\rho_R(\hat{f}_S), \hat{G}_R \Big\vert \hat{f}_S \in \mathcal{A}_S; \rho \in \mathbb{R}\Big\rangle.
    \label{eq:invAlgebra}
\end{equation}
$\mathcal{A}_{\rm inv}$ is thus generated by the relational observables describing $S$ relative to $R$, as well as the reorientations of $R$ itself (generated by $\hat{G}_R$).
As discussed later in Sect.~\ref{sect:PNswitching}, different QRFs define isomorphic gauge-invariant algebras.

Henceforth, we will be interested in the representation $\mathcal{A}_{\rm phys}$ of the relational observable algebra $\mathcal{A}_{\rm inv}$ on the perspective-neutral Hilbert space $\mathcal{H}_{\rm phys}$. Note that on $\mathcal{H}_{\rm phys}$, $\hat{G}_R=-\hat{G}_S$ and the latter is contained among the relational observables, as $\hat{O}^\rho_R(\hat{G}_S)=\hat{G}_S$ (provided $\hat{G}_S\in\mathcal{A}_S$, as will be the case in most of this work). Thus, in contrast to $\mathcal{A}_{\rm inv}$, the algebra $\mathcal{A}_{\rm phys}$ is entirely generated by the relational observables. In our setting (see \cite{DeVuyst:2024uvd} for a comparison with the co-invariant method), the representation of these relational observables on $\mathcal{H}_{\rm phys}$ is given by multiplying the conditional part of the relational observable with the coherent group averaging operator,
\begin{equation}\label{eq:relobsphys}
    \hat{\mathcal{O}}^\rho_R(\hat{f}_S):=\Pi\left((\dyad{\rho}\otimes \mathds{1}_S) \hat{f}_S\right)\,,
\end{equation}
because $\hat{O}^\rho_R(\hat{f}_S)\Pi = \Pi\left((\dyad{\rho}\otimes \mathds{1}_S)\hat{f}_S\right)\Pi=\hat{\mathcal{O}}^\rho_R(\hat{f}_S)\,\Pi$ and $\Pi$ maps to $\mathcal{H}_{\rm phys}$. 

We note that the same would be true had we instead chosen another QRF $R'$ inside the composite system $RS$, provided $\hat{C}$ also takes the form \eqref{eq:Constraint} with respect to that choice of QRF. In other words, the same operators in $\mathcal{A}_{\rm phys}$ can be interpreted as relational observables both relative to $R$, as well as to $R'$. In this sense, we may also view $\mathcal{A}_{\rm phys}$ abstractly as a perspective-neutral algebra. The way in which we interpret its elements as relational observables depends on the choice of QRF, see \cite{Hoehn:2023ehz,delaHamette:2021oex} for more discussion of this.

As an aside, algebras of this broad type in \eqref{eq:invAlgebra} have recently gained interest within the algebraic approach to defining subregions in perturbative quantum gravity \cite{Chandrasekaran:2022cip,Jensen:2023yxy,DeVuyst:2024pop,DeVuyst:2024uvd,Faulkner:2024gst,AliAhmad:2023etg,KirklinGSL,Witten:2021unn,Witten:2023xze,Klinger:2023auu,Kudler-Flam:2023qfl,Aguilar-Gutierrez:2023odp,Gomez:2023wrq}. In those cases, $\hat{G}_S \notin \mathcal{A}_S$ (due to $\hat{G}_S$\ generating an outer automorphism), and $\mathcal{A}_S$ forms a type III$_1$ von Neumann algebra\footnote{A von Neumann algebra is an operator algebra which is equal to its double commutant $\mathcal{A} = \mathcal{A}''$.} for the regional QFT degrees of freedom (including gravitons), thus admitting no trace and entropies, while $\mathcal{A}_{\rm inv}$ is of type II, therefore admitting a trace and entropies. We refer the interested reader to \cite{DeVuyst:2024pop, DeVuyst:2024uvd}, where the connection with QRFs is explained in  detail.

\subsubsection{Page--Wootters construction and ``jumping'' into a QRF perspective}
\label{sect:PW}

What does it mean to ``jump'' into $R$'s internal perspective on $S$? There are broadly two equivalent ways of doing so \cite{Trinity,TrinityRel,delaHamette:2021oex,Hoehn:2023ehz}: the first is by remaining purely at the gauge-invariant level and choosing $R$ as the reference for building relational observables that may then be evaluated in physical states $\ket{\psi_{\rm phys}}\in\mathcal{H}_{\rm phys}$ via the manifestly gauge-invariant physical inner product $(\cdot|\cdot)$ in \eqref{eq:PhysicalIP}. For example, relational time evolution, in the case that $\hat C$ is a Hamiltonian constraint, would then be encoded in letting the parameter $\rho$ in $\hat{\mathcal{O}}^\rho_R(\hat{f}_S)$ run and describe how $\hat{f}_S$ evolves relative to $R$. Note that this is not in contradiction with $\mathcal{H}_{\rm phys}$ and $\mathcal{A}_{\rm phys}$ constituting a perspective-neutral structure; they simply encode all possible choices of $R$ and the relational description relative to any QRF can be formulated gauge-invariantly.

The second way of ``jumping'' into $R$'s perspective is by simply gauge fixing its orientation. This amounts to aligning the internal and external reference frames. In this way, only the $S$ degrees of freedom remain and the resulting description can be interpreted as a description of $S$ relative to $R$. For example, for spatial translations, where $R$ may be a reference particle, one may simply define its position as the origin \cite{Vanrietvelde:2018pgb,Vanrietvelde:2018dit}. It turns out that this gauge fixing is equivalent to (or rather is a generalization of) the Page--Wootters formalism \cite{Page:1983uc,1984IJTP...23..701W} and unitarily equivalent to the manifestly gauge-invariant description summarized above \cite{Trinity} (see also \cite{TrinityRel,delaHamette:2021oex,Chataignier:2024eil} for generalizations).

Let us now summarize the (generalized) Page-Wootters formalism for gauge fixing via conditioning on a frame orientation \cite{Trinity}. This defines a unitary reduction map $\mathcal{R}_R(\rho):\mathcal{H}_{\rm phys}\to \mathcal{H}_{S|R}$ from the perspective-neutral Hilbert space to one describing $S$ relative to $R$ alone and it is given by 
\begin{equation}
    \mathcal{R}_R (\rho) := \bra{\rho} \otimes \mathds{1}_S.
    \label{eq:PWreduction}
\end{equation}
Its inverse is given by\begin{equation}\label{eq:PWinverse}
    \mathcal{R}_R^{\dag}(\rho)[.] := \Pi ((\ket{\rho} \otimes \mathds1_S) [.])\,,
\end{equation}
i.e.\ one appends again the frame orientation states that was ``projected out'' and averages coherently over the group generated by $\hat C$.
The relational states  $\ket*{\psi_{S|R}(\rho)}=\mathcal{R}_R(\rho)\ket{\psi_{\rm phys}} \in \mathcal{H}_{S\vert R}$ describe the global physical state from the perspective of $R$ when it is in orientation $\rho$.
 These relational states make up a projected Hilbert space of the original system Hilbert space
\begin{equation}
    \mathcal{H}_{S\vert R} := \Pi_{\vert R} \mathcal{H}_S, \qquad \Pi_{\vert R} := \expval*{\Pi}{\rho},
    \label{eq:PWprojector}
\end{equation}
where any reference state $|\rho \rangle$\ can be used leading to the same result. The projector $\Pi_{\vert R}$ essentially picks out those states of $\mathcal{H}_S$ which have a $\hat{G}_S$ eigenvalue corresponding to a state with the opposite eigenvalue of $\hat{G}_R$ in $\mathcal{H}_R$, so it projects onto the overlap $\sigma_S \cap (- \sigma_R)$. (This is in fact how this projector was originally defined in~\cite{Trinity}.) Hence, when the spectrum of the reference is the full real line $\sigma_R = \mathbb{R}$, then $\Pi_{\vert R} = \mathds{1}_S$.

It is easy to check, using the invariance of physical states and the covariance of $R$'s orientation states, that the reduced states transform covariantly under reorientations of the QRF
\begin{equation}
    \label{eq:covariantTransf}
    \begin{split}
    \ket{\psi_{S|R}(\rho)}&=U_S(\rho-\rho')\ket{\psi_{S|R}(\rho')}\\
    &=e^{-i(\rho-\rho')\hat{G}_S /\hbar}\ket{\psi_{S|R}(\rho')}\,.
    \end{split}
\end{equation}

Finally, relational observables
gauge fix via the reduction map to \cite{Trinity}
\begin{equation}\label{eq:fSR}
    \mathcal{R}_R(\rho)\,\hat{\mathcal{O}}_R^\rho(\hat{f}_S)\,\mathcal{R}^\dag_R(\rho)=\Pi_{|R}\,\hat{f}_S\,\Pi_{|R}=:\hat{f}_{S|R}\,,
\end{equation}
so that the relational observable
algebra $\mathcal{A}_{\rm phys}$ reduces on $\mathcal{H}_{S | R}$ to
\begin{equation}
\label{eq:reducedAlgebra}
    \mathcal{A}_{S\vert R} := \mathcal{R}_R(\rho)\mathcal{A}_{\rm phys}\mathcal{R}^\dag_R(\rho)=\Pi_{\vert R} \,\mathcal{A}_S\,\Pi_{\vert R}\,.
\end{equation}
Hence, for ideal QRFs, relational observables assume their bare undressed form in the reduced picture with $\hat{f}_{S|R}=\hat f_S$. Due to the unitarity of the reduction map, expectation values are preserved \cite[Thm.~4]{Trinity}
\begin{align}\label{eq:PWExpVal}
    &\left(\psi_{\rm phys}\big|\hat{\mathcal{O}}^\rho_R(\hat{f}_S)\big|\psi_\text{phys}\right) 
    = \left(\psi_{\rm phys}\big|\hat{{O}}^\rho_R(\hat{f}_S)\big|\psi_\text{phys}\right) \nonumber \\
    &\quad = \expval*{\,\hat{f}_{S}\,}{\psi_{S| R}(\rho)}\,,
\end{align}
where the first two expressions are evaluated in the physical inner product~\eqref{eq:PhysicalIP}, and the right hand side is evaluated in the inner product on $\mathcal{H}_{S|R}$.

We will discuss ``jumping'' from one QRF perspective to another, i.e.\ QRF transformations in Sect.~\ref{sect:switchingIdeal}, making the perspective-neutral nature of the gauge-invariant description manifest.

\subsection{Algebraic quantum reference frames}

\label{sec:AlgebraicMethod}

Let us now turn to the algebraic approach to QRFs.
This construction, originally proposed in \cite{QDSR}, works directly with the kinematical algebra $\mathcal{A}$, which is assumed to contain the constraint $\hat{C}$. The objective is to use a kinematical reference frame observable\footnote{There is a slight difference in terminology in these two constructions. In \cite{Trinity,delaHamette:2021oex} the term ``clock'' (or more generally QRF) refers to a factor subsystem with the associated factor Hilbert space $\mathcal{H}_R$\ and factor algebra of observables $\mathcal{A}_R$, whereas in \cite{QDSR} ``clock'' (or more generally QRF) refers to a kinematical observable whose values correspond to the reading of time.} $\hat{Z}=\hat{Z}^*\in \mathcal{A}$ to cast the constrained system $(\mathcal{A}, \hat{C})$\ as an unconstrained system with a symmetry flow $(\mathcal{F}, D_\rho)$, where $\mathcal{F}$\ is a $*$-subalgebra of  $\mathcal{A}$ and $D_\rho$\ is a (possibly orientation-dependent) $*$-algebra derivation on $\mathcal{F}$\ induced by the adjoint action of $\hat{C}$,\footnote{That is $D_\rho(\hat{f})=\frac{1}{i\hbar} [\hat{f}, \hat{C}]$. However, $\mathcal{F}$\ captures unrestricted degrees of freedom so that $\hat{C}\notin\mathcal{F}$. Therefore, this derivation may be external to $\mathcal{F}$.} which is responsible for generating the coordinate transformation on $\mathcal{F}$. The assumption that $\hat{C} \in \mathcal{A}$\ has important consequences for the possible choices of $\mathcal{A}$. In the atypical case where $\hat{C}$\ is bounded, we can choose $\mathcal{A}$\ to be the $C^*$-algebra of all bounded kinematical operators. In general, however, $\mathcal{A}$\ must include some unbounded operators (for example, the constraint of~\eqref{eq:Constraint} where $\sigma_R=\mathds{R}$). In the rest of this subsection we introduce key concepts used in the algebraic construction of \cite{QDSR}, outline its structure and overview its main results. We refer interested readers to the more in-depth discussion in the original, as well as in \cite{AlgebraicPoT} and \cite{AlgebraicClocks}.

\subsubsection{Physical states}

Concretely, reduction in \cite{QDSR} proceeds by selecting an appropriate subset of kinematical states. Here, a \emph{kinematical state} is a complex linear functional $\omega: \mathcal{A} \rightarrow \mathbb{C}$ which is normalized $\omega(\mathds{1})=1$. We will use $\Gamma$ to denote the space of all such states. In line with the algebraic treatment of quantum mechanics \cite{Strocchi}, a state $\omega$\ on an algebra $\mathcal{A}$\ is \emph{positive} if $\omega(\hat{a}^*\hat{a})\geq0$\ for all $\hat{a}\in \mathcal{A}$. A positive state allows one to define a probability distribution on the spectrum of each element of $\mathcal{A}$. Positivity is therefore key to interpreting states as giving predictions for physical measurements. However, for a constraint with ``$0$'' in the continuous part of its spectrum, it is generally not possible to construct a state out of its solution that is positive on the entire kinematical algebra. We will therefore not impose positivity on $\Gamma$: only some of its elements will have a physically meaningful interpretation.

Following \cite{QDSR}, we define the subalgebra of (strong) \emph{Dirac observables} as the commutant $\mathcal{A}_{\rm obs}:=\{ \hat{O} \in \mathcal{A}:[\hat{O}, \hat{C}] = 0 \}\subset \mathcal{A}$ (this is the direct analogue of $\mathcal{A}_\text{inv}$ in~\eqref{eq:invAlgebra}). Since elements of this subalgebra  are gauge-invariant, they should be well-defined in any reduction of the constrained system, and we will require physically meaningful states to be positive on $\mathcal{A}_{\rm obs}$. Furthermore, in line with the Dirac constraint quantization condition, we will require that physically meaningful states solve the constraint. We define the \emph{constraint surface} as $\Gamma_{C}:=\{ \omega \in \Gamma: \omega(\hat{a}\hat{C})=0 \ \forall \hat{a}\in \mathcal{A} \}\subset \Gamma$, which consists of linear functionals that are right-multiplicative\footnote{Algebraically, the choice to use right- rather than left-multiplicative null states of $\hat{C}$\ is arbitrary. As we will see later, this choice nicely aligns with the convention where the Hilbert space inner product is complex-linear with respect to the argument on the right and complex-anti-linear with respect to the argument on the left.} and zero-value with respect to $\hat{C}$. Thus, all elements of the right ideal $\mathcal{A}\hat{C}$\ are constrained to have zero value, and we will call $\mathcal{A}\hat{C}$\ the subalgebra of \emph{constraint elements}.

Every element $\hat{a}\in \mathcal{A}$\ generates a flow $S_{\hat{a}}$\ on the space of states via the commutator. Given any $\omega\in \Gamma$, the flow creates a one-parameter family of states $S_{\hat{a}}(\lambda)\omega$, $\lambda \in \mathbb{R}$, that satisfies
\begin{equation}
    \begin{split}
        \label{eq:AlgebraicSFlow}
    i\hbar \frac{d}{d \lambda} S_{\hat{a}}(\lambda) \omega (\hat{b}) &= S_{\hat{a}}(\lambda) \omega ([\hat{b}, \hat{a}]) \ , \\ {\rm and} \ S_{\hat{a}}(0) \omega &= \omega \ .
    \end{split}
\end{equation}
For a constraint element $\hat{a}\hat{C}$\ the corresponding flow is
\begin{equation}
    \begin{split}
    \label{eq:AlgebraicGaugeFlow}
    i\hbar \frac{d}{d \lambda} S_{\hat{a}\hat{C}}(\lambda) \omega (\hat{b}) &= S_{\hat{a}\hat{C}}(\lambda) \omega ([\hat{b}, \hat{a}\hat{C}]) \\&\approx_{\Gamma_{\hat{C}}}  S_{\hat{a}\hat{C}}(\lambda) \omega (\hat{a}[\hat{b}, \hat{C}]) \ ,
    \end{split}
\end{equation}
where ``$\approx_{\Gamma_{C}}$'' indicates equality on the constraint surface $\Gamma_{C}$. By setting $\hat{b}=\hat{f}\hat{C}$, for some $\hat{f} \in \mathcal{A}$, it is not difficult to see that these flows preserve the constraint surface. Further, on $\Gamma_{C}$, the values assigned to Dirac observables $\hat{O} \in \mathcal{A}_{\rm obs}$\ are invariant along the flows. By analogy with the flows induced by classical (first-class) constraints on the classical constraint surface, we refer to all these flows as \emph{gauge}.\footnote{That a quantum constraint induces a gauge flow on the space of its solution states may seem counter-intuitive. After all, $\hat{C} |\psi \rangle = 0$\ immediately implies $e^{-i \lambda\hat{C}/\hbar} |\psi \rangle = |\psi \rangle$. However, an algebraic state $\omega$\ is the analog not of a Hilbert space state $|\psi\rangle$ but of an expectation value $\langle \psi| . | \psi \rangle$. For zero in the continuous part of the spectrum of $\hat{C}$, a null eigenstate of $\hat{C}$\ is not normalizable and does not define an expectation value on $\mathcal{A}$\ via the kinematical inner product.  Instead, one may resort to the physical inner product \eqref{eq:PhysicalIP}. Typically expectation values derived from it are only applied to operators that preserve the space of solutions to the constraint, rather than to the entirety of $\mathcal{A}$. We will see in Sect.~\ref{sect:generalGaugeFreedom} that one can reproduce this gauge flow at the Hilbert space level by extending the physical inner product to non-invariant observables.}

The gauge flows above define an equivalence relation: for $\omega, \omega' \in \Gamma$\ we say that $\omega' \thicksim_{C} \omega$\ if there are $\hat{a}_1, \hat{a}_2, \ldots \hat{a}_N \in \mathcal{A}$\ and $\lambda_1, \lambda_2, \ldots \lambda_N \in \mathbb{R}$ such that
\begin{equation}
\omega' = S_{\hat{a}_1\hat{C}}(\lambda_1)S_{\hat{a}_2\hat{C}}(\lambda_2) \ldots S_{\hat{a}_N\hat{C}}(\lambda_N) \omega \ .
\end{equation}
The collection of all gauge-equivalent states defines a gauge orbit
\begin{align}
\label{eq:algebraicGaugeOrbit}
    [\omega]_{C} :=\{\omega' \in \Gamma : \omega' \thicksim_{C} \omega \} \ . 
\end{align}
From our earlier observations, if $\omega \in \Gamma_{C}$, then its orbit lies entirely within the constraint surface $[\omega]_{C} \subset \Gamma_{C}$, and all states on the orbit assign identical values to all Dirac observables $\omega'(\hat{O})=\omega(\hat{O})$\ for all $\hat{O}\in \mathcal{A}_{\rm obs}$ for any $\omega'\in [\omega]_{C}$. We now come to a key definition of the algebraic construction of \cite{QDSR}. The space of \emph{physical states} is the space of gauge orbits that are positive on the algebra of Dirac observables
\begin{align} \label{eq:AlgPhysicalStates}
    \Gamma_{\rm phys} := \left\{ [\omega]_{C} \in \Gamma_{C}/\!\thicksim_{C}\ : \omega(\hat{O}^* \hat{O}) \geq 0,\ \forall \hat{O} \in \mathcal{A}_{\rm obs} \right\} \ .
\end{align}

As a side remark, note that we could have started with a different definition of gauge-equivalence class
\begin{equation}
[\omega]_{\rm obs} := \{ \omega' \in \Gamma : \omega'(\hat{O}) = \omega(\hat{O}),~\forall \hat{O} \in \mathcal{A}_{\rm obs} \} \ .
\end{equation}
For $\omega \in \Gamma_{C}$, clearly $[\omega]_{C} \subseteq [\omega]_{\rm obs}$, however the reverse is not in general true. Strong Dirac observables, beyond those constructed out of $\hat{C}$\ and $\mathds{1}$, may not be in the chosen operator algebra $\mathcal{A}$\ and, in the case of temporal reference frames for chaotic systems, may in fact not exist~\cite{Dittrich:2016hvj} . Here we have chosen to define physical states as gauge orbits rather than states on some complete set of gauge-invariant observables.

\subsubsection{Quantum reference frames through gauge fixing}
\label{sect:algebraicClock}

Definition~(\ref{eq:AlgPhysicalStates}) looks difficult to implement in practice: in order to construct a gauge orbit $[\omega]_{C}$\ one would need to integrate a potentially infinite number of gauge flows generated by $\mathcal{A}\hat{C}$. Fortunately, this definition is well-suited to analyzing gauge-fixing conditions---additional conditions on states that are not preserved by the gauge flows, and therefore ``break'' them.

Gauge-fixing in \cite{QDSR} is accomplished by fixing the value assigned to some kinematical reference observable $\hat{Z}$. For any such observable, its commutant $Z' := \{ \hat{a} \in \mathcal{A}: [\hat{a}, \hat{Z}] = \hat{0} \}\subset \mathcal{A}$\ contains all kinematical observables that can be specified simultaneously with $\hat{Z}$. The quotient $Z'/(\hat{Z}-\rho)Z'$\ can be interpreted as the algebra of observables compatible with $\hat{Z}$\ having value ``$\rho$'' (it is assumed that the spectrum of $\hat{Z}$\ is purely continuous and equal to $\mathbb{R}$). An \emph{algebraic reference frame} has been defined in \cite{QDSR} to consist of
\begin{itemize}
    \item the reference variable $\hat{Z}=\hat{Z}^*\in \mathcal{A}$, where the values $\omega(\hat{Z})$\ represent orientation readings;
    \item a \emph{fashionable algebra} $\mathcal{F} \subset Z'$, that is isomorphic to $Z'/(\hat{Z}-\rho)Z'$\ for every ``$\rho$''. 
\end{itemize}
For orientation $\rho\in \mathbb{R}$, the associated gauge-fixing conditions are
\begin{align} \label{eq:AlgebraicGauge}
    \omega( (\hat{Z}-\rho) \hat{a} ) = 0 \ , \ \ \text{for all}\ \hat{a}\in \mathcal{A} \ .
\end{align}
These conditions completely fix the gauge freedom if for each state $\omega\in \Gamma_{C}$\ that satisfies~(\ref{eq:AlgebraicGauge}) no non-vanishing gauge flow $S_{\hat{a}\hat{C}}$\ preserves all of the conditions~(\ref{eq:AlgebraicGauge}). If the gauge freedom is indeed completely fixed, then each $\omega$\ is the (locally) unique representative of the corresponding gauge orbit $[\omega]_{C}$.

We now list key conditions that are used in \cite{QDSR} to ensure the definition of physical states~(\ref{eq:AlgPhysicalStates}) is consistent and that the reference frame gauge~(\ref{eq:AlgebraicGauge}) completely fixes all gauge flows. First, there are some conditions on the constraint itself
\begin{itemize}
    \item $\hat{C}^*=\hat{C}$;
    \item $\hat{C}$\ has no inverse in $\mathcal{A}$;
    \item $\hat{a}\hat{C}=\hat{0}$\ only for $\hat{a}=\hat{0}$.
\end{itemize}
These conditions essentially mean that $\hat{C}$\ is a self-adjoint operator with zero in the purely continuous part of the spectrum. Further there are conditions on the algebraic decomposition between frame and constraint degrees of freedom
\begin{itemize}
    \item $\hat{Z}^*=\hat{Z}$;
    \item The set $Z'\cup \{\hat{C}\}$\ algebraically generates $\mathcal{A}$;
    \item $Z'\cap \mathcal{A} \hat{C} = \{\hat{0}\}$.
\end{itemize}
These may seem quite restrictive, but, as we will see from concrete examples, they are not difficult to satisfy if $\hat{Z}$\ is part of a one-component canonical subsystem. The most restrictive condition is placed on the commutator
\begin{equation}
[\hat{Z}, \hat{C}]=i\hbar \ .
\end{equation}
Furthermore, provided states satisfy some additional conditions, the algebraic gauge fixing~(\ref{eq:AlgebraicGauge}) can be applied to constraints of the form $\hat{C} = \hat{N} \hat{C}'$, where the right factor $\hat{C}'$\ rather than the original constraint $\hat{C}$\ is canonically conjugate to the reference variable $\hat{Z}$. This allows one to treat some models that include interactions between the system and the reference frame, as well as certain cases where the QRF is degenerate (see Sect.~\ref{sect:degenerateFrames}).

When the system is completely algebraically gauge-fixed in this way, for each value $\rho \in \mathbb{R}$\ we can extend a positive algebraic state $\tilde{\omega}: \mathcal{F} \rightarrow \mathbb{C}$\ to a unique \emph{almost-positive} state $\omega_{Z|\rho}$\ on $\mathcal{A}$ such that
\begin{itemize}
    \item $\omega_{Z|\rho}(\hat{f})=\tilde{\omega}(\hat{f})$\ for all $\hat{f}\in \mathcal{F}$, in particular $\omega_{Z|\rho}(\hat{f}^*\hat{f})\geq 0$\ and $\omega(\mathds{1})=1$;
    \item It is a right solution of the constraint $\omega_{Z|\rho} \in \Gamma_{C}$, so that $\omega_{Z|\rho}(\hat{a}\hat{C})=0$\ for all $\hat{a}\in \mathcal{A}$;
    \item It is a left multiplicative state of the reference observable $\omega_{Z|\rho}(\hat{Z}\hat{a})=\rho\, \omega_{Z|\rho}(\hat{a})$ for all $\hat{a}\in \mathcal{A}$; 
    \item It is positive on Dirac observables $\omega(\hat{O}^*\hat{O})\geq0$\ for all $\hat{O}\in \mathcal{A}_{\rm obs}$;
    \item The flow induced by the adjoint action of the constraint itself, preserves all of the above conditions while uniformly shifting the reference orientation $S_{\hat{C}}(\rho')\omega_{Z|\rho}(\hat{Z})=\rho+\rho'$. 
\end{itemize}

An almost-positive state is thus a normalized linear functional on $\mathcal{A}$\ that, while not positive on the entirety of $\mathcal{A}$, is positive on two important subalgebras: Dirac observables $\mathcal{A}_{\rm obs}$, and kinematical observables compatible with the reference observable $\mathcal{F}$. The values assigned by the almost-positive representative $\omega_{Z|\rho}\in[\omega]_{C}$ of a physical state~\eqref{eq:algebraicGaugeOrbit} are then interpreted as the expectation values of the observed system's degrees of freedom $\hat{f} \in \mathcal{F} \subset \mathcal{A}$\ given the QRF's orientation is $\rho$.

\subsubsection{Algebraic treatment of an ideal QRF}\label{sec:AlgebraicIdealExample}

It is instructive to see how gauge-fixing conditions described in the previous section are satisfied by the constraint associated with a single ideal QRF. Sect.~\ref{sect:equivalence}, \ref{sect:switchingIdeal}, and~\ref{sect:rel} consider all three approaches in this precise scenario, where the constraint governing frame transformations is
\begin{equation}
    \label{eq:idealConstraint}
    \hat{C} = \hat{p}_R + \hat{G}_S \ .
\end{equation}
Here we assume that the system generator $\hat{G}_S=\hat{G}_S^*\in \mathcal{A}_S$\ commutes with the reference frame transformation generator $\hat{p}_R=\hat{p}_R^*\in \mathcal{A}_R$. The latter has a non-degenerate spectrum equal to $\mathbb{R}$\ and therefore possesses a canonically conjugate frame orientation observable $[\hat{q}_R, \hat{p}_R]=i\hbar $. The QRF subalgebra $\mathcal{A}_R\subset \mathcal{A}$\ consists of complex polynomials in $\hat{q}_R$\ and $\hat{p}_R$. The system subalgebra $\mathcal{A}_S\subset \mathcal{A}$ is the trivial extension of some sufficiently large algebra of operators on $\mathcal{H}_S$\ to the full kinematical Hilbert space $\mathcal{H}$. We do not make any further assumptions about $\mathcal{A}_S$.

Applying the algebraic method to this constraint we can verify that the conditions listed in Sect.~\ref{sect:algebraicClock} are satisfied. First, we check the  conditions that must be satisfied by the constraint element 
\begin{itemize}
    \item $\hat{C}^*=\hat{C}$\ is true by definition of $\hat{C}$;
    \item $\hat{a}\hat{C}=\hat{0}$\ implies $\hat{a}\hat{p}_R = - \hat{a}\hat{G}_S$. Suppose $\hat{a}$\ is non-zero and the highest power of $\hat{p}_R$\ in the expression for $\hat{a}$\ in terms of products of elements of $\mathcal{A}_S$\ and $\mathcal{A}_R$ is $N$. Then the highest power for $\hat{p}_R$\ in $\hat{a}\hat{p}_R$\ is $(N+1)$, while that for $(-\hat{a}\hat{G}_S)$\ is $N$, whereby the two cannot be equal unless $\hat{a}=\hat{0}$. Thus, $\hat{C}$\ is not a divisor of zero within $\mathcal{A}$\ as required;
    \item $\hat{C}$\ does not have an inverse in $\mathcal{A}$\ because $\hat{p}_R$\ does not have an inverse in $\mathcal{A}_R$.  Explicitly $\hat{C}^{-1}(\hat{p}_R+\hat{G}_S)=\mathds{1}$\ implies $\hat{C}^{-1}\hat{p}_R=\mathds{1}-\hat{C}^{-1}\hat{G}_S$\ and we can repeat the argument about the highest power of $\hat{p}_R$\ above.
\end{itemize}
In line with the original definition in \cite{QDSR}, we define our algebraic reference frame to consist of the reference observable $\hat{Z}=\hat{q}_R$, and the corresponding fashionable algebra $\mathcal{F} = \mathcal{A}_S$. Indeed, $(\hat{q}_R, \mathcal{A}_S)$ satisfies the conditions of an algebraic reference frame:
\begin{itemize}
    \item By construction $\hat{Z}^*=\hat{Z}$;
    \item It is straightforward to see that the commutant $Z'=\left\{\sum_{n=0}^N \hat{a}_{S\, n} \hat{Z}^n : \hat{a}_{S\, n} \in \mathcal{A}_S\right\}$. Clearly, $\mathcal{A}_S \subset Z'$\ as required. Furthermore, for any $\rho\in \mathbb{R}$, let $[\hat{a}]_\rho$\ denote the coset of $\hat{a}\in Z'$ \ relative to the ideal $(\hat{Z}-\rho)Z'$. Writing a general element of $Z'$
\begin{align}
\sum_{n=0}^N \hat{Z}^n\hat{a}_{S\, n} = &\sum_{n=0}^N \rho^n\hat{a}_{S\, n} \\+ &\sum_{n=1}^N \hat{a}_{S\, n} \sum_{m=1}^n {n \choose m} (\hat{Z}-\rho)^m \ , \nonumber
\end{align}
we see that $\left[ \sum_{n=0}^N \hat{Z}^n\hat{a}_{S\, n}\right]_\rho = \left[ \sum_{n=0}^N \rho^n\hat{a}_{S\, n}\right]_\rho$\ and the map $\alpha_\rho: Z'/(\hat{Z}-\rho)Z' \rightarrow \mathcal{A}_S$\ defined by $\left[ \sum_{n=0}^N \hat{Z}^n\hat{a}_{S\, n}\right]_\rho \mapsto  \sum_{n=0}^N \rho^n\hat{a}_{S\, n}$\ is a ($*$-algebra) isomorphism for each $\rho$:
\end{itemize}
Finally, we verify the properties that need to be satisfied by the constraint and the reference frame together.
\begin{itemize}
    \item By construction $[\hat{Z}, \hat{C}]=i\hbar $;
    \item Noting that $Z'$\ contains no elements that contain products of $\hat{p}_R$, while any element of the subalgebra $\mathcal{A}\hat{C}$\ has the form $\hat{a}\hat{p}_R+\hat{a}\hat{G}_S$,  and, unless $\hat{a} = \hat{0}$, must contain at least one such non-zero term, we see that $Z'\cap \mathcal{A}\hat{C} = \{\hat{0}\}$;
    \item A general element of $\mathcal{A}$\ can be generated by products of elements of $Z'$\ and powers of $\hat{p}_R=\hat{C} - \hat{G}_S$. Since $\hat{G}_S\in Z'$\ one can equivalently generate the whole algebra from the set $Z'\cup\{\hat{C}\}$.
\end{itemize}

\subsection{Semiclassical effective quantum reference frames}
\label{sect:effective}

Finally, we turn to the older effective approach to QRFs, \cite{EffectiveConstr,EffectiveConstrRel,EffectivePoT1,EffectivePoT2,EffectivePoTChaos}, which was originally only formulated for clocks, i.e.\ temporal QRFs, and so this work can be viewed as slightly extending it to other kinds of QRFs. It describes the physics through the dynamics of the expectation values and their higher-order moments on a quantum phase space by means of a perturbative expansion in $\hbar$ which is then cut off at the order of interest. These
\emph{canonical effective equations} were first developed for ordinary quantum mechanical systems in \cite{EffectiveEq1, EffectiveEq2}.\footnote{ The original construction in~\cite{EffectiveEq1, EffectiveEq2} assumed that the generators of the observable algebra $\mathcal{A}$\ satisfy canonical commutation relations. This was extended to generators forming a (semisimple) Lie algebra is~\cite{SemiclassicalLie} (see~\cite{Bojowald:2014uaa} for an application to a cosmological model).} They were applied to systems with a Hamiltonian constraint in \cite{EffectiveConstr, EffectiveConstrRel, Bojowald:2009jk} and to more general cosmological models with multiple clock (i.e.\ temporal QRF) choices in \cite{EffectivePoTChaos,EffectivePoT1, EffectivePoT2}; in particular, these latter references constitute the first ones in the literature exploring transformations between QRFs, albeit in a semiclassical setting. Effective temporal QRFs have since been used for defining relational dynamics and exploring semiclassical fluctuations (mostly) in quantum cosmology~\cite{Bojowald:2016fac, Amaral:2016hud, Brizuela:2019elx, Brizuela:2021yln, Marchetti:2020umh, Gielen:2021vdd, Balsells:2025iac}. However, as will be clear from our discussion throughout this manuscript, it is also well-suited for constructing other types of QRFs.

Historically, the algebraic approach of \cite{QDSR} was directly inspired by the features of the effective semiclassical treatment, so the initial setup of the two approaches is nearly identical, with differences emerging in the description of the state space. In particular, the effective approach builds a convenient quantum phase space with Poisson structure that permits one to treat gauge flows in the usual classical manner, except that the quantum phase space has more dimensions (and constraints) because of the independent fluctuation variables. A core motivation for this approach was to sidestep explicit physical Hilbert space constructions in constrained systems, which can be notoriously arduous depending on the properties of the constraint. The effective approach is thus a particularly convenient tool for extracting (semiclassical) physics when an explicit physical Hilbert space formulation is not known, such as in the closed Friedman universe with massive scalar field, which is chaotic \cite{EffectivePoTChaos}.

\subsubsection{Effective quantum phase space}

We start with the same algebra of kinematical operators $\mathcal{A}$ as in the algebraic approach, except in the effective case, we additionally assume that it is generated by a finite set of generators $\{\hat{y}_i\}_{i=1, 2, \ldots, N}$\ and the identity $\mathds{1}$ that satisfy $\hat{y}_i^* = \hat{y}_i$\ and form a Lie algebra with respect to the associative commutator (see \cite{SemiclassicalLie})
\begin{align}\label{eq:GeneratorLieAlgebra}
[\hat{y}_i, \hat{y}_j] := \hat{y}_i\hat{y}_j - \hat{y}_j\hat{y}_i = \sum_{k=0}^N i\hbar \tensor{\alpha}{_{ij} ^k} \hat{y}_k \ ,
\end{align}
for some structure constants $\tensor{\alpha}{_{ij} ^k} \in \mathbb{R}$, where $\hat{y}_0:=\mathds{1}$. As in the algebraic construction above, we consider the quantum phase space $\Gamma$\ consisting of all normalized complex linear functions on $\mathcal{A}$ which are not necessarily positive.

The effective approach deals directly with the geometry of $\Gamma$. To that end we note that every element $\hat{a} \in \mathcal{A}$\ defines a scalar function $\langle \hat{a} \rangle$\ on $\Gamma$\ in a natural way $\langle \hat{a} \rangle (\omega) := \omega(\hat{a})$. Because the states $\omega \in \Gamma$\ are linear, this assignment of functions to operators is also linear $\langle \hat{a} + \hat{b} \rangle = \langle \hat{a} \rangle + \langle \hat{b} \rangle$. In the case of the polynomial algebra $\mathcal{A} = \mathbb{C} [\hat{y}_i; \mathds{1} ]$, a useful coordinate basis is formed by the functions associated with symmetrized monomials $\left\langle \hat{y}_1^{n_1} \ldots  \hat{y}_N^{n_N} \right\rangle_\text{Weyl}$, where the subscript ``Weyl'' stands for the totally symmetrised (or ``Weyl-'') ordering of individual generators inside the ``bracket''. The space of states $\Gamma$\ inherits a Poisson bracket from the associative commutator on $\mathcal{A}$\
\begin{equation}
    \acomm*{\expval*{\hat{a}}}{\expval*{\hat{b}}} = \frac{1}{i \hbar} \expval*{\comm*{\hat{a}}{\hat{b}}}, \quad \forall \hat{a},\hat{b} \in \mathcal{A}.
    \label{eq:quantumPoisson}
\end{equation}
which is well-defined on the above coordinate functions and extends to arbitrary smooth functions of those by linearity and the Leibniz rule. One can immediately see that the flow along the Poisson vector field $X_{\langle \hat{a} \rangle}$\ generated by function $\langle \hat{a} \rangle$\ is exactly the same as the flow $S_{\hat{a}}(\lambda)$\ defined by~(\ref{eq:AlgebraicSFlow}). For a system with a single constraint $\hat{C} \in \mathcal{A}$ we pass to the constraint surface $\Gamma_{C}$ defined, just like in the algebraic case, by setting all constraint functions $\langle \hat{a} \hat{C} \rangle = 0$. These constraint functions form a closed Poisson algebra
\begin{equation}
\{ \langle \hat{a} \hat{C} \rangle, \langle \hat{b} \hat{C} \rangle \} = \frac{1}{i\hbar} \langle [ \hat{a},  \hat{b} \hat{C}] \hat{C} \rangle + \frac{1}{i\hbar} \langle \hat{a} [\hat{C} ,  \hat{b} ] \hat{C} \rangle.
\end{equation}
Thus, the Poisson brackets between constraints vanish on $\Gamma_{C}$\ so that the corresponding flows preserve the constraint surface $X_{\langle \hat{b}\hat{C} \rangle}( \langle \hat{a}\hat{C} \rangle ) \approx_{\Gamma_{C}} 0$. These flows are treated as gauge so that the physical states are defined exactly as in~(\ref{eq:AlgPhysicalStates}).

Viewed geometrically, one has an infinite-dimensional state space $\Gamma$\ with a Poisson bracket which needs to be reduced by applying an infinite number of constraint conditions $\langle \hat{a} \hat{C} \rangle = 0$, and factoring out an infinite number of corresponding gauge flows $\{., \langle \hat{a} \hat{C} \rangle \}$.  The distinct feature of the effective approach lies in using the semiclassical approximation to truncate this system to one with a finite number of degrees of freedom. We start by changing to a different coordinate basis on $\Gamma$, namely the expectation values of the generators $y_i:=\langle \hat{y}_i \rangle$, which we will call ``classical variables'' (since $\hat{y}_i$\ should be the quantization of a classically (over-)complete set of phase-space functions), and their moments which we will call ``quantum variables''
\begin{equation}\label{eq:QuantumVariables}
    \Delta(y_1^{n_1}y_2^{n_2}...y_N^{n_N}) := \left\langle  \widehat{\Delta y}_1^{n_1} \widehat{\Delta y}_2^{n_2} \ldots \widehat{\Delta y}_N^{n_N}\right\rangle_\text{Weyl} \ ,
\end{equation}
where $\widehat{\Delta y}_m:=(\hat{y}_m - \langle \hat{y}_m \rangle)$\ are the ``moment operators'', and we have again chosen to completely symmetrize their ordering. The sum $\sum_m n_m$\ is called the \emph{order} of the moment. We note that all first-order moments are identically zero, the lowest-order moments are therefore of order two. As the sole exception to the notation of~\eqref{eq:QuantumVariables}, second-order variances will be denoted as $(\Delta y_i)^2$ instead of $\Delta (y_i^2)$ to match previous literature and to make the interpretation of these quantities as uncertainties more explicit. The combination of classical and quantum variables forms an alternative set of coordinates that can be transformed to and from the symmetrized monomial coordinate functions. For example, $
(\Delta y_i)^2 = \langle \hat{y}_i^2 \rangle - \langle \hat{y}_i \rangle^2$, and conversely $\langle \hat{y}_i^2 \rangle = (\Delta y_i)^2 + y_i^2$.

It is clear that the ``classical variables'' obey the same Poisson relations as their classical counterparts. Thus, we can view the $N$-dimensional submanifold described by these variables (and setting all quantum variables to zero) as the classical phase space embedded in the quantum phase space. The rest of the infinite number of dimensions of the quantum phase space is then accounted for by all the quantum variables describing fluctuations. This infinite dimensionality of the quantum phase space is, of course, to be expected: for example, suppose that the total system is given classically by a two-dimensional phase space $\mathbb{R}^2$ with one canonical pair. The corresponding Hilbert space is $L^2(\mathbb{R})$, which is infinite-dimensional. The quantum phase space a priori encodes all of this information, including that contained in mixed states. Similarly, the reason for imposing infinitely many constraints on the quantum phase space is that a Hilbert space condition such as $\hat{C}\ket{\psi}=0$ means that not only $\langle\hat{C}\rangle=0$, but also that all its independent fluctuations vanish.

We now restrict to the semiclassical region of $\Gamma$\ that corresponds to ``sharply peaked'' states, in which the basic generators $\hat{y}_i$\ have small variances and higher-order moments. Formally we assume a hierarchy in powers of $\sqrt{\hbar}$. Let $\mathscr{O}$ denote the semiclassical order such that $\mathscr{O}(\hbar^n) = 2n$, then we assume 
\begin{equation}
    \mathscr{O}(\expval{\hat{y}_i}) = 0, \quad \text{and} \ \ \mathscr{O}(\Delta( y_1^{n_1} \ldots y_N^{n_N})) =  \sum_{m = 1}^N n_m \, .
    \label{eq:order}
\end{equation}
The order of a product of two or more quantum variables is the sum of their orders, and the leading order of a sum of functions of quantum variables is equal to that of the lowest-order term in the sum.
This hierarchy allows us to write down semiclassical expansions and consistently truncate our degrees of freedom and dynamical equations up to some order resulting in a finite-dimensional system. To that end, we write the constraint conditions in terms of the classical and quantum variable basis and discard terms of order larger than $M$. We will use $\mathscr{T}_M$ to denote truncation at order $M$, defined for a function $f$\ of classical and quantum variables~\eqref{eq:QuantumVariables} as (see~\cite{SemiclassicalLie})
\begin{equation}\label{eq:TruncOperator}
    \mathscr{T}_M(f) = 0, \quad \text{when } \mathscr{O}(f) > M,
\end{equation} 
and for any quantity allowing a series expansion we can write down the expansion in increasing order as\footnote{In this expression, partial derivatives w.r.t.~algebra generators are performed as if they were ordinary independent numerical variables.} 
\begin{widetext}
\begin{equation}
    \expval*{P(\hat{a}_i)}_\text{Weyl} = P(a_i) + \sum_{m_1,...,m_n}^{+\infty} \frac{1}{m_1!...m_n!} \left(\frac{\partial^{m_1 + ... + m_n}}{\partial \hat{a}_1^{m_1} ... \partial \hat{a}_n^{m_n}} P(\hat{a}_i) \right) \eval_{\hat{a}_i = a_i} \Delta(a_1^{m_1}...a_n^{m_n}).
    \label{eq:effExpand}
\end{equation}
\end{widetext}
There is one additional subtlety in counting the semiclassical order: for an arbitrary $\hat{C} = f(\hat{y}_i)$, where $f$\ is some ordered polynomial, we have
\begin{equation}
\langle \hat{C} \rangle = f(y_i) + (\text{terms of order}\ \hbar) \ .
\end{equation}
Here $f(y_i)=:C_{\rm class}$\ is the classical expression for the constraint in terms of the classical variables. On the constraint surface, where we set $\langle \hat{C} \rangle =0$, we explicitly treat $C_{\rm class}$ as having semiclassical order two.

As an example, let us consider truncation at semiclassical order $M=2$. The degrees of freedom at this order are the $N$\ classical variables $\langle\hat{y}_i \rangle$\ and ${N+1 \choose N-1} = \frac{1}{2}N(N+1)$\ moments (for explicit counting see \cite{SemiclassicalLie}). Up to this order we need to include $N+1$\ constraints 
\begin{equation} \label{eq:EffectiveConstraintFunctions}
    \expval*{\hat{C}} = 0, \quad \text{and} \left \langle \widehat{\Delta y}_i \hat{C}\right\rangle = 0 \ ,
\end{equation}
which are to be expressed in terms of the classical and quantum variables. Assuming some regularity conditions, these constraint functions are functionally independent in the semiclassical region of $\Gamma$\ near the classical constraint surface (again see \cite{SemiclassicalLie}). The number of local degrees of freedom left once the constraints are imposed is equivalent to that of a system with one fewer generating element. The precise number of independent gauge flows generated by the constraints~(\ref{eq:EffectiveConstraintFunctions}) depends on the Lie algebra structure~(\ref{eq:GeneratorLieAlgebra}) and is generically fewer than the number of constraint functions due to the degeneracy of the Poisson bracket~(\ref{eq:quantumPoisson}).

\subsubsection{Reference gauge }\label{sect:clockGauge}

In all constrained systems that have been treated in detail using the effective approach, reduction was accomplished by fixing the gauge flows. The procedure relies on the generator algebra containing a canonical pair of generators $[\hat{y}_1, \hat{y}_2]=i\hbar$, which commute with all other generators $\{\hat{y}_i\}_{i=3, \ldots N}$, such that $[\hat{y}_1, \hat{C}] \neq \hat{0}$. Here $\hat{y}_1$\ will play the same role as $\hat{Z}$\ in the algebraic approach of Sect.~\ref{sec:AlgebraicMethod}. Combining these conditions requires the polynomial form of $\hat{C}$\ to contain $\hat{y}_2$. Gauge-fixing is accomplished by setting all moments of $\hat{y}_1$\ that do not involve $\hat{y}_2$\ to zero, for $n_1 \neq 0$
\begin{align}\label{eq:EffectiveGaugeCond}
    \Delta(y_1^{n_1}y_3^{n_3}y_4^{n_4}...y_N^{n_N}) = 0 \ ,
\end{align}
while the constraint functions~(\ref{eq:EffectiveConstraintFunctions}) themselves are used to eliminate $\langle \hat{y}_2 \rangle$\ and moments containing $\hat{y}_2$. For $\hat{C}$\ linear in $\hat{y}_2$ \cite{EffectiveConstr} and for a general class of constraints quadratic in $\hat{y}_2$\ \cite{EffectiveConstrRel, EffectivePoT1, EffectivePoT2, EffectivePoTChaos} analyzed at order $\hbar$\ it was found that the above gauge-fixing conditions eliminate all gauge flows except one. For the case where $\hat C$ is a Hamiltonian constraint, the remaining gauge flow was interpreted as the flow encoding the dynamics in $\hat{y}_1$-gauge, by parametrizing it relationally with time $y_1=\langle \hat{y}_1 \rangle$ (or, more precisely, with the real part of $y_1$ \cite{EffectivePoT1,EffectivePoT2}).
Similarly, when $\hat C$ generates a spatial (or some other kind of) gauge symmetry, the surviving gauge flow will be relationally parametrized by the (real part of the) QRF orientation variable $y_1$ and encodes the relations between the remaining degrees of freedom and the QRF in $\hat{y}_1$-gauge. For example, for a spatial translation, this would encode all the positions of the remaining degrees of freedom relative to the QRF.

Now, let us explicitly consider the leading semiclassical order effective treatment of the system with a single ideal QRF \cite{EffectiveConstr, EffectiveConstrRel} governed by the constraint~\eqref{eq:idealConstraint}. Within the effective treatment we make the additional assumption that the system transformation generator $\hat{G}_S$ is (or can be approximated by) a polynomial in the generators $\hat{x}_i$, $i=1, \ldots, N$, associated with the system algebra $\mathcal{A}_S$. 
Following Sect.~\ref{sec:AlgebraicIdealExample} there is a canonically conjugate reference orientation observable $\hat{q}_R$. Together, they form the basis $\{\hat{y}_i\} = \{ \hat{q}_R, \hat{p}_R; \hat{x}_i \}$, where $[\hat{p}_R, \hat{x}_i]=0=[\hat{q}_R, \hat{x}_i]$.
The tower of effective constraints (\ref{eq:EffectiveConstraintFunctions}) at order two can be explicitly written in terms of the classical variables $y_i$, moments $(\Delta y_i)^2$, and variances 
$\Delta (y_i y_j)$. Denoting $C_{y_i} \equiv \expval*{\widehat{\Delta y}_i \hat{C}}$, we have
\begin{align}
\label{eq:constraintTower}
        C &= G_S + p_R, \nonumber\\
        C_{p_R}  &= \left( \Delta p_R \right)^2  +  \mathscr{T}_2 \Delta \left( p_R G_S \right) , \nonumber\\
        C_{q_R}  &= \Delta \left( q_R p_R \right) + \mathscr{T}_2 \Delta \left( q_R G_S \right) + \frac{1}{2} i \hbar , \\
        C_{x_i} &=  \Delta \left( p_R x_i \right) + \mathscr{T}_2 \Delta( x_i G_S ) + \frac{1}{2} \mathscr{T}_2 \expval*{\comm*{\hat{x}_i}{\hat{G}_S}} \ , \nonumber\\ &\ i=1, \ldots, N\ .\nonumber
\end{align}
For compactness, we defined $\Delta(y_i G_S) := \expval*{\widehat{\Delta y}_i \widehat{\Delta G}_S}_{\rm Weyl}$. Depending on the functional form of $\hat{G}_S$, these quantities can contain higher-order contributions, so we use $\mathscr{T}_2$-maps to truncate at order two. For the case $y_i = x_i$ we need to account for the commutator $\comm*{\hat{x}_i}{\hat{G}_S}$\ in order to implement Weyl ordering, which again may contain higher-order terms that we drop upon truncation.

The reference frame gauge is achieved by implementing the conditions $\Delta(q_R y_i) = 0$ (\ref{eq:EffectiveGaugeCond}) with $y_i \neq p_R$, and by using the effective constraints 
to solve for $p_R$ and $\Delta(p_R y_i)$, with $y_i \neq q_R$. 
This gauge choice sets the fluctuations of the frame to zero, similarly to how the frames in the Schr\"odinger states in the Page--Wootters formalism \cite{Trinity} no longer fluctuate upon invoking the conditioning (gauge fixing) map $\mathcal{R}_R(\rho)$ in \eqref{eq:PWreduction}.
Note that $\Delta(q_R p_R)$ is not gauge-fixed to zero and is in fact determined by setting $C_{q_R}=0$\ and imposing~\eqref{eq:EffectiveGaugeCond}. We have $\expval*{\hat{q}_R \hat{p}_R} = \expval*{\hat{q}_R ( -\hat{G}_S)} = \expval*{\hat{q}_R} \expval*{ -\hat{G}_S} = \expval*{\hat{q}_R} \expval*{\hat{p}_R}$, so that
\begin{align}
 \Delta(q_R p_R) = \expval*{\hat{q}_R \hat{p}_R} - \expval*{\hat{q}_R} \expval*{\hat{p}_R} - \frac{1}{2} i \hbar = - \frac{1}{2} i \hbar.
    \label{eq:gaugeFixing}
\end{align}
As a consequence, such a gauge-fixed state cannot be positive on the entire kinematical algebra $\mathcal{A}$. In particular, the self-adjoint element $\frac{1}{2}(\hat{p}_R\hat{q}_R+\hat{q}_R\hat{p}_R)$\ has a complex expectation value
\begin{equation}
\frac{1}{2}\langle \hat{p}_R\hat{q}_R+\hat{q}_R\hat{p}_R \rangle = q_Rp_R + \Delta(q_Rp_R) = q_Rp_R - \frac{1}{2} i \hbar \ .
\end{equation}
This does not pose a problem with physical interpretation of the gauge-fixed state, however, as we only impose positivity on the independent $S$ degrees of freedom (the evolving degrees of freedom when $\hat{C}$ is a Hamiltonian constraint \cite{EffectivePoTChaos,EffectivePoT1, EffectivePoT2}).

Finally, we note that the effective constraint $C_{q_R} = \expval*{\widehat{\Delta q}_R \hat{C}}$ generates no flow on the surface where both constraint and gauge conditions are imposed. For any $\hat{a} \in \mathcal{A}$
\begin{align}
    &\acomm*{\expval*{\hat{a}}}{\expval*{\widehat{\Delta q}_R \hat{C}}}
    \\ &\quad = \frac{1}{i\hbar}\expval*{ \widehat{\Delta q}_R \comm*{\hat{a}}{\hat{C}}} + \frac{1}{i \hbar} \expval*{ \comm*{\hat{a}}{\widehat{\Delta q}_R} \hat{C} } = 0. \nonumber
\end{align}
In the final expression, the second term vanishes because it is a linear combination of the effective constraints~\eqref{eq:constraintTower}, while the first term vanishes because it is a linear combination of the gauge-fixing  conditions, after using the constraint to remove any $\hat{p}_R$ dependence in $[\hat{a}, \hat{C}]$: 
$\expval*{\hat{b}\hat{p}_R } = \expval*{-  \hat{b}\hat{G}_S}$ once~(\ref{eq:EffectiveConstraintFunctions}) is enforced.

\subsubsection{Comparison with algebraic reference gauge fixing}\label{sssec_algeffequiv}

We conclude this subsection by noting that, in the cases treatable by the algebraic approach introduced in Sect.~\ref{sec:AlgebraicMethod}, the effective gauge-fixing  conditions~(\ref{eq:EffectiveGaugeCond}) are equivalent to the algebraic gauge-fixing conditions~(\ref{eq:AlgebraicGauge}). Identifying $\hat{Z}=\hat{y}_1$, we see that, within the polynomial algebra assumed in the effective treatment, the commutant $Z'$\ is generated by $\hat{y}_1$\ and $\{\hat{y}_i\}_{i=3, \ldots N}$. Further identifying $\rho=\langle \hat{y}_1 \rangle$, the conditions
\begin{widetext}
\begin{equation}
    \begin{split}
        \Delta(y_1^{n_1}y_3^{n_3}...y_N^{n_N}) &= \left\langle \left(\hat{y}_1 - \langle \hat{y}_1 \rangle\right)^{n_1} \prod_{m=3}^N (\hat{y}_m - \langle \hat{y}_m \rangle)^{n_m}\right\rangle_\text{Weyl}
\\
&= \left\langle (\hat{Z} - \rho )^{n_1} \left( \prod_{m=3}^N (\hat{y}_m - \langle \hat{y}_m \rangle)^{n_m} \right)_\text{Weyl} \right\rangle = 0 \ ,
    \end{split}
\end{equation}
\end{widetext}
are equivalent to $\langle (\hat{Z} - \rho )\hat{a} \rangle = 0$\ for all $\hat{a} \in Z'$. This extends to the entire kinematical algebra, since in the algebraic treatment we demand that $Z'\cup\{\hat{C}\}$\ algebraically generates $\mathcal{A}$. In particular, for any $\hat{a}\in \mathcal{A}$, there are $\hat{a}_1 \in Z'$\ and $\hat{a}_2\in \mathcal{A}$\ such that $\hat{a} = \hat{a}_1 + \hat{a}_2\hat{C}$. Thus, on the constraint surface
\begin{equation}
\langle (\hat{Z} - \rho )\hat{a} \rangle = \langle (\hat{Z} - \rho )\hat{a}_1 \rangle + \langle (\hat{Z} - \rho )\hat{a}_2 \hat{C} \rangle = 0\,,
\end{equation}
where the first term vanishes due to the gauge conditions~(\ref{eq:EffectiveGaugeCond}) and the second one due to the constraint conditions~(\ref{eq:EffectiveConstraintFunctions}). Thus, whenever the full algebraic gauge-fixing of \cite{QDSR} can be accomplished, the effective treatment is its semiclassical truncation. However, the class of models analyzed effectively to order $\hbar$ in \cite{EffectivePoT2,EffectivePoTChaos} includes some that do not currently have an exact algebraic (or, in the case of \cite{EffectivePoTChaos}, Hilbert space) treatment.

\subsection{Example: a translationally invariant system}
\label{sect:example}

To give the reader more familiarity with the three approaches, we will apply them to the model considered in \cite{Vanrietvelde:2018pgb}: 
$N$ interacting particles in one dimension with Hamiltonian 
\begin{equation}
    \label{eq:translationInvariantHam}
    \hat{H} = \sum_{i \in I} \frac{\hat{p}_i^2}{2m_i} + V(\{\hat{q}_i - \hat{q}_j\}),
\end{equation}
where the potential $V$ depends only on the relative positions, and with index set $I = \{A,B,C,...\}$\ containing $N$\ labels for the individual particles. Due to global translational invariance, this model has a momentum constraint of the form 
\begin{equation}\label{eq:momcon}
    \hat{C} = \sum_{i\in I } \hat{p}_i,
\end{equation}
which means the corresponding QRFs will be spatial (and which can be derived from a Lagrangian \cite{Vanrietvelde:2018pgb}).

In this model, each individual particle's kinematical algebra $\mathcal{A}_i$ is generated by the corresponding canonical pair $\{\hat{q}_i,\hat{p}_i\}$, and each particle can serve as an ideal reference frame for the remaining particles ($\sigma_i \equiv \text{Spec}(\hat{p}_i) = \mathbb{R}$).\footnote{One could also use composite variables as an ideal reference frame, for example, suitable linear combinations such as the center of mass degrees of freedom.} Let us start with the PN treatment of the system. In particular, due to the tensor product nature of the Hilbert space, a kinematical state can be written as \eqref{eq:kinematicalState}
\begin{equation} 
    \ket{\psi_\text{kin}} = \int_{\mathds{R}} \dd{p}_A \int_{\mathds{R}} \dd{p}_{\bar{A}} \psi_{\text{kin}}(p_A,p_{\bar{A}}) \ket{p_A}_A \ket{p_{\bar{A}}}_S.
\end{equation}
We have separated particle $A$, which will be used as the reference frame for the remaining particles $\bar{A} := I \setminus \{A\}$ making up the system $S$.\footnote{We use the shorthand $\dd{p}_{\bar{A}}$ to denote $\dd{p}_B \dd{p}_C...$} This state can  be mapped to a physical state using \eqref{eq:projector} 
\begin{equation}\label{eq:ExamplePhysStates}
    \ket{\psi_{\rm phys}} = \int_{\mathds{R}} \dd{p}_{\bar{A}} \psi_{\rm kin}(-\sum_{i \in \bar{A}} p_i, p_{\bar{A}}) \ket*{-\sum_{i \in \bar{A}} p_i}_A \ket{p_{\bar{A}}}_S.
\end{equation}

The orientation states for frame $A$, or indeed any particle, are simply the position states found through the Fourier transform \eqref{eq:clock states}
\begin{equation}
    \ket{q}_i = \int_{\mathds{R}} \dd{p} e^{-i q p / \hbar} \ket{p}_i,
\end{equation}
which forms a complete orthogonal basis and for which the unitary transformations generated by $\hat{G}_i \equiv \hat{p}_i$ act covariantly \eqref{eq:covariance}. Since these frames are ideal, the orientation operator is nothing but the position operator \eqref{eq:time operator} $\hat{R}_i \equiv \hat{q}_i$. By virtue of this, the relational observables associated with particle positions take the simple form \eqref{eq:idealDirac}
\begin{equation}\label{eq:idQRFposrel}
     \dirac{\hat{q}_i}{j}{\rho_j} = \hat{q}_i-\hat{q}_j + \rho_j \ ,
\end{equation}
which are the relative distances between pairs of particles. They measure the position of particle $i$ conditional on particle $j$ being in position $\rho_j$. Indeed, this can best be seen classically from the fact that Eq.~\eqref{eq:idQRFposrel} is gauge-invariant, thus having the same value in all gauges. Now gauge fixing the frame position to $\hat{q}_j=\rho_j$, yields $\hat{O}^{\rho_j}_j(\hat{q}_i)\big|_{\hat{q}_j=\rho_j}=\hat{q}_i$. Thus the relational observable measures the position of $i$ conditional on $j$ being in position $\rho_j$. 

Using the fact that $\comm*{\hat{p}_i}{\hat{C}} = 0$, relational momentum variables are identical to the individual particles' momenta
\begin{equation}
    \dirac{\hat{p}_i}{j}{\rho_j} = \hat{p}_i \ .
\end{equation}
Choosing $A$ as the origin of our reference frame corresponds to setting $j=A$\ and $\rho_A = 0$, giving us one particular way of generating the invariant algebra $\mathcal{A}_{\rm inv}$ \eqref{eq:invAlgebra}  
\begin{equation}
    \mathcal{A}_{\rm inv} = \langle \hat{q}_i - \hat{q}_A,\hat{p}_i, \hat{p}_A|i \in \bar{A} \rangle.
\end{equation}
Picking any other particle to serve as our reference frame leads to a different description of the same algebra.

The Page--Wootters construction of a QRF perspective (Sect.~\ref{sect:PW}) proceeds by conditioning the physical state on a particular orientation, here position, of the frame. This amounts to creating the relational states\footnote{Note that if $\psi_{\rm kin}$ were proportional to the identity, one would have $\ket{\psi_{S|A}}
\propto \otimes_{i \in \bar{A}} \ket{-\rho_A}_i$.}
\begin{align}
\label{eq:ExamplePageWoottersState}
    &\ket{\psi_{S|A}} = (\bra{\rho_A}_A \otimes \mathds{1}_S) \ket{\psi_{\rm phys}}  \\
    &\quad = 2 \pi \hbar \int_{\mathds{R}} \dd{p}_{\bar{A}} \psi_{\rm kin}(-\sum_{i \in \bar{A}} p_i, p_{\bar{A}}) e^{i \rho_A \sum_{i \in \bar{A}} p_i / \hbar} \ket{p_{\bar{A}}}_S. \nonumber
\end{align}
With this choice, the relevant reduced algebra is
\begin{equation}
    \label{eq:redAlgebraEx}
    \mathcal{A}_{S | A} = \langle \hat{q}_i, \hat{p}_i | i \in \bar{A}\rangle,
\end{equation}
which consists of all bare canonical pairs except for the particle we picked to serve as our QRF.
\\
\\
\noindent
In the algebraic approach of Sect.~\ref{sec:AlgebraicMethod}, one works directly with the gauge-generating constraint (\ref{eq:momcon}), avoiding the explicit construction of physical states~\eqref{eq:ExamplePhysStates}. The kinematical degrees of freedom of the model are captured by generalized algebraic states, $\omega \in \Gamma$, on the complex polynomial algebra $\mathcal{A}$\ generated by $\hat{q}_i$, $\hat{p}_i$. The constraint is enforced by restricting to states such that $\omega(\hat{A}\hat{C} ) = 0$\ for all $\hat{A} \in \mathcal{A}$, which defines the constraint surface $\Gamma_C\subset \Gamma$. With an eye toward using particle $A$\ as the reference, we can write these constraint conditions as
\begin{equation} \label{eq:ExampleAlgEliminateA}
\omega(\hat{A}\hat{p}_A ) = - \omega \left( \hat{A} \sum_{i\in \bar{A}}\hat{p}_i \right) \ ,
\end{equation}
using them to eliminate the values assigned by states to all polynomials involving $\hat{p}_A$. The states on $\Gamma_C$\ are generally still subject to an infinite number of gauge flows generated by elements of $\mathcal{A}\hat{C}$. For example, along the flow generated by $\hat{p}_A\hat{C}$\ on $\Gamma_C$\ the value of $\hat{q}_B$\ changes according to~\eqref{eq:AlgebraicGaugeFlow}
\begin{align} 
\label{eq:ExampleAlgFlow}
i\hbar\frac{d}{d\lambda} \omega (\hat{q}_B) &= \omega \left( \left[ \hat{q}_B, \hat{p}_A\hat{C}\right] \right) \\ &= i\hbar\, \omega(\hat{p}_A) \approx_{\Gamma_C} -i\hbar\, \omega \left( \hat{A} \sum_{i\in \bar{A}}\hat{p}_i \right) \ .\nonumber
\end{align}
Importantly, the above flows preserve the values assigned by the states on $\Gamma_C$\ to Dirac observables. For example, $[\hat{p}_B, \hat{C}] =0$, so that along any gauge flow
\begin{align}
    i\hbar\frac{d}{d\lambda} \omega (\hat{p}_B) &= \omega \left( \left[ \hat{p}_B, \hat{A}\hat{C}\right] \right)
    \\ &= \omega \left( \left[ \hat{p}_B, \hat{A}\right] \hat{C}\right) \approx_{\Gamma_C} 0 \ . \nonumber
\end{align}

Dirac observables can in principle be used to characterize physically distinct states on $\Gamma_C$, however, the algebraic approach was explicitly designed to not assume that a sufficiently large Dirac observable algebra can be realized as a subalgebra of $\mathcal{A}$. Instead, a ``neutral'' perspective is assumed to exist in the form of the space of gauge orbits on the constraint surface $\Gamma_\text{phys}=\Gamma_C/\thicksim_C$. These orbits are the direct analogs of the physical states of the PN approach. In practice, we do not attempt to actually integrate the gauge flows such as~\eqref{eq:ExampleAlgFlow}, which would indeed be more complicated than constructing physical states. Instead the orbit space is sampled by restricting to a relational gauge. Since constraint~\eqref{eq:momcon} is of the form~\eqref{eq:idealConstraint}, the discussion of Sect.~\ref{sec:AlgebraicIdealExample} directly applies. Picking particle $A$ as the reference, we identify $\hat{Z}=\hat{q}_A$\ as the reference variable, its commutant $Z'$\ consists of polynomials generated by $\hat{q}_A$\ and $\{ \hat{q}_i, \hat{p}_i\}_{i\in \bar{A}}$. It is natural to use the algebra generated by just $\{ \hat{q}_i, \hat{p}_i\}_{i\in \bar{A}}$\ as the corresponding fashionable algebra $\mathcal{F}$.\footnote{See Sect.~III-D-3 of \cite{AlgebraicPoT} for a discussion of the freedom remaining in the choice of $\mathcal{F}$\ once the reference variable is fixed.} The gauge choice is completed by fixing the value of the reference variable $q_A=\rho$, which corresponds to additional conditions on the algebraic states $\omega(\hat{q}_A \hat{A})=\rho\, \omega(\hat{A})$\ for all $\hat{A} \in \mathcal{A}$.

The result of this construction is that the system, when viewed from within the $A$-frame, has the degrees of freedom characterized completely by the subalgebra $\mathcal{F}$, which is equivalent to the reduced algebra of the PN approach~\eqref{eq:redAlgebraEx}. (See Sect.~\ref{sec:frameDynamics} for the discussion of projection of the physical Hamiltonian flow into the gauge-fixed reference frame perspective.) Any state on $\mathcal{F}$\ corresponds to a unique gauge-fixed state on the entire algebra $\mathcal{A}$. We think of this state as the gauge-sampled representative of its corresponding orbit. This should be analogous to the sampling of a physical Hilbert state~\eqref{eq:ExamplePhysStates} by the Page-Wootters method~\eqref{eq:ExamplePageWoottersState}, and in Sect.~\ref{sect:equivalence} we establish this link concretely. The value of the reference variable $q_A$\ can be shifted by the residual symmetry transformation generated by $\hat{C}$, the flow of which restricted to $\mathcal{F}$\ is generated by $\displaystyle \left( -\sum_{i\in \bar{A}}\hat{p}_i \right)$. However, in its original form the algebraic construction possessed no direct way for transforming between gauge choices associated with a pair of \emph{different} subsystems (for example from particle $A$-frame to particle $B$-frame). Sect.~\ref{sect:switchingIdeal} leans on the PN approach to create an algebraic method for switching between ideal reference frames.
\\
\\
\noindent
As already mentioned in Sect.~\ref{sect:effective}, the effective approach can, in essence, be thought of as the semiclassical formulation of the algebraic approach. Here the space of states is characterized by a particular collection of expectation values that they assign to the elements of $\mathcal{A}$: the classical variables $q_i:=\langle \hat{q}_i \rangle$, $p_i = \langle \hat{p}_i \rangle$, and the quantum variables of~\eqref{eq:QuantumVariables} comprised of all the mixed moments $\Delta(q_A^{n_A} p_A^{m_A} q_B^{n_B} p_B^{m_B}...)$. The dynamics on this phase space is generated via the quantum Poisson bracket of~\eqref{eq:quantumPoisson} by the quantum Hamiltonian function
\begin{equation}
    \label{eq:HQ}
    H_Q := \langle \hat{H} \rangle = \sum_{i \in I} \frac{p_i^2 + (\Delta p_i)^2}{2m_i} + \expval*{V(\{\hat{q_i} - \hat{q}_j\})},
\end{equation}
where the last term can be expanded similarly when the potential is known. For instance, if the potential were quadratic $V \sim (\hat{q}_i -\hat{q}_j)^2$, one would have the following terms appearing
\begin{equation}
    \expval{V} \sim (q_i - q_j)^2 + (\Delta q_i)^2 + (\Delta q_j)^2 - 2\Delta (q_i q_j).
\end{equation}
The constraint~\eqref{eq:momcon} gives rise to an infinite number of quantum constraint functions: the expectaion value of the constraint
\begin{equation} \label{eq:ExampleEffectiveC}
    C = \langle \hat{C} \rangle =  \sum_{i \in I} p_i,
\end{equation}
as well as an additional constraint function $\langle \hat{A} \hat{C} \rangle$ for each linearly independent element $\hat{A} \in \mathcal{A}$. It is convenient to use moment operators to systematically generate the tower of constraint functions order-by-order within the semiclassical hierarchy. For example, at order $\hbar$, we get $2N$\ independent constraint functions in addition to $C$\ of~\eqref{eq:ExampleEffectiveC}
\begin{align}
    C_{q_i} &:= \langle \widehat{\Delta q_i} \hat{C} \rangle =  \Delta(q_i p_i) + \frac{1}{2}i \hbar + \sum_{j \neq i} \Delta(q_ip_j) ,  \nonumber \\ 
    \quad C_{p_i} &:= \langle \widehat{\Delta p_i} \hat{C} \rangle = (\Delta p_i)^2 + \sum_{j \neq i} \Delta(p_i p_j).
\end{align}
If we were to truncate the system to this order (corresponding to $\mathscr{T}_2$\ in \eqref{eq:TruncOperator}) then this set of $2N+1$ effective constraints contains all of the remaining independent constraints.

Picking $A$ as our reference frame, we can solve these constraints to eliminate $p_A$\ and its moments in exact analogy to~\eqref{eq:ExampleAlgEliminateA}
    \begin{align}
        &p_A = - \sum_{i \in \bar{A}} p_i, \\
        &\Delta(p_A q_i) = - \sum_{j \in \bar{A}} \Delta(q_i p_j) - \frac{1}{2} i \hbar \ , \quad i \neq A, \nonumber\\ 
        &\Delta(p_A p_i) = - \sum_{j \in \bar{A}} \Delta(p_i p_j). \nonumber
    \end{align}
The algebraic $A$-gauge conditions truncated to this order correspond to setting $q_A=\rho$\ and $\Delta(q_A y_i) = 0$ where $y_i \neq p_A$.

\section{Equivalence of approaches for ideal quantum reference frames}
\label{sect:equivalence}

In this section we directly compare the three QRF approaches introduced in the previous section: the PN framework, the algebraic approach and the effective semiclassical treatment. The three methods still have different domains of applicability and, before we proceed with our comparison, we determine that their current overlap is precisely the case of the ideal reference frame~\eqref{eq:con1}, \eqref{eq:idealConstraint}, cf.~Sect.~\ref{sssec_orientations}. The equivalence we will establish pertains to this case and requires aligning a few assumptions made across the three approaches. In particular, as discussed in Sect.~\ref{sec:AlgebraicMethod}, the algebraic method for constructing a QRF setup assumes that there is some algebra $\mathcal{A}$\ of operators on the kinematical Hilbert space $\mathcal{H}$\ that contains the gauge generating constraint $\hat{C}=\hat{C}^*$\ and a reference orientation observable operator $\hat{Z}=\hat{Z}^*$, such that
\begin{equation}
    \label{eq:ZCconjugate}
    [\hat{Z}, \hat{C}] = i\hbar \ .
\end{equation}
Now, let us assume, following the PN framework construction of Sect.~\ref{sect:trinity} that $\mathcal{H}=\mathcal{H}_R \otimes \mathcal{H}_S$, and that $\hat{Z} \in \mathcal{A}_R$. By taking the trace of~(\ref{eq:ZCconjugate}) over the system degrees of freedom with an arbitrary normalized system state, we conclude that there must be some operator $\hat{\tilde{G}}_R$ on $\mathcal{H}_R$\ that is canonically conjugate to the restriction of  $\hat{Z}$\ to $\mathcal{H}_R$. (Since $\hat{Z} \in \mathcal{A}_R$, this restriction is trivial.) Trivially extending $\hat{G}_R =  \hat{\tilde{G}}_R\otimes \mathds{1}_S\in \mathcal{A}_R$\ we have
\begin{equation}
[\hat{Z}, \hat{G}_R] =  i\hbar  \ .
\end{equation}
We now define $\hat{G}_S:=\hat{C}-\hat{G}_R$\ to put the general algebraically treatable constraint into the same form as~(\ref{eq:Constraint}). Clearly, the existence of a canonical conjugate frame orientation observable $\hat{Z}$ means that the algebraic approach requires $\sigma_R=\mathbb{R}$, so that the frame is ideal. On the other hand, unlike our presentation of the PN framework in Sect.~\ref{sect:trinity}, the algebraic treatment allows $\hat{G}_S$ to explicitly depend on $\hat{Z}$. This is in principle also permitted in the PN framework, e.g.\ see \cite{Hoehn:2023axh} but not yet fully developed. Additionally, all three approaches apply to some factorizable constraints 
\begin{equation} \label{eq:factorcon}
\hat{C} = \hat{N} (\hat{G}_R+\hat{G}_S) \ \,,
\end{equation}
where $\hat N$ is some factor that may or may not have $0$ in its spectrum. For example, it may be the lapse factor in a Hamiltonian constraint, or, as we will discuss in Sect.~\ref{sect:degenerateFrames}, a factor in a constraint involving degenerate QRFs, such as relativistic internal clocks.
In order for such a constraint to be treatable by all three methods, we need $[\hat{N},\hat{G}_R+\hat{G}_S]=0$\ in addition to the same ideal frame conditions for the factor constraint $(\hat{G}_R+\hat{G}_S)$\ as before.

For the rest of this section (as well as in Sect.~\ref{sect:switchingIdeal} and~\ref{sect:rel}) we focus on the current common domain of the three approaches, namely, ideal QRFs, governed by constraint~\eqref{eq:idealConstraint}. In this context, we first establish the equivalence between the PN and the algebraic approaches in Sect.~\ref{sect:algebraicIdealClock}, and then use the existing (semiclassical) connection between the algebraic and the effective approaches discussed in Sect.~\ref{sssec_algeffequiv} to relate the latter to the PN framework in Sect.~\ref{sect:effectiveIdealClock}. We look at the relation between the three approaches within the concrete example of the free Newtonian particle in Sect.~\ref{sect:NewtonianParticle}. In Sect.~\ref{sect:degenerateFrames}, we extend our equivalence argument to the case of degenerate ideal QRFs, such as relativistic internal clocks.

\subsection{Linking the algebraic approach with the PN framework}
\label{sect:algebraicIdealClock}

Here, we shall establish the equivalence between the algebraic and PN approaches in the following sense: the PN framework provides a concrete Hilbert space realization of the more abstract algebraic approach, thereby encoding equivalent physical data.

The end result of the algebraic QRF construction of Sect.~\ref{sec:AlgebraicMethod} in the case of an ideal QRF~\eqref{eq:idealConstraint} is the observed system subalgebra $\mathcal{F}=\mathcal{A}_S\subset \mathcal{A}$\ together with the flow generated by $\hat{C}$\ that transforms the observed system $S$\ along with the changes of orientation of the reference system $R$. The view of $S$ from the perspective of frame $R$\ when its orientation is $\rho$ is captured by the almost-positive states $\omega_{R|\rho}$.\footnote{{For convenience we slightly change notation here compared to Sect.~\ref{sec:AlgebraicMethod}. Algebraic gauge-fixed states will from now on be labeled using the subsystem label (e.g.\ $R$) rather than using the reference orientation observable (e.g.\ $q_R$).}}. These states are required to be positive on $\mathcal{A}_S$ but are otherwise allowed to take arbitrary values on its elements. However, the values they assign to all other elements of $\mathcal{A}$\ are completely fixed by constraint conditions and gauge choice
\begin{equation}
    \label{eq:algebraicFrame}
    \omega_{R \vert \rho}(\hat{a} \hat{C}) = 0, \quad \omega_{R \vert \rho}(\hat{q}_R \hat{a}) = \rho \omega_{R \vert \rho}(\hat{a}), 
\end{equation}
for all $\hat{a} \in \mathcal{A}$. Given a system observable $\hat{f}_S\in \mathcal{A}_S$, the value $\omega_{R \vert \rho}(\hat{f}_S)$\ is the expectation value of measurement $f_S$\ when $R$ is in orientation $\rho$.

Sect.~\ref{sect:trinity} provides two methods for capturing the same $R$-perspective information within the PN framework: (1) by constructing a relational Dirac observable $\hat{O}_R^{\rho}(\hat{f}_S)$~\eqref{eq:diracObservable}  and evaluating its expectation value in a physical state and (2) by evaluating the expectation value of the non-invariant system observable using relational Page-Wootters states obtained by projection~\eqref{eq:PWreduction}. As originally shown in~\cite{Trinity} these two ways of jumping into $R$'s perspective are equivalent, with both methods leading to identical relational expectation values~\eqref{eq:PWExpVal}. Below we show that the same expectation value is also obtained if one uses an algebraic almost-positive state corresponding to frame $R$.

Given a kinematical system observable $\hat{f}_S \in \mathcal{A}_S$ and an algebraic solution-state to the constraint with the frame properties~\eqref{eq:algebraicFrame} (see Sect.~\ref{sect:algebraicClock}), 
the following equation holds (see appendix~\ref{app:equivalenceAlgebraic} for details)
\begin{equation}
    \omega_{R \vert \rho'}(\hat{O}_R^{\rho}(\hat{f}_S)) = \omega_{R \vert \rho'}( e^{-i(\rho' - \rho) \hat{G}_S / \hbar} \hat{f}_S e^{i(\rho' - \rho) \hat{G}_S / \hbar}).
\end{equation}
Note that in the above expression the orientation reading of the algebraic state $\rho'$ is not necessarily equal to the conditioning $\rho$ in the definition of the Dirac observable. When we do set $\rho'=\rho$, we immediately get
\begin{equation}
    \omega_{R \vert \rho}(\hat{O}_R^{\rho}(\hat{f}_S)) = \omega_{R\vert \rho}(\hat{f}_S).
    \label{eq:equivalenceAlgebraic}
\end{equation}
We have therefore established that the expectation value of a kinematical system observable $\hat{f}_S$ evaluated in an algebraically gauge-fixed state $\omega_{R\vert \rho}$ coincides with the value assigned by that state to the corresponding relational
Dirac observable $\hat{O}_R^{\rho}(\hat{f}_S) \in \mathcal{A}_\text{inv}$. At a formal level, comparing this with~\eqref{eq:PWExpVal}, this expectation value is also equivalent to the one obtained via Page-Wootters reduced states within the PN framework.

To see this last equivalence explicitly, we propose a concrete Hilbert space realization of the algebraic reference frames 
\begin{equation}
    \omega_{R \vert \rho}(.) := \expval{(\dyad{\rho} \otimes \mathds{1}_S) (.)}{\psi_\text{phys}}_\text{kin},
    \label{eq:stateReal}
\end{equation}
where $\ket{\psi_\text{phys}} \in \mathcal{H}_\text{phys}$ is a physical state that corresponds to a solution of the constraint. This expression coincides with conditioning the physical state on the ``bra'' side using~\eqref{eq:PWreduction} and extending the resulting state back to the kinematical Hilbert space using the definite reference state $|\rho\rangle$. Clearly, because of the ``projector'' $\dyad{\rho}$ on the left of the argument and with a physical state on the right, 
this realization obeys both of the algebraic frame and constraint properties (\ref{eq:algebraicFrame}).

Using an algebraic state of the form~\eqref{eq:stateReal}, we show in appendix~\ref{app:equivalenceAlgebraic} that (\ref{eq:equivalenceAlgebraic}) slightly generalizes and holds in the same form also for non-ideal QRFs. 
Although the expectation value in (\ref{eq:equivalenceAlgebraic}) is with respect to a gauge-fixed state; using the concrete Hilbert space realization (\ref{eq:stateReal}), 
we can cast it as a physical expectation value. The key is to note that while $\ket{\psi_\text{phys}}$ is a physical state, 
$(\dyad{\rho} \otimes \mathds{1}_S ) \ket{\psi_\text{phys}} \equiv \ket{\psi_\text{kin}}$ is a kinematical state that group averages to the same physical state, i.e.\
$ \Pi \ket{\psi_\text{kin}} = \ket{\psi_\text{phys}}$. Explicitly, we have
\begin{align}
    &\Pi \ket{\psi_\text{kin}} \nonumber \\ &\quad = \frac{1}{2 \pi \hbar} \int_\mathbb{R}  \dd{\rho'} \left(\dyad{\rho + \rho'}{\rho} \otimes \mathds{1}_S \right) e^{-i\rho' \hat{G}_S / \hbar} \ket{\psi_\text{phys}} \nonumber \\
    &\quad = \frac{1}{2 \pi \hbar} \int_\mathbb{R} \dd{\rho'} (\dyad{\rho + \rho'}{\rho + \rho'} \otimes \mathds{1}_S) \ket{\psi_\text{phys}} \nonumber \\ &\quad = \ket{\psi_\text{phys}} \ ,
\end{align}
where we used $\hat{G}_S \ket{\psi_\text{phys}} = - \hat{G}_R \ket{\psi_\text{phys}}$, shifted $\rho' \to \rho' - \rho$, and used the resolution of the identity (\ref{eq:idResolution}). 
From the definition of the physical inner product (\ref{eq:PhysicalIP}) and~\eqref{eq:PWExpVal} we find
\begin{align}
\label{eq:algDirac} 
    &(\psi_\text{phys} \vert \dirac{\hat{f}_S}{R}{\rho} \vert \psi_\text{phys})= \mel{\psi_\text{kin}}{\,\Pi\, \dirac{\hat{f}_S}{R}{\rho}}{\psi_\text{kin}} \nonumber\\ 
    &= \mel{\psi_\text{phys}}{(\dyad{\rho} \otimes \mathds{1}_S) \dirac{\hat{f}_S}{R}{\rho}}{\psi_\text{phys}} \\&= \omega_{R \vert \rho}(\dirac{\hat{f}_S}{R}{\rho}) . \nonumber
\end{align}
We used the fact that $\dirac{\hat{f}_S}{R}{\rho}$ being a strong Dirac observable commutes with the projector $\Pi$. 

Finally, invoking again~\eqref{eq:PWExpVal}, we  arrive at the equalities
\begin{align}\label{eq:Phys-to-PWexp}
   \omega_{R \vert \rho}(\dirac{\hat{f}_S}{R}{\rho}) &= \left(\psi_{\rm phys}\big|\hat{{O}}^\rho_R(\hat{f}_S)\big|\psi_\text{phys}\right) \nonumber\\
    &= \expval*{\,\hat{f}_S\,}{\psi_{S| R}(\rho)}
     \\
    &=\mel{\psi_\text{phys}}{(\dyad{\rho} \otimes \mathds{1}_S) \hat f_S }{\psi_\text{phys}} \nonumber\\ &=\omega_{R \vert \rho}(\hat{f}_{S})\,. \nonumber
\end{align}
(In the third expression above, the expectation value is taken within the perspectival Hilbert space, with $\hat{f}_{S}$ denoting the observable's action on $\mathcal{H}_S$.) Hence, we established that the Page--Wootters conditioning is the Hilbert space-based realization (\ref{eq:stateReal}) of gauge-fixed states of the algebraic approach (and slightly generalizes them to encompass non-ideal frames).

\subsection{Linking the semiclassical effective approach with the PN framework}
\label{sect:effectiveIdealClock}

In the effective semiclassical construction, the algebraic quantum state is replaced by the expectation values and moments it assigns to kinematical observables. The description of $S$\ from the perspective of $R$\ consists of the expectation values and moments of the observables of $S$\ up to some truncation order $M$, which are restricted to correspond to a positive state on $\mathcal{A}_S$\ but are otherwise free to take any values. The values of all other classical and quantum variables are then completely determined by the constraint functions $\langle\hat{a} \hat{C} \rangle = 0$ and gauge conditions~\eqref{eq:EffectiveGaugeCond}. To find the relational expectation value of a system observable $\hat{f}_S$, we would expand $\langle \hat{f}_S \rangle$\ in expectation values and moments as in~\eqref{eq:effExpand} to order $M$.

As established in Sect.~\ref{sssec_algeffequiv}, before truncation is performed, effective constraint functions and gauge conditions are exactly equivalent to their algebraic counterparts~\eqref{eq:algebraicFrame}. It then follows from~\eqref{eq:equivalenceAlgebraic} that
\begin{equation}
    \langle \hat{O}_R^{\rho}(\hat{f}_S) \rangle = \langle \hat{f}_S \rangle \ ,
\end{equation}
when both sides are expanded in classical and quantum variables as in~\eqref{eq:effExpand} and the constraint and $R$-gauge conditions are imposed. It further follows that the equality still holds when both sides are truncated at semiclassical order $M$, giving us an effective version of agreement with the PN framework.

There is, however, a priori one important caveat. A truncated effective system possesses its own notion of invariant observables, namely the functions $h_\text{obs}$ that have a vanishing Poisson bracket~\eqref{eq:quantumPoisson} with the truncated system of constraints up to order $M$, when evaluated on the truncated constraint surface, that is, 
\begin{equation}\label{eq:effectiveObservables}
\mathscr{T}_M \left\{ \mathscr{T}_M(h_\text{obs}), \mathscr{T}_M( \langle \hat{a} \hat{C} \rangle ) \right\} \approx_{\Gamma_C}  0 \ ,  
\end{equation}
for any $\hat{a} \in \mathcal{A}$. We therefore need to explicitly verify that truncated expressions for the relational Dirac observables satisfy~\eqref{eq:effectiveObservables}, yielding invariant functions of the truncated effective QRF.

As noted earlier, these observables are strong Dirac observables \cite{Trinity} and from (\ref{eq:quantumPoisson}) it immediately follows that $\acomm*{\expval*{\dirac{\hat{f}_{S}}{R}{\rho}}}{\expval*{\hat{C}}} = 0$. Furthermore, we find
\begin{align}\label{eq:effStrongDirac}
    &\acomm*{\expval*{\hat{O}_R^{\rho}(\hat{f}_S)}}{\expval*{\hat{a}\hat{C}}} 
    = \frac{1}{i \hbar} \expval*{\comm*{\hat{O}_R^{\rho}(\hat{f}_S)}{\hat{a} \hat{C}}}  \\
    &= \frac{1}{i \hbar} \expval*{\comm*{\hat{O}_R^{\rho}(\hat{f}_S)}{\hat{a}} \hat{C}}
    + \frac{1}{i \hbar}  \expval*{\hat{a} \comm*{\hat{O}_R^{\rho}(\hat{f}_S)}{\hat{C}}} \ , \nonumber
\end{align}
where the second term in the final expression is identically zero while the first term vanishes on the constraint surface.
 Thus, the functions $\expval*{\dirac{\hat{f}_{S}}{R}{\rho}}$ are gauge invariant on the untruncated quantum phase space $\Gamma$.

In order to show that this is still true after truncation up to the corresponding order, we use the result from~\cite{SemiclassicalLie} that the truncation operation commutes with the quantum Poisson bracket. That is, for any pair of functions $f$, $g$ of generator expectation values $\langle \hat{y}_i\rangle$\ and their moments~\eqref{eq:QuantumVariables} we have
\begin{equation}
    \mathscr{T}_M \acomm*{\mathscr{T}_M (f)}{\mathscr{T}_M (g) } 
    = \mathscr{T}_M \acomm*{f}{g}.
    \label{eq:truncPoisson}
\end{equation}
We have just shown that the right-hand side in (\ref{eq:truncPoisson}) vanishes for the Poisson bracket between a Dirac observable and all effective constraints. 
Therefore, we find that the Poisson bracket between truncated Dirac observables and truncated constraints on the left-hand side also vanishes, when evaluated on the constraint surface, to within the order of truncation. 
In other words, the following diagram commutes up to terms of semiclassical order $M+1$
\begin{widetext}
\begin{equation}
    \begin{tikzcd}[column sep=6em, row sep=3em]
        \expval*{\dirac{\hat{f}_{S}}{R}{\rho}}, \expval*{\hat{C}} \arrow[shift right]{r}{\mathscr{T}_M} \arrow[swap]{d}{\acomm{.}{.}} & \mathscr{T}_M \expval*{\dirac{\hat{f}_{S}}{R}{\rho}}, \mathscr{T}_M \expval*{\hat{C}} \arrow{d}{\acomm{.}{.}} \\
        \acomm*{\expval*{\dirac{\hat{f}_{S}}{R}{\rho}}}{\expval*{\hat{C}}} \arrow[shift right, swap]{r}{\mathscr{T}_M} & \acomm*{\mathscr{T}_M \expval*{\dirac{\hat{f}_{S}}{R}{\rho}}}{\mathscr{T}_M \expval*{\hat{C}}} \approx {\mathscr{T}_M} \acomm*{\expval*{\dirac{\hat{f}_{S}}{R}{\rho}}}{\expval*{\hat{C}}}
    \end{tikzcd}
\end{equation}
\end{widetext}
Hence, within the truncated system, the truncated expectation values of relational Dirac observables $\mathscr{T}_M \expval*{\dirac{\hat{f}_{S}}{C}{\rho}}$ remain invariant  up to the order of truncation, providing a concrete link between effective QRFs and the PN framework.

\subsection{Example: free Newtonian particle}
\label{sect:NewtonianParticle}

To showcase the interplay between the three approaches in a concrete setting, let us consider a system of two particles governed by the following Hamiltonian constraint 
\begin{equation}
    {\hat{C}_H} = \hat{p}_C + \frac{1}{2} \hat{p}_S^2 \,,
\end{equation}
which was treated according to the PN framework in \cite{Hohn:2018toe}. Because the constraint is also the Hamiltonian, the reference frame is temporal: we will refer to it as the \emph{clock} $C$ and use notation that is typical in discussion of relational dynamics \cite{Trinity,Smith2019quantizingtime,giovannettiQuantumTime2015}; e.g., rather than $\rho$ we now use an evolution label $\tau$.

We take the system and clock algebras to be generated by complex polynomials in canonical variables $[\hat{q}_S, \hat{p}_S] = i\hbar$\ and $[\hat{t}_C, \hat{p}_C] = i\hbar$, respectively. We immediately see that both $\hat{p}_S$\ and $\hat{p}_C$\ are strong Dirac observables. Let us examine the expressions and expectation values of the relational Dirac observables $\dirac{\hat{f_S}}{C}{\tau}$. For example, using~\eqref{eq:idealDirac} (see also \cite{Hohn:2018toe}),
\begin{align}
    \label{eq:particleDirac}
    \dirac{\hat{q}_S}{C}{\tau} &= e^{-i (\hat{t}_C - \tau) \hat{p}^2_S/(2\hbar) }\, \hat{q}_S e^{i(\hat{t}_C - \tau) \hat{p}^2_S/(2\hbar) } \nonumber \\
    &=  \hat{q}_S - (\hat{t}_C - \tau)\hat{p}_S,
\end{align}
which indeed commutes with the constraint. 

Now, let $\omega_{C \vert \tau}$ be an algebraic solution-state {satisfying} \eqref{eq:algebraicFrame}. Using \eqref{eq:particleDirac}, 
\begin{align}\label{examplealg}
    \omega_{C \vert \tau}(\dirac{\hat{q}_S}{C}{\tau}) &= \omega_{C \vert \tau}(\hat{q}_S) - \omega_{C \vert \tau}((\hat{t}_C - \tau)\hat{p}_S) \nonumber \\ 
    &=  \omega_{C \vert \tau}(\hat{q}_S), 
\end{align}
as expected from \eqref{eq:equivalenceAlgebraic}.

For the effective approach we expand constraints and observables in quantum and classical variables as in~\eqref{eq:effExpand}. For example,
\begin{equation}\label{eq:NewtonianC_H}
    {C_H} = p_C+\frac{1}{2} p_S^2 + \frac{1}{2} (\Delta p_S)^2 
\end{equation}
and
\begin{equation} \label{eq:Newtonian_Oq_S}
\left\langle \dirac{\hat{q}_S}{C}{\tau} \right\rangle = q_S - (t_C - \tau)p_S - \Delta(t_C p_S) \ .
\end{equation}
Now fixing the gauge according to \eqref{eq:EffectiveGaugeCond} means setting $\Delta(t_C p_S)=0$ and further fixing $t_C=\tau$, we  obtain the effective analog of~\eqref{examplealg}
\begin{equation}
   \left\langle \dirac{\hat{q}_S}{C}{\tau} \right\rangle = q_S\,.
\end{equation}

Since the above expressions are not truncated, and since $[\dirac{\hat{q}_S}{C}{\tau}, \hat{C}_H]=0$, we are guaranteed to find $\{\langle \dirac{\hat{q}_S}{C}{\tau} \rangle, C_H\}=0$\ for the Poisson bracket defined by~\eqref{eq:quantumPoisson}. As a particular example of relation~\eqref{eq:truncPoisson}, we can check that the bracket still vanishes after the $M=0$\ truncation (i.e.\ at the classical order), which is the only non-trivial truncation for~\eqref{eq:NewtonianC_H} and~\eqref{eq:Newtonian_Oq_S}. The truncated expressions are
\begin{equation}
\begin{split}
    \mathscr{T}_0 \expval*{\dirac{\hat{q}_S}{C}{\tau}}  &= q_S - (t_C - \tau) p_S \ , \ \ \\ \text{and} \ \ \mathscr{T}_0 C_H &= p_C+\frac{1}{2} p_S^2 \ .
\end{split}
\end{equation}
Given that the Poisson bracket between classical variables is identical to the classical canonical Poisson relations, it is now straightforward to verify $\mathscr{T}_0 \{ \mathscr{T}_0 \langle \dirac{\hat{q}_S}{C}{\tau} \rangle, \mathscr{T}_0C_H \}=0$ as we expected based on~\eqref{eq:truncPoisson}.

\subsection{Dynamics within a reference frame}
\label{sec:frameDynamics}

As discussed at the start of Sect.~\ref{sect:trinity}, constraints associated with reference frame transformations fall into two broad categories: temporal, encoding reparametrization invariance of time evolution, and spatial (broadly speaking), encoding redundancy of the kinematical description of the system's configuration. While constructing QRFs, both types of constraints can be treated in the same way, as we have in fact done up to this point. An important difference does arise when one defines time evolution of the constrained theory either on the physical Hilbert space or within a frame perspective.

In the case of a temporal constraint, typical for gravitational theories, the constraint itself is also the Hamiltonian of the total system $\hat{C} = \hat{H}$ (see the example of Sect.~\ref{sect:NewtonianParticle}). The Hamiltonian therefore has a vanishing action on physical states, which, in this case, correspond to entire time-evolution orbits of the system. Change over time only emerges when one selects a particular reference system, the orientations of which will parametrize the flow of time. In the PN framework, one can track change over the time read by clock $C$\ within the observed system $S$\ encoded in a physical state by evaluating one-parameter families of relational Dirac observables $\hat{\mathcal{O}}^\tau_C(\hat{f}_S)$, as clock configuration $\tau$\ varies. Alternatively, one can jump into the clock frame perspective: the one-parameter family of $C$-frame states $|\psi_{S|C}(\tau) \rangle = \mathcal{R}_C (\tau) |\psi_\text{phys} \rangle$\ represents the time evolution of $S$\ relative to $C$\ encoded in $|\psi_\text{phys} \rangle$. In both cases, time evolution is generated by the true (i.e. non-vanishing) system Hamiltonian $\hat H_S$. The crux is that if there are multiple clocks, the physical time evolution generator depends on the choice of that clock \cite{Trinity,TrinityRel,Chataignier:2024eil,DeVuyst:2024uvd}. In the algebraic case, time evolution of $S$\ relative to clock $C$\ is similarly given by the one-parameter family of gauge-fixed states $\omega_{C|\tau} = S_{\hat{C}} (\tau-\tau_0) \omega_{C|\tau_0}$\ on $\mathcal{A}_S$, using the flow defined by~\eqref{eq:AlgebraicSFlow}. Analogously, on the truncated semiclassical phase space of the effective theory time evolution relative to $C$\ is given by the unique combination of Poisson flows $X_{\langle \hat{a}\hat{C}\rangle}(.)\equiv \{.,\langle \hat{a}\hat{C}\rangle\} $\ that preserves conditions~\eqref{eq:EffectiveGaugeCond} and uniformly increments the reading of the clock $q_c\equiv\langle \hat{q}_C \rangle$. Because time evolution governed by a Hamiltonian constraint is discussed at length elsewhere, we will not devote any more space to it here but instead refer the interested reader to in-depth discussions in~\cite{Trinity, TrinityRel,Chataignier:2024eil} for the PN framework, in~\cite{AlgebraicPoT} for the algebraic approach, and in~\cite{EffectiveConstrRel, EffectivePoT1, EffectivePoT2,EffectivePoTChaos} for the effective semiclassical approach.

When the constraint is non-temporal and the constrained system does possess a separate true Hamiltonian $\hat{H}$\ governing time evolution, as in the translation-invariant example of Sect.~\ref{sect:example}, one may wonder how this evolution projects into the perspective of a reference frame. Once again, in the case of the PN framework this procedure is discussed in detail elsewhere (see for example~\cite{Hoehn:2023ehz, Vanrietvelde:2018pgb,Vanrietvelde:2018dit}). However, the primary literature on algebraic and effective QRFs is focused on implementing the Hamiltonian constraint and does not discuss time evolution in non-temporal QRFs. The purpose of the rest of this subsection is to fill this gap within the literature.

In particular, we assume that the Hamiltonian is a strong Dirac observable $[\hat{H}, \hat{C}]=0$, otherwise additional restrictions may be needed in order to have a well-defined physical time-evolution flow. For the purpose of applying the algebraic method, we will further assume $\hat{H} \in \mathcal{A}$ (although it is also possible to work with an outer derivation on $\mathcal{A}$, as long as it annihilates the constraint).\footnote{Specifically, for Type III von Neumann algebras, the Hamiltonian need not be in the algebra.}  As a Dirac observable, the Hamiltonian has a well-defined action on the physical Hilbert space of the PN framework and can be used to generate Schr\"odinger- or Heisenberg-picture time evolution directly on $\mathcal{H}_\text{phys}$\ or $\mathcal{A}_\text{phys}$. When jumping into the frame $R$\ perspective, the Hamiltonian operator is projected like any other Dirac observable using~\eqref{eq:fSR}. In the ideal reference frame scenario, the Hamiltonian, like any other element of $\mathcal{A}$, consists of products between polynomials in $\hat{q}_R$\ and $\hat{p}_R$\ and some elements $\hat{f}_{S\, i} \in \mathcal{A}_S$. Let $\hat{H} \equiv H(\hat{q}_R, \hat{p}_R; \hat{f}_{S\, i})$\ be ordered so that all $\hat{q}_R$-s are on the left while the $\hat{p}_R$-s are on the right of each term in the sum, then the $R$-perspective reduced Hamiltonian on $\mathcal{H}_{S | R}$ is given by
\begin{equation}
    \label{eq:reducedH}
    \hat{H}_{R|\rho} :=\mathcal{R}_R (\rho) \hat{H} \mathcal{R}^\dag_R (\rho) = H(\rho,-\hat{G}_S,\hat{f}_{S\, i}) \, 
\end{equation}
The latter expression follows directly from definitions~\eqref{eq:PWreduction} and~\eqref{eq:PWinverse}, and the properties of the physical projection $\hat{C} \Pi|\psi \rangle=0$\ and~\eqref{eq:PWprojector}. 
By construction, $e^{-i\hat{H}_{R|\rho}t/\hbar} = \mathcal{R}_R(\rho) e^{-i\hat{H}t/\hbar} \mathcal{R}^\dag_R(\rho)$, so that the projected Hamiltonian generates the projection of the time evolution onto the states of $S$\ relative to $R$. That is, for $ | \psi_{S|R}(\rho) \rangle = \mathcal{R}_R(\rho) |\psi_\text{phys} \rangle$, we have
\begin{equation}
    e^{-i\hat{H}_{R|\rho}t/\hbar} | \psi_{S|R}(\rho) \rangle = \mathcal{R}_R(\rho) \left( e^{-i\hat{H}t/\hbar} |\psi_\text{phys} \rangle \right)\ .
\end{equation}

For an example, we consider again the model of Sect.~\ref{sect:example} and compute the form of the Hamiltonian~\eqref{eq:translationInvariantHam} when it is projected into a given particle's reference frame. For $R = A$, with $\rho_A=0$, we find 
\begin{widetext}
\begin{align}\label{eq:ExampleFrameAHam}
    \hat{H}_{A|0} &=  \sum_{i \neq A} \frac{\hat{p}_i^2}{2m_i} + \frac{1}{2m_A} (- \sum_{i \neq A} \hat{p}_i)^2 + \left. V(\{\hat{q}_i - \rho_A \}_{i \neq A}) \right|_{\rho_A=0} \nonumber \\
    &= \frac{1}{2} \sum_{i \neq A} \left( \frac{1}{m_i} + \frac{1}{m_A} \right) \hat{p}_i^2 + \sum_{\substack{i \neq j \\ i,j \neq A}} \frac{\hat{p}_i \hat{p}_j}{m_A} + V(\{\hat{q}_i\}_{i \neq A}).
\end{align}
\end{widetext}
Note that the potential only depends on the $\hat{q}_i - \rho_A$ since other relative position observables can be written as linear combinations: $\hat{q}_i - \hat{q}_j = \hat{q}_i - \rho_A - (\hat{q}_j - \rho_A)$. This result matches \cite[Eq.~(15)]{Vanrietvelde:2018pgb}.

In the algebraic and effective approaches the method for reducing time evolution is somewhat different, but yields an equivalent result for an ideal QRF. Specifically, while the space of physical states, $\Gamma_\text{phys}$\ of~\eqref{eq:AlgPhysicalStates}, is not explicitly constructed, time evolution generated by $\hat{H}$\ is well defined when evaluated on Dirac observables $[\hat{O}, \hat{C}]=0$, via
\begin{equation} \label{eq:AlgebraicFrameEvolution}
i\hbar\frac{d}{dt} \omega ( \hat{O} ) = \omega ( \comm*{\hat{O}}{\hat{H}}) \ . 
\end{equation}
This flow is gauge-independent in the sense that expressions on both sides of the differential equation~\eqref{eq:AlgebraicFrameEvolution} are invariant under the flows~\eqref{eq:AlgebraicGaugeFlow} generated by the constraint when the state is a solution of the constraint, that is $\omega(\hat{a}\hat{C})=0$. The above flow is only unique up to the addition of arbitrary gauge flows: the Hamiltonian $\hat{H}' = \hat{H} + \hat{h}\hat{C}$\ for any $\hat{h} \in \mathcal{A}$\ generates an equivalent evolution of Dirac observables when evaluated in solution states.

When reducing the Hamiltonian to gauge-fixed states of a particular reference frame, one has to select the unique equivalent Hamiltonian $\hat{H}'$\ that preserves the algebraic gauge conditions  in~\eqref{eq:algebraicFrame}. (Constraint conditions are automatically preserved since $[\hat{H}', \hat{a}\hat{C}]=\left([\hat{H}, \hat{a}] + [\hat{h}, \hat{a}\hat{C}] + [\hat{h}\hat{C}, \hat{a}] \right) \hat{C}$\ vanishes on solution states.) From the algebraic decomposition conditions for a reference frame listed in Sect.~\ref{sect:algebraicClock}, it follows that any element of $\mathcal{A}$\ can be uniquely written as a sum $\hat{a} = \hat{a}_1 + \hat{a}_2\hat{C}$, for some $\hat{a}_1, \hat{a}_2 \in \mathcal{A}$, such that $[\hat{a}_1, \hat{q}_R]=0$. Decomposing the original Hamiltonian in this way 
\begin{equation}
\hat{H} = \hat{H}_{R} + \hat{h}_0 \hat{C} \ ,
\end{equation}
we see that $\hat{H}_R=\hat{H} - \hat{h}_0\hat{C}$\ is precisely the equivalent Hamiltonian that generates the flow which preserves frame $R$\ gauge conditions. Explicitly, using $[\hat{H}_R, \hat{q}_R]=0$,
\begin{align}
i\hbar \frac{d}{dt} \omega \left( (\hat{q}_R - \rho) \hat{a} \right)  &= \omega \left( \left[ (\hat{q}_R - \rho) \hat{a}, \hat{H}_R \right]\right) \\ &= \omega \left( (\hat{q}_R - \rho) \left[ \hat{a}, \hat{H}_R \right]\right) \ , \nonumber
\end{align}
which vanishes on the states satisfying gauge conditions of frame $R$. 

Let us examine the form of $\hat{H}_R$\ for an ideal reference frame. Writing $\hat{H} \equiv H (\hat{q}_R, \hat{p}_R; \hat{f}_{S\, i})$\ with the same ordering as before, we note that any power of $\hat{p}_R$ can be written in the form
\begin{equation}
\hat{p}_R^n = \left(\hat{C}-\hat{G}_S \right)^n = \left( -\hat{G}_S \right)^n + \hat{a} \hat{C} \ ,
\end{equation}
for some $\hat{a}\in\mathcal{A}$, where we used the fact that $[\hat{G}_S, \hat{C}]=0$. Therefore, writing all powers of $\hat{p}_R$\ inside the expression for $\hat{H}$\ in this form, 
\begin{equation}
\hat{H} = H(\hat{q}_R, -\hat{G}_S; \hat{f}_{S\, i}) + \hat{b} \hat{C} \ ,
\end{equation}
for some $\hat{b}\in\mathcal{A}$. Since $[H(\hat{q}_R, -\hat{G}_S; \hat{f}_{S\, i}),\hat{q}_R]=0$, evidently $\hat{H}_R = H (\hat{q}_R, -\hat{G}_S; \hat{f}_{S\, i})$\ is the correct generator of time evolution as seen from the perspective of frame $R$. Evolving the value of an arbitrary relational observable $\hat{a}_S\in\mathcal{A}_S$\ in a gauge-fixed state using $\hat{H}_R$\ we get
\begin{align}
i\hbar \frac{d}{dt} \omega \left( \hat{a}_S \right) &= \omega \left( \left[ \hat{a}_S, H(\hat{q}_R, -\hat{G}_S; \hat{f}_{S\, i}) \right] \right) \nonumber \\ &=  \omega \left( \left[ \hat{a}_S, H(\rho, -\hat{G}_S; \hat{f}_{S\, i}) \right] \right) \nonumber \\ &=  \omega \left( \left[ \hat{a}_S, \hat{H}_{R|\rho} \right] \right) \ ,
\end{align}
where we used the fact that $\hat{q}_R$\ can be moved all the way to the left, since $[\hat{a}_S, \hat{q}_R]=0$, and applied the $R$-gauge conditions. The result is that the $R$-gauge time evolution in the algebraic approach is generated by a reduced Hamiltonian of the same form as~\eqref{eq:reducedH} of the PN construction. In particular, for the example of Sect.~\ref{sect:example}, the generator of time evolution in the particle-$A$\ frame would be precisely the Hamiltonian of~\eqref{eq:ExampleFrameAHam}.

Finally, let us briefly outline the analogous construction in the effective semiclassical approach. The most general time evolution flow that is compatible with gauge freedom encoded in the constraint $\hat{C}$, is generated via the Poisson bracket~\eqref{eq:quantumPoisson} by 
\begin{equation}
H' = \langle \hat{H}\rangle + \alpha_0 \langle \hat{C} \rangle + \sum_i\alpha_i C_i \ ,
\end{equation}
where $C_i$\ are the various additional quantum constraint functions up to the appropriate order (such as~\eqref{eq:EffectiveConstraintFunctions}) and $\alpha_i$\ are arbitrary functions of the classical and quantum variables. In order to project the Hamiltonian flow within a particular reference frame one looks for a combination of $\alpha_i$-s such that the flow $\{., H'\}$\ preserves the $R$-gauge conditions~\eqref{eq:EffectiveGaugeCond}. Using the equivalence between effective and algebraic constraints, gauge flows, and gauge conditions discussed in Sect.~\ref{sect:effective}, we can deduce that, in the case of an ideal QRF, $H' = \langle \hat{H}(\hat{q}_R, -\hat{G}_S; \hat{f}_{S\, i}) \rangle$\ provides the desired flow, which on the gauge-fixing surface is equivalent to the flow generated by the expectation value $\langle \hat{H}_{R|\rho} \rangle$\ of~\eqref{eq:reducedH}.

\subsection{Degenerate quantum reference frames}
\label{sect:degenerateFrames}

As noted around~\eqref{eq:factorcon}, the currently developed form of all three approaches to QRFs can also be applied to some factorizable constraints. (For details, see discussions in \cite{TrinityRel,Hohn:2018toe,Hohn:2018iwn,DeVuyst:2024uvd} for the PN formalism, in \cite{QDSR,AlgebraicPoT} for the algebraic one, and in \cite{EffectiveConstrRel,EffectivePoT1,EffectivePoT2,EffectivePoTChaos} for the effective one.) Let us now explore their relation for an interesting family of such cases.

So far we have assumed that the reorientation generator $\hat{G}_R$ in~\eqref{eq:Constraint} has a non-degenerate continuous spectrum. We will now permit it to have a degenerate spectrum, provided the degeneracy is independent of the eigenvalues (except for possibly a zero-measure set), the prime example being $\hat{G}_R\propto\hat{p}_R^2$ for some canonical momentum $\hat p_R$, which has a two-fold degenerate spectrum (except for the zero-measure set $p_R=0$). 
In this case, $R$'s Hilbert space decomposes into a direct sum over its $m\in\mathbb{N}$ degeneracy sectors, $\mathcal{H}_R=\bigoplus_{\lambda=1}^m\mathcal{H}_{R}^\lambda=L^2(\sigma_R)\otimes\mathbb{C}^m$, where $\sigma_R$ is again the spectrum of $\hat{G}_R$. Denote by $\Pi_\lambda$ the orthogonal projector on $\mathcal{H}_R$ onto its $\lambda$-sector $\mathcal{H}^\lambda_R$. Clearly, $\sum_\lambda\,\Pi_\lambda=\mathds1_R$, and so we may decompose the coherent group averaging operator~\eqref{eq:projector} as
\begin{equation}
   \Pi=\sum_{\lambda=1}^m\tilde\Pi_\lambda\,,\quad\text{where}\quad \tilde\Pi_\lambda:=(\Pi_\lambda \otimes \mathds{1}_S) \Pi\,,
\end{equation}
which means that the direct sum structure carries over to the physical Hilbert space  \cite{TrinityRel,DeVuyst:2024uvd}
\begin{equation}
     \mathcal{H}_\text{phys} = \bigoplus_{\lambda = 1}^m \mathcal{H}_\lambda,
\end{equation}
and $\tilde\Pi_\lambda: \mathcal{H}_\text{kin} \to \mathcal{H}_\lambda$.

The relevant gauge-invariant subalgebra at the kinematical level is then \cite{DeVuyst:2024uvd} 
\begin{equation}    
    \label{eq:degenerateInvAlg}
    \mathcal{A}_\text{inv} = \left(\mathcal{A}_R\otimes \mathcal{A}_S \right)^{\hat{C}} \otimes \mathcal{B}(\mathds{C}^m)\,,
\end{equation}
where $\mathcal{A}_R$ contains the same QRF operators on $L^2(\sigma_R)$ as in the non-degenerate case. That is,
the invariant algebra now consists of the invariant algebra for the case of a QRF with non-degenerate reorientation generator spectrum $\sigma_R$ given in \eqref{eq:invAlgebra}, tensored with a matrix algebra encoding the sector dependency and therefore containing sector-swapping operators. Similarly, its representation on $\mathcal{H}_{\rm phys}$ is given by 
\begin{equation}
    \mathcal{A}_{\rm phys}=\mathcal{A}_{\rm phys}^{\rm nd}\otimes\mathcal{B}(\mathbb{C}^m)\,,
\end{equation}
where $\mathcal{A}_{\rm phys}^{\rm nd}$ is the algebra of relational observables for the corresponding non-degenerate QRF discussed above~\eqref{eq:relobsphys} and the cross-sector algebra is unaffected.

For definiteness, we now specialize to double degeneracy, which is physically the most interesting class. In this case, we may still invoke an ideal QRF orientation observable. We will exploit this to argue that the three approaches encode the same relational information per degeneracy sector in this case.

\subsubsection{Doubly degenerate frame reorientation operators}\label{sssec_doubledeg}

For the remainder of this section we will consider constraints of the form,
\begin{equation}\label{eq:DegenerateC}
    \hat{C} = \hat{p}_R^2 - \hat{G}_S  \equiv \hat{C}_{+} \hat{C}_{-},
\end{equation}
where $\comm*{\hat{p}_R}{\hat{G}_S} = 0$, $\hat{G}_S$\ is for simplicity assumed to be positive-semi-definite, and $\hat{C}_\pm = \hat{p}_R \pm \sqrt{\hat{G}_S}$. To make contact with the above, denote by $\Pi_+,\Pi_-$ the projectors onto the subspaces of $\mathcal{H}_R=L^2(\mathbb{R})$ with $p_R\geq0$ and $p_R<0$, respectively. This leads to a decomposition into two sectors, $\mathcal{H}_+$ and $\mathcal{H}_-$, of the physical Hilbert space.
One can interpret the choice of the sector as left vs right movers if $\hat{C}$\ generates spatial translations, or as positive vs negative frequency modes if $\hat{C}$\ is a Hamiltonian constraint. A well-known example of the latter is the Hamiltonian constraint of a non-interacting relativistic particle.

Consider now the $+$-sector. We may then consider the constraint  on $\mathcal{H}_{\rm kin}^+:=\Pi_+\left(\mathcal{H}_{\rm kin}\right)$, where it takes the form $\hat{C}=\hat{N}\hat{C}_-$, with $\hat{N}=\hat{C}_+$ and $\hat{N}$ introduces no additional zero-solutions.\footnote{On the $+$-sector, $\hat{N}$ can only have zero in its spectrum for $p_R=G_S'=0$, a case which also yields zero for $\hat{C}_-$.} Since clearly $[\hat C_+,\hat C_-]=0$, $\hat{C}$ takes the form~\eqref{eq:factorcon} on $\mathcal{H}^+_{\rm kin}$, and we may equivalently describe it in terms of the factor $\hat{C}_-$ alone. The corresponding physical Hilbert space is then $\mathcal{H}_+$. In fact, we may equivalently consider $\hat{C}_-$ as the constraint on the \emph{full} kinematical Hilbert space $\mathcal{H}_{\rm kin}$, again yielding $\mathcal{H}_+$ as the corresponding physical Hilbert space. At this point, $R$ has become an ideal QRF and we are entitled to employ the exact same techniques as before in each of the three approaches. By complete analogy, one obtains $\mathcal{H}_-$ as the physical Hilbert by considering $\hat{C}_+$ on $\mathcal{H}_{\rm kin}$ and, similarly, $R$ is an ideal QRF in this case. This will ensure sector-by-sector equivalence of the three approaches, provided that focusing on $\hat{C}_\pm$ captures the same information sector-wise as working with the full $\hat{C}$.

In the PN framework, this indeed holds true. The above discussion shows that one obtains the same physical Hilbert space, sector-wise, regardless of whether one works with $\hat{C}_\pm$ or $\hat{C}$. However, there is a difference in the type of QRF orientation observable for $R$: working with $\hat{C}_\pm$ we may invoke an ideal QRF as noted above yielding a covariant QRF orientation POVM for $\hat{C}_\pm$, but not for $\hat{C}$. For example, this is the route taken in \cite{Hohn:2018toe,Hohn:2018iwn} and the one we will also pursue here. On the other hand, working with $\hat{C}$ instead, one may construct a covariant QRF orientation POVM that is necessarily fuzzy due to the boundedness of $\hat{p}_R^2$, thus yielding a non-ideal QRF. This is the route taken in \cite{TrinityRel,DeVuyst:2024uvd}. Despite these differences in the type of orientation observable for the QRF $R$, the two options yield the same relational observable algebras $\mathcal{A}^{\rm nd}_{\rm phys}$, sector-wise. To see this, first note that, for each degeneracy sector, an operator is a Dirac observable w.r.t.\ the pertinent $\hat{C}_\pm$ if and only if it is a Dirac observable w.r.t.\ the full constraint $\hat{C}$.\footnote{Indeed, suppose that $\hat{O}_\pm$ is a Dirac observable on $\mathcal{H}_\pm$, i.e.\ $[\hat{O}_\pm,\hat{C}_\mp]$ vanishes on $\mathcal{H}_\pm$ (as indicated in the main text, we focus here on observables leaving each degeneracy sector invariant). Then we also have that $[\hat O_\pm,\hat C]=\hat{C}_\pm[\hat O_\pm,\hat{C}_\mp]+[\hat{O}_\pm,\hat{C}_\pm]\hat{C}_\mp$ vanishes on $\mathcal{H}_\pm$ because both $[\hat{O}_\pm,\hat{C}_\mp]$ and $\hat{C}_\mp$ do. Similarly, if $[\hat O_\pm,\hat{C}]$ vanishes on $\mathcal{H}_\pm$, then so does $[\hat{O}_\pm,\hat{C}_\mp]$.} Second, with each choice of QRF orientation operator in the two cases, we dress the pertinent system algebra $\mathcal{A}_S$ to obtain the corresponding relational observables~\eqref{eq:diracObservable}. While the concrete relational observables are in general different for any given $\hat f_S\in\mathcal{A}_S$, the algebras that they form are the same. This is because we may now define the Page-Wootters reduction map $\mathcal{R}_R(\rho)$ sector-wise w.r.t.\ either choice of QRF orientation observable for $R$. Crucially, both are still unitaries sector-wise due to the arguments in Sect.~\ref{sect:PW}, and for both the reduction relation~\eqref{eq:fSR} holds. Due to our assumption that $\hat{G}_S$ is positive-semi-definite, we have that $\Pi_{|R}=\mathds{1}_S$ in either case, which means that for both choices of orientation observable $\mathcal{A}_{\rm phys}^{\rm nd}\simeq\mathcal{A}_S$.
For example, if $\mathcal{A}_S=\mathcal{B}(\mathcal{H}_S)$ then we get the full algebra of bounded operators on $\mathcal{H}_{\rm phys}$ in either case because $\mathcal{H}_S\simeq\mathcal{H}_\pm$. One argues similarly when $\mathcal{A}_S$ is the algebra of complex polynomials in some generators. Hence, without loss of information, we may proceed sector-wise in identical manner to the previous sections describing the PN framework by invoking the ideal QRF orientation observable and using $\hat{C}_\pm$.\footnote{To encompass also cross-sector observables, one may consider the direct sum of the Page-Wootters reduction maps across the sectors \cite{DeVuyst:2024uvd}.} 

A similar method of factorization and separate treatment of each degeneracy sector is also used within the algebraic approach. For a constraint of the form~\eqref{eq:DegenerateC}, one would select one of its commuting factors $\hat{C}_\pm$\ and implement the algebraic constraint and gauge-fixing conditions of Sect.~\ref{sec:AlgebraicMethod} using the chosen factor in place of the original constraint $\hat{C}$. As discussed in~\cite{QDSR} and~\cite{AlgebraicPoT}, the substitution for the factor is valid as long as the Dirac observables of the original constraint are invariant along the gauge flows~\eqref{eq:AlgebraicGaugeFlow} generated by its factor. This condition can either be verified directly by finding a complete set of Dirac observables of $\hat{C}$\ or enforced by placing restrictions on the solution states one considers (see~\cite[Sect.~IV]{AlgebraicPoT} for a detailed discussion). 

In previous discussions of the effective approach, the degeneracy treatment is \emph{a priori} different  \cite{EffectiveConstrRel} (see also \cite{EffectivePoT1,EffectivePoT2}). Instead of factorizing the constraint operator, Ref.~\cite{EffectiveConstrRel} constructs the full tower of effective constraints \eqref{eq:EffectiveConstraintFunctions} and truncates at the desired order. One uses the constraints to eliminate the classical and quantum variables containing $\hat{p}_R$. Due to the two-fold degeneracy of the generator $\hat{p}_R^2$, one encounters a sign choice when performing this procedure, leading to a pair of disjoint constraint surfaces which are quantum extensions of the two classical constraint surfaces defined by $C_\pm = 0$.\footnote{See also the classical discussion about constraint factorization and disjoint constraint surfaces in \cite{Hohn:2018toe,Hohn:2018iwn}.} 
At this point, gauge-fixing conditions can be imposed on each disjoint surface. Crucially, in Sect.~\ref{sec:degenerateEffective} below, we show that, for the constraints considered here and to order $\hbar$\ in the semiclassical expansion, the above procedure is, in fact, equivalent to first factorizing the constraint and then treating each factor constraint using effective semiclassical methods separately.

Thus, since the individual degeneracy sectors within each approach behave identically to the ideal non-degenerate frame scenario~\eqref{eq:idealConstraint}, all results derived for non-degenerate ideal QRFs -- in particular the equivalence between the three approaches -- still hold sector-by-sector. 

If we want to simultaneously consider the full state space, which is the direct sum of all sectors, then we need to also take into account the sector swaps residing in $\mathcal{B}(\mathds{C}^2)$ \eqref{eq:degenerateInvAlg}. Since these operators have not been considered in the algebraic approach yet, we refrain from asserting the equivalence conclusively for the case of including the cross-sector operators and leave this as an interesting future investigation.

\subsubsection{Constraint factorization in the effective approach for the degenerate case}
\label{sec:degenerateEffective}

Here we show that, to order $\hbar$, working with the full constraint or its two factors individually is equivalent also in the effective approach. This then guarantees the sector-wise equivalence among the three approaches also for the doubly degenerate case. 

For compactness of notation, we define the square root operator $\hat{H} := \sqrt{\hat{G}_S}$. The constraint~\eqref{eq:DegenerateC} factorizes as before 
\begin{equation}
    \label{eq:degFactor}
    \hat{C} = (\hat{p}_R - \hat{H}) (\hat{p}_R + \hat{H}) \equiv \hat{C}_+ \hat{C}_-.
\end{equation}
The square root $\hat{H}$ can be expanded about $\expval*{\hat{G}_S}$ to any semiclassical order, for instance, to linear order in $\hbar$ one finds 
\begin{widetext}
\begin{equation}
    \hat{H} := \sqrt{\langle \hat{G}_S \rangle +  \widehat{\Delta G}_S}
    = \sqrt{\langle \hat{G}_S \rangle } + \frac{1}{2} \frac{1}{\sqrt{\langle \hat{G}_S \rangle }} \widehat{\Delta G}_S - \frac{1}{8} \frac{1}{\langle \hat{G}_S \rangle ^{3/2}} ( \widehat{\Delta G}_S )^2 +O( \hbar^{3/2} ) \ ,
\end{equation}
\end{widetext}
so that 
\begin{equation}
    H = \expval*{\hat{H}} =  \sqrt{\langle \hat{G}_S \rangle } - \frac{1}{8} \frac{1}{\langle \hat{G}_S \rangle ^{3/2}} ( \Delta G_S )^2 \ .
\end{equation}
Similar expressions can be found for the moments involving $\hat{H}$, such as $\left( \Delta H\right)^2$.
Since $\hat{G}_S$ is a combination of the system generators $\hat{x}_i$, in order to find expressions for $H$, $\left( \Delta H\right)^2$, and so on, in terms of the expectation values $x_i=\langle \hat{x}_i\rangle$ and the corresponding quantum variables~\eqref{eq:QuantumVariables} one would further expand quantities such as $(\Delta G_S)^2$ in terms of $x_i$\ and moments to the desired order. For compactness, we omit this explicit expansion.

To order $\hbar$ the effective system of constraints contains 
\begin{equation}
    \label{eq:DegenerateEffC}
    \langle \hat{C} \rangle = p_R^2 - H^2 + \left( \Delta p_R \right)^2 - \left( \Delta H \right)^2 = 0 \ ,
\end{equation}
and one constraint per generator of the reference and system algebras $\expval*{({\hat{y}_i - \expval*{\hat{y}_i}})\hat{C}}$. For our purposes, it will be sufficient to consider $\langle \hat{C} \rangle$\ and the following two constraints to order $\hbar$
\begin{subequations}
    \begin{align}
    \langle (\hat{p}_R - \langle \hat{p}_R \rangle )\hat{C} \rangle &= 2p_R\left( \Delta p_R \right)^2 - 2H\Delta (p_R H)=0  \ ,    \label{eq:DegenerateEffCp_R}
    \\
    \langle (\hat{H} - \langle \hat{H} \rangle )\hat{C} \rangle &= 2p_R\Delta (p_R H) - 2H\left( \Delta H \right)^2=0 \ . \label{eq:DegenerateEffCH}
    \end{align}
\end{subequations}
Of course~\eqref{eq:DegenerateEffCH} is a linear combination of the ``basic'' constraints~\eqref{eq:EffectiveConstraintFunctions} associated with the generators of $\mathcal{A}_S$: $\langle (\hat{H} - \langle \hat{H} \rangle )\hat{C} \rangle = \sum_j f_{ij} \langle (\hat{x}_i - \langle \hat{x}_i \rangle )\hat{C} \rangle$, where $f_{ij}$ are some functions of the system $S$\ variables.  

Eqs.~(\ref{eq:DegenerateEffC}--\ref{eq:DegenerateEffCH}) allow us to entirely eliminate the three reference degrees of freedom $p_R$,  $\left( \Delta p_R \right)^2$, and $\Delta (p_R H)$ in terms of the quantities associated with the observed system $S$. Combining~\eqref{eq:DegenerateEffCp_R} and~\eqref{eq:DegenerateEffCH} we quickly get
\begin{equation}
    p_R^2 \left( \Delta p_R \right)^2 = H^2 \left( \Delta H \right)^2 \ .
\end{equation}
Using~\eqref{eq:DegenerateEffC} to eliminate $p_R$\ we get a quadratic equation in $\left( \Delta p_R \right)^2$
\begin{equation} \label{eq:DegenerateEffQuadratic}
\left( H^2+\left( \Delta H \right)^2-\left( \Delta p_R \right)^2\right) \left( \Delta p_R \right)^2 = H^2 \left( \Delta H \right)^2 \ .
\end{equation}
We truncate this equation to order $\hbar$\ by dropping products between 2nd order moments to find\footnote{In fact, the full quadratic equation~\eqref{eq:DegenerateEffQuadratic} has two real solutions one of which is incompatible with the assumed semiclassical hierarchy and the other reduces to this solution when truncated at order $\hbar$. See~\cite{EffectiveConstrRel} for details.}
\begin{equation}\label{eq:DegenerateEffMom1}
\left( \Delta p_R \right)^2 = \left( \Delta H \right)^2 \ .
\end{equation}
Substituting~\eqref{eq:DegenerateEffMom1} into~\eqref{eq:DegenerateEffC} gives us two solutions for $p_R$
\begin{equation}\label{eq:DegenerateEffMomCfactor}
p_R = \pm H \ .
\end{equation}
Substituting both into either~\eqref{eq:DegenerateEffCp_R} or~\eqref{eq:DegenerateEffCH} yields
\begin{equation}\label{eq:DegenerateEffMom2}
\Delta (p_R H) = \pm \left( \Delta H \right)^2 \ .
\end{equation}
The two sets of solutions that we obtain are entirely equivalent to what we would get if we used one of the factors $\hat{C}_\pm$\ in place of the full constraint. Explicitly, to order $\hbar$:
    \begin{align}
        \langle \hat{C}_\pm \rangle &= p_R  \pm H  = 0  \ ,
        \\
        \langle (\hat{p}_R - \langle \hat{p}_R \rangle )\hat{C}_\pm \rangle &=  \left( \Delta p_R \right)^2 \pm \Delta (p_R H) = 0 \ , \nonumber
        \\
        \langle (\hat{H} - \langle \hat{H} \rangle )\hat{C}_\pm \rangle &= \Delta (p_R H) \pm \left( \Delta H \right)^2 = 0 \ , \nonumber
    \end{align}
The above system is completely equivalent to the set of solutions~(\ref{eq:DegenerateEffMom1}--\ref{eq:DegenerateEffMom2}).

At order $\hbar$\ choosing subsystem $R$\ as the reference frame corresponds to the following gauge conditions \eqref{eq:EffectiveGaugeCond}
\begin{subequations}
    \begin{alignat}{2}
        \phi_0 &:= \left( \Delta q_R \right)^2 &&= 0,
        \\
        \phi_i &:= \Delta (q_R y_i) && = 0 ,
    \end{alignat}
\end{subequations}
with one $\phi_i$ for each generator $\hat{y}_i \in \mathcal{A}_S$.
The flow induced by $\langle \hat{C}_\pm \rangle$\ via the quantum Poisson bracket~\eqref{eq:quantumPoisson} preserves this gauge while advancing the gauge parameter $q_R=\langle \hat{q}_R\rangle$. To see this explicitly we note that $\{q_R, p_R\}=1$\ is the only non-vanishing Poisson bracket between $p_R$\ and the basic effective coordinates (see~\cite{EffectiveEq1}), so that, in particular, $\{ \left( \Delta q_R \right)^2, p_R\}=\{ \Delta (q_R y_i) , p_R\} = 0$. Since reference system variables commute with system $S$\ variables we also have $\{\left( \Delta q_R \right)^2, H\}=0$. Finally, because $[ \hat{q}_R, \hat{H}]=0$, we also have, to order $\hbar$,  
\begin{equation}
\{ \Delta (q_R y_i), H\}= \sum_j f_{ij} \Delta (q_R y_j)  \ ,
\end{equation}
where $f_{ij}$\ are some functions of the expectation values of system $S$\ generators and their moments, so that the expression vanishes on the gauge-fixed surface within the constraint surface. Therefore, $\{C_\pm, q_R\}=1$, while $\{C_\pm, \phi_i\} = 0$\ on the surface defined by imposing both the quantum constraint and the reference gauge conditions.

\section{Quantum reference frame transformations}
\label{sect:switchingIdeal}

Some of the most interesting questions in the study of QRFs arise in scenarios where multiple reference frames are available. After all, one is essentially always free to make a different split into system and reference. While the descriptions relative to distinct reference frames will be different, the principle of covariance tells us that the physical laws should remain the same under such transformations and the question is whether this remains true also under changes of internal QRF. Different choices of QRF decompose the total composite system of interest (of which each QRF is a subsystem itself) in different ways into gauge-invariant subsystems. At the level of observable algebras, this means that different choices of QRF ascribe distinct subalgebras of the relational observable algebra to a given kinematical subsystem $S$. This is known as subsystem relativity in the literature \cite{AliAhmad:2021adn,Hoehn:2023ehz,delaHamette:2021oex,DeVuyst:2024pop,DeVuyst:2024uvd,Castro-Ruiz:2021vnq,Araujo-Regado:2025ejs}. In other words, what one calls a subsystem is an observer-dependent concept, and so are the quantities associated with a subsystem like its temperature and entropy. These relational subalgebras describing $S$ relative to distinct choices of QRF need not even necessarily be isomorphic when the QRFs have vastly different properties. 

In this section, we first review the PN framework machinery for transforming between QRFs in~\ref{sect:PNswitching}, before using the links between the three approaches established in Sect.~\ref{sect:equivalence} to derive frame transformations in the algebraic case (in Sect.~\ref{sec:algebraicSwitching}) and to link them with the older semiclassical transformations \cite{EffectivePoT1,EffectivePoT2,EffectivePoTChaos} of the effective case (in Sect.~\ref{sec:effectiveSwitching}), which we then also generalize. For formulating the QRF transformations in the latter two approaches, we restrict to the case of a pair of ideal references frames $A$\ and $B$, so that the constraint has the form
\begin{equation}
    \label{eq:constraintIdeal}
    \hat{C} = \hat{p}_A + \hat{p}_B + \hat{G}_S.
\end{equation}
It is not difficult to convince oneself that the results of  Sect.s~\ref{sect:algebraicIdealClock} and \ref{sect:effectiveIdealClock} remain valid when two 
ideal frames are present as one could think of the second frame as being part of the system.

We defer the application of QRF transformations to our model previously discussed in Sect.~\ref{sect:example} to Sect.~\ref{sect:rel}.

\subsection{QRF transformations within the PN framework}
\label{sect:PNswitching}

Let us first review switching between QRFs that may be non-ideal in the PN framework \cite{Trinity,TrinityRel,Chataignier:2024eil,delaHamette:2021oex}. We assume the kinematical Hilbert space and the constraint to have the form 
\begin{equation}
    \mathcal{H}_\text{phys} = \mathcal{H}_A \otimes \mathcal{H}_B \otimes \mathcal{H}_S, \quad \ \hat{C} =  \hat{G}_A + \hat{G}_B + \hat{G}_S ,
\end{equation}
with generators $\hat{G}_i$\ commuting with each other and $\hat{G}_A,\hat{G}_B$ possessing arbitrary continuous but non-degenerate spectra $\sigma_A,\sigma_B$. The gauge-invariant picture is straightforward: we can again map kinematical states to physical states by means of the group averaging operator \eqref{eq:projector}. The total gauge-invariant algebra on the kinematical Hilbert space can now be written in two equivalent (isomorphic) ways by building relational Dirac observables with respect to~$A$ or with respect to~$B$ \cite{DeVuyst:2024uvd}
\begin{align}
    \mathcal{A}_\text{inv} &= \left\langle \hat{O}^\rho_B(\hat{f}_{AS}),  \hat{G}_B \vert\hat{f}_{AS} \in \mathcal{A}_{AS}; \rho \in \mathbb{R}   \right\rangle \nonumber \\
    &= \left\langle \hat{O}^\rho_A(\hat{f}_{BS}), \hat{G}_A \vert \hat{f}_{BS} \in \mathcal{A}_{BS}; \rho \in \mathbb{R}   \right\rangle, \label{eq:invariantAlgebra}
\end{align}
where we used the same notation as in~\eqref{eq:invAlgebra}, and additionally introduce $\mathcal{A}_{iS} := \langle \mathcal{A}_i \cup \mathcal{A}_S\rangle$  as the subalgebra generated by $\mathcal{A}_i$ and $\mathcal{A}_S$ only. 
In what follows, we will at times be interested in only the relational subalgebras of $\mathcal{A}_{\rm inv}$, which correspond to describing the system $S$ relative to either QRF on the full kinematical Hilbert space\footnote{Alternatively, one could have fixed $\rho$ and add (bounded functions of) the generator of reorientations $\hat{G}_i$ to the algebra to get the relational observables for all values of $\rho$. The result would be a (subalgebra of) the crossed product \cite{DeVuyst:2024pop,DeVuyst:2024uvd}.}
\begin{align}
    \mathcal{A}^{\rm inv}_{S|A} &:= \left\langle\hat{O}^\rho_A(\hat{f}_S)\vert \hat{f}_{S} \in \mathcal{A}_S ; \rho \in \mathbb{R}   \right\rangle, \label{eq:AinvSA}\\
    \mathcal{A}^{\rm inv}_{S|B} &:= \left\langle \hat{O}^\rho_B(\hat{f}_S)\vert \hat{f}_{S} \in \mathcal{A}_S ; \rho \in \mathbb{R}   \right\rangle.\label{eq:AinvSB}
\end{align}    
It may be checked that
\begin{equation}
    \mathcal{A}^{\rm inv}_{S|A}\cap\mathcal{A}^{\rm inv}_{S|B} = \mathcal{A}_S^{\hat C}=\mathcal{A}_S^{\hat G_S}\,,\label{eq:alginter}
\end{equation}
i.e.\ the two algebras overlap exactly in those operators of the system $S$ that are already gauge-invariant without dressing to either of $A$ or $B$ \cite{DeVuyst:2024pop,DeVuyst:2024uvd,Hoehn:2023ehz,Araujo-Regado:2025ejs}. These are precisely all the internal relational observables of $S$ (which may contain other internal QRFs) and these are invariant under reorientations of either frame. This is at the heart of the subsystem relativity alluded to above.
In some cases, this overlap may be trivial (i.e.\ the identity component of the algebra).\footnote{For example, one such case is when $\mathcal{A}_S$ is a type III$_1$ algebra in perturbative quantum gravity, as explored in \cite{DeVuyst:2024pop,DeVuyst:2024uvd}.}

Assuming $\hat{G}_i\in\mathcal{A}_i$, $i=A,B,S$, on the physical Hilbert space, the representation of $\mathcal{A}_{\rm inv}$ becomes
\begin{align}
    \label{eq:Aphys}
    \mathcal{A}_{\rm phys}&=\big\langle\hat{\mathcal{O}}^\rho_B(\hat f_{AS})\,\big|\,\hat{f}_{AS}\in \mathcal{A}_{AS};\rho\in\mathbb{R}\big\rangle \nonumber \\&=\big\langle\hat{\mathcal{O}}^\rho_A(\hat f_{BS})\,\big|\,\hat{f}_{BS}\in \mathcal{A}_{BS};\rho\in\mathbb{R}\big\rangle\,.
\end{align}    

In the context of two QRFs, the Page-Wootters reduction map \eqref{eq:PWreduction} trivially extends to 
\begin{equation}
    \mathcal{R}_A(\rho) := \bra{\rho}_A \otimes \mathds{1}_B \otimes \mathds{1}_S,
\end{equation}
and similarly for $\mathcal{R}_B(\rho)$. Likewise, the definition of the projector~\eqref{eq:PWprojector} then extends to $\Pi_{\vert A}$ projecting onto the subspace $\mathcal{H}_{BS \vert A}$ of $\mathcal{H}_B \otimes \mathcal{H}_S$, spanned by the eigenstates of $\hat{G}_B+\hat{G}_S$ with eigenvalues in the range $\sigma_{B + S} \cap (-\sigma_A)$, and similarly with $\Pi_{|B}$. Exploiting the invertibility of the PW reduction maps on the physical Hilbert space, we can build a QRF transformation from $A$-perspective to $B$-perspective, $V_{A \to B}(\rho_A,\rho_B): \mathcal{H}_{BS \vert A} \to  \mathcal{H}_{AS \vert B}$, simply by \cite{Trinity,delaHamette:2021oex}
\begin{align}
\label{eq:temporalChange}
    &V_{A \to B}(\rho_A,\rho_B) := \mathcal{R}_B(\rho_B) \circ \mathcal{R}^{\dag}_A(\rho_A) \\&\quad = \int_{\mathbb{R}} \dd{\rho'} (\ket{\rho_A + \rho'}_A \otimes\bra{\rho_B - \rho'}_B \otimes \mathds{1}_S) e^{-i \rho' \hat{G}_S }. \nonumber
\end{align}
This transformations maps the reduced states (defined as in Sect.~\ref{sect:PW}) as
\begin{equation}\label{eq:PNstatetransf}
    \ket{\psi_{AS|B}(\rho_B)}=V_{A\to B}(\rho_A,\rho_B)\,\ket{\psi_{BS|A}(\rho_A)}\,.
\end{equation}
Note that these take the form of quantum coordinate transformations, mapping from one perspective via the perspective-neutral Hilbert space $\mathcal{H}_{\rm phys}$ to the new perspective, in resemblance with coordinate transformations on a manifold.

In order to compare with the algebraic and effective approaches below, let us now restrict $A,B$ to be ideal QRFs, with $\hat{G}_i=\hat{p}_i$, $i=A,B$, as in~\eqref{eq:constraintIdeal}, so that $\Pi_{|A}=\mathds1_B \otimes \mathds1_S$, $\Pi_{|B}=\mathds1_A \otimes \mathds1_S$, and (cf.~\eqref{eq:reducedAlgebra}) 
\begin{equation}
\begin{split}
    \mathcal{A}_{BS|A}&=\mathcal{R}_A(\rho)\,\mathcal{A}_{\rm phys}\,\mathcal{R}^\dag_A(\rho)=\mathcal{A}_{BS}, \\
    \mathcal{A}_{AS|B}&=\mathcal{R}_B(\rho)\,\mathcal{A}_{\rm phys}\,\mathcal{R}_B^\dag(\rho)=\mathcal{A}_{AS}\,,\label{eq:redalgAB}
\end{split}
\end{equation}
each consisting of the bare operators, and explore the effect of the QRF transformations on observables. 
\begin{widetext}
We use the fact that \cite{Vanrietvelde:2018pgb,Vanrietvelde:2018dit,Trinity,Giacomini:2017zju}
\begin{equation}
\label{eq:idealtransfp}
\begin{split}
    V_{A\to B}(\rho_A,\rho_B)\left(\hat q_B \otimes \mathds1_S\right)\,V^\dag_{A\to B}(\rho_A,\rho_B)&= - \hat{q}_A \otimes \mathds{1}_S + \rho_A + \rho_B \\
    V_{A\to B}(\rho_A,\rho_B)\left(\hat p_B \otimes \mathds1_S\right)\,V^\dag_{A\to B}(\rho_A,\rho_B)&= - \hat{p}_A \otimes \mathds{1}_S - \mathds{1}_A \otimes \hat{G}_S.
\end{split}
\end{equation}
An arbitrary element of $\mathcal{A}_{BS|A}$ is (up to closure) a sum of products of operators $\hat{f}_B\otimes\hat{f}_S$, so let us focus on these. Assuming $\hat{f}_B \equiv f_B (\hat q_B,\hat p_B)$ can be expanded in powers of $\hat q_B,\hat p_B$, we find

\begin{align}\label{eq:PNobstrans}
    V_{A\to B}(\rho_A,\rho_B)&\left(\hat{f}_B\otimes\hat{f}_S\right)\,V^\dag_{A\to B}(\rho_A,\rho_B) = \\
    &= f_B( (\rho_B+\rho_A)-\hat q_A \otimes \mathds{1}_S,-\hat p_A\otimes\mathds{1}_S -\mathds{1}_A\otimes \hat G_S)\,\hat{O}_A^{\rho_A}(\hat{f}_S) \ .
    \nonumber
\end{align}
Here we treat $\hat{O}^{\rho_A}_A(\hat f_S)$ as an operator on the factor space $\mathcal{H}_A\otimes \mathcal{H}_S$. It is the \emph{kinematical} form (i.e.\ using incoherent group averaging) of relational observables in~\eqref{eq:diracObservable}, which, in the ideal case, reduces to the expression~\eqref{eq:idealDirac} that can be evaluated on the factor space $\mathcal{H}_A\otimes \mathcal{H}_S$. Note that $\hat O^{\rho_A}_A(\hat f_S)$ and $\hat f_B$ commute. Together with~\eqref{eq:PNstatetransf} it is then clear that expectation values transform as
\begin{align}
\label{eq:idealexpectransf}
    \bra{\psi_{BS|A}(\rho_A)}\,&\left(\hat f_B\otimes\hat f_S\right)\,\ket{\psi_{BS|A}(\rho_A)}=  \\&= \bra{\psi_{AS|B}(\rho_B)}\, f_B( (\rho_B+\rho_A)-\hat q_A \otimes \mathds{1}_S,-\hat p_A \otimes\mathds{1}_S-\mathds{1}_A\otimes\hat G_S)\,\hat{O}^{\rho_A}_A(\hat f_S)\,\ket{\psi_{AS|B}(\rho_B)}\,.
    \nonumber
\end{align}
We will find the same relations in the algebraic and effective approaches below.
\end{widetext}

Finally, as regards subsystem relativity, note that the subalgebra $\mathds1_B \otimes \mathcal{A}_S$ of $\mathcal{A}_{BS|A}$ in~\eqref{eq:redalgAB} is the representation of $\mathcal{A}^{\rm inv}_{S|A}$ in~\eqref{eq:AinvSA} on the physical Hilbert space but expressed in $A$-perspective. Similarly, $\mathds1_A \otimes  \mathcal{A}_S$ is the representation of $\mathcal{A}_{S|B}^{\rm inv}$ in $B$-perspective. Setting $\hat{f}_B=\mathds1_B$ in~\eqref{eq:PNobstrans}, we see that $\mathds1_B \otimes \mathcal{A}_S$ does not map into $\mathds1_A \otimes \mathcal{A}_S$ under QRF transformations, reflecting the fact that the two algebras are distinct, cf.~\eqref{eq:alginter}. This leads to a host of QRF-dependent effects \cite{Hoehn:2023ehz,DeVuyst:2024pop,DeVuyst:2024uvd,Araujo-Regado:2025ejs,Giacomini:2017zju,Cepollaro:2024rss,Suleymanov:2025nrr}.

\subsection{Constructing frame transformations for the algebraic approach} \label{sec:algebraicSwitching}

We again focus on the case where both reference frames are ideal, as in \eqref{eq:constraintIdeal}. From the perspective of $A$, the observed system algebra $\mathcal{A}_{BS}$\ is generated by $\mathcal{A}_B\cup\mathcal{A}_S$, while from the perspective of $B$, the observed system is described by $\mathcal{A}_{AS}$\ and generated by $\mathcal{A}_A\cup\mathcal{A}_S$. Given a physical state $[\omega]_{C}$~\eqref{eq:algebraicGaugeOrbit}, the two descriptions correspond to two distinct almost positive states $\omega_{A|\rho_A}, \omega_{B|\rho_B}\in [\omega]_{C}$, sampling the physical state, each satisfying its own frame's gauge conditions~\eqref{eq:algebraicFrame}. How are the expectation values assigned by $\omega_{A|\rho_A}$\ and $\omega_{B|\rho_B}$\ related to each other? Given $\hat{f}_{BS}\in \mathcal{A}_{BS}$, relation~\eqref{eq:equivalenceAlgebraic} links its evaluation to a relational Dirac observable
\begin{equation}\label{eq:A_Frame_Obs}
     \omega_{A \vert \rho_A}( \hat{f}_{BS}) =
    \omega_{A \vert \rho_A}(\dirac{\hat{f}_{BS}}{A}{\rho_A}) \ .
\end{equation}
Since $\hat{O}_A^{\rho_A}(\hat{f}_{BS})$ is a strong Dirac observable,  its value does not change along the gauge orbit that constitutes the algebraic physical state $[\omega]_{C}$, yielding the same result for both $A$-gauge and $B$-gauge sampling of $[\omega]_{C}$:
\begin{equation}
    \omega_{A \vert \rho_A}(\dirac{\hat{f}_{BS}}{A}{\rho_A}) =  \omega_{B \vert \rho_B}(\dirac{\hat{f}_{BS}}{A}{\rho_A}).
    \label{eq:changeOrbit}
\end{equation}
Combining~\eqref{eq:A_Frame_Obs} and~\eqref{eq:changeOrbit} we get the key relation for transforming algebraic data from $B$\ to $A$
\begin{equation}
    \omega_{A \vert \rho_A}( \hat{f}_{BS}) =  \omega_{B \vert \rho_B}(\dirac{\hat{f}_{BS}}{A}{\rho_A}). 
    \label{eq:algFrameChange}
\end{equation}
The transformation to $A$'s perspective is then achieved by evaluating the expression on the right-hand side of~\eqref{eq:algFrameChange} in terms of elements of $\mathcal{A}_{AS}$\ which describe the observed system from $B$'s perspective. 

Before we take a closer look at~\eqref{eq:algFrameChange}, we note that, once again,  using the concrete Hilbert space realization~(\ref{eq:stateReal}) of gauge-fixed algebraic states, this relation holds in exactly the same form also for non-ideal QRFs. This follows from the fact that we derived~\eqref{eq:A_Frame_Obs} in~\eqref{eq:Phys-to-PWexp} for states in~\eqref{eq:stateReal} without assuming an ideal QRF, while~\eqref{eq:changeOrbit} only assumes the two states belong to the same gauge orbit and clearly also holds in the non-ideal case.

Let us now return to $A,B$ being ideal QRFs, so that we do not need to assume states of the form~\eqref{eq:stateReal}. To make the frame transformation implied by~(\ref{eq:algFrameChange}) more explicit, we consider $\hat{f}_{BS} = \hat{f}_B \hat{f}_S$, where $\hat{f}_B$\ is ordered so that each of its monomial terms has all of the powers of $\hat{q}_B$\ on the left. We note that any polynomial in $\hat{q}_B$\ and $\hat{p}_B$\ can be brought to this ordering using the canonical commutation relation. Furthermore, the transformation for an arbitrary element of $\mathcal{A}_{BS}$\ follows, since (up to closure) it is a linear combination of terms of the form $\hat{f}_B\hat{f}_S$. As shown in detail in appendix~\ref{sect:algebraicFrameSwitchting}, this gives us the following equation for transforming between a pair of ideal frames
\begin{widetext}
\begin{equation}
    \omega_{A \vert \rho_A}(\hat{f}_{B}  \hat{f}_S)
    = \omega_{B \vert \rho_B}(  f_B(\rho_B + \rho_A - \hat{q}_A, -\hat{p}_A - \hat{G}_S) \hat{O}_A^{\rho_A}(\hat{f}_S))  .
    \label{eq:algTransf}
\end{equation}
\end{widetext}
This is precisely the algebraic incarnation of~\eqref{eq:idealexpectransf} from the PN framework.
This identity allows us to compute the value assigned by a state in $A$-gauge to any observable of the form $\hat{f}_B \hat{f}_S$ in terms of the values assigned by the $B$-gauge sampling of the same physical state.

What originally was an element $f_B(\hat{q}_B, \hat{p}_B)$ on the left-hand side is replaced by the same element but with 
its arguments transformed as on the right-hand side. In particular, setting $f_S = \mathds{1}$, and $\hat{f}_B$ to $\hat{q}_B$ and $\hat{p}_B$ respectively, we find 
\begin{equation}
\begin{split}
    \omega_{A \vert \rho_A}(\hat{q}_B) &= \omega_{B \vert \rho_B}(\rho_B + \rho_A - \hat{q}_A), \\
    \omega_{A \vert \rho_A}(\hat{p}_B) &= \omega_{B \vert \rho_B}(- \hat{p}_A - \hat{G}_S).
    \label{eq:changeClockObs}
\end{split}
\end{equation}
This holds for all states, so we read off the transformations $(\hat{q}_B)_{A|\rho_A} \to (\rho_B + \rho_A - \hat{q}_A)_{B|\rho_B}$ and $(\hat{p}_B)_{A|\rho_A} \to (-\hat{p}_A - \hat{G}_S)_{B|\rho_B}$ which agree
with~\eqref{eq:idealtransfp} of the PN framework. Here, the shorthand notation $(.)_{A|\rho_A}$ simply means `evaluation in $A$-gauge', including $q_A = \rho_A$.

For any purely-$S$ observable, where $\hat{f}_B = \mathds{1}$, relation~\eqref{eq:algTransf} looks exactly the same as~(\ref{eq:algFrameChange}). Expanding it using~\eqref{eq:idealDirac}, we find 
\begin{align}
\label{eq:changeGenerator}
    &\omega_{A \vert \rho_A}(\hat{f}_S) = \omega_{B \vert \rho_B}(\dirac{\hat{f}_S}{A}{\rho_A})\nonumber
    \\
    &= \omega_{B \vert \rho_B}( e^{-i(\hat{q}_A - \rho_A) \hat{G}_S / \hbar} \hat{f}_S e^{i(\hat{q}_A - \rho_A) \hat{G}_S / \hbar}) \nonumber \\
    &= \omega_{B \vert \rho_B} ( \hat{f}_S) \\
    &+ \omega_{B \vert \rho_B} \left( e^{-i(\hat{q}_A - \rho_A) \hat{G}_S / \hbar} \comm*{\hat{f}_S}{e^{i(\hat{q}_A - \rho_A) \hat{G}_S / \hbar}} \right). \nonumber
\end{align}
The same expression, of course, also follows from~\eqref{eq:idealexpectransf} in the PN framework.
We see that the expectation values assigned to pure-$S$ observables differ between the two perspectives by  additional terms whenever $\hat{f}_S$ does not commute with the system generator $\hat{G}_S$. Note that if $\hat f_S$ commutes with $\hat{G}_S$, then it lies exactly in the overlap~\eqref{eq:alginter} of the relational observable algebras describing $S$ relative to $A,B$. Only in that case do the expectation value expressions taken in different reference frames agree.

\subsection{Effective semiclassical frame transformations}\label{sec:effectiveSwitching}

The information one needs to relate when switching between effective QRFs is similar to the algebraic case. Physical states correspond to entire gauge orbits generated by the constraint functions on the quantum phase space defined by classical and quantum variables of the truncated system. Descriptions from $A$'s and $B$'s perspectives correspond to choosing different sample points along those gauge orbits \cite{EffectivePoT1,EffectivePoT2,EffectivePoTChaos}. Let us first reproduce the QRF transformations of the previous two subsections in the effective approach, before linking them with the older method of semiclassical frame changes described in \cite{EffectivePoT1,EffectivePoT2,EffectivePoTChaos}, which contains equivalent information.

Earlier in Sect.~\ref{sect:effectiveIdealClock}, we established that the expectation values of Dirac observables are gauge-invariant functions on the quantum phase space, and that this still holds true to the appropriate semiclassical order once truncation is performed.  Hence, we can use the expectation values of relational Dirac observables~\eqref{eq:idealDirac} when changing frames just like we did for the algebraic frame transformation in Sect.~\ref{sec:algebraicSwitching}.  In particular, both sides of relation~\eqref{eq:algFrameChange} can be expanded in expectation values of generators and their moments and then truncated at some order $M$
\begin{equation}\label{eq:truncatedSwitching}
\mathscr{T}_M\left\langle \hat{f}_{BS} \right\rangle_{A\vert \rho_A} =  \mathscr{T}_M\left\langle \dirac{\hat{f}_{BS}}{A}{\rho_A} \right\rangle_{B\vert \rho_B} \ , 
\end{equation}
where the subscript ``$A|\rho_A$'' indicates that $A$-gauge conditions are imposed with $q_A=\rho_A$, and similarly for $B$. Furthermore, because the gauge and constraint conditions in the effective approach are identical to those of the algebraic approach (see Sect.~\ref{sssec_algeffequiv}), the argument from appendix~\ref{sect:algebraicFrameSwitchting} leading up to~\eqref{eq:algTransf} also applies in the effective case giving us a truncated version of that relation
\begin{widetext}
\begin{equation}\label{eq:truncatedSwitchingProduct}
\mathscr{T}_M\left\langle\hat{f}_{B}  \hat{f}_S\right\rangle_{A\vert \rho_A} 
    = \mathscr{T}_M\left\langle  f_B(\rho_B + \rho_A - \hat{q}_A, -\hat{p}_A - \hat{G}_S)\hat{O}_A^{\rho_A}(\hat{f}_S) \right\rangle_{B\vert \rho_B} ,
\end{equation}
\end{widetext}
which can be specialized to truncated versions of~\eqref{eq:changeClockObs} and~\eqref{eq:changeGenerator}. Thus, also in the effective approach one finds agreement with~\eqref{eq:idealexpectransf} of the PN framework (up to the chosen order $M$).

We note that, within the effective construction, $A$\ and $B$\ perspectives possess preferred sets of coordinates. Let $\{\hat{x}_i\}_{i=1, \ldots, N}$ be the basic observables that generate the system algebra $\mathcal{A}_S$, then $\mathcal{A}_{BS}$\ is generated by $\{\hat{y}_i\}_{i=1, \ldots, N+2} = \{\hat{q}_B, \hat{p}_B; \hat{x}_j\}$, while $\mathcal{A}_{AS}$\ is generated by $\{\hat{z}_i\}_{i=1, \ldots, N+2} = \{\hat{q}_A, \hat{p}_A; \hat{x}_j\}$. The preferred coordinates for describing physics relative to $A$\ are then the classical and quantum variables associated with the generators of $\mathcal{A}_{BS}$, namely $\{y_i\}$\ and $\{\Delta\left(y_1^{n_1} y_2^{n_2}\ldots \right)\}$, similarly for $B$'s perspective we should use $\{z_i\}$\ and $\{\Delta\left(z_1^{n_1} z_2^{n_2}\ldots \right)\}$. We can now use~\eqref{eq:truncatedSwitching} and~\eqref{eq:truncatedSwitchingProduct} to create explicit coordinate transformations for generator expectation values
\begin{equation} \label{eq:effectiveExpValSwitching}
   {(y_i)_{A\vert \rho_A} } = \left\langle \dirac{\hat{y}_i}{A}{\rho_A}\right\rangle_{B\vert \rho_B} \,,
\end{equation}
for their second-order moments
\begin{align}
    \label{eq:effectiveMomentSwitching}
    &\Delta(y_i y_j)_{A\vert \rho_A} = \expval*{(\dirac{\hat{y}_i}{A}{\rho_A}\dirac{\hat{y}_j}{A}{\rho_A})_\text{Weyl}}_{B\vert \rho_B} \nonumber \\ &\quad - \expval*{\dirac{\hat{y}_i}{A}{\rho_A}}_{B\vert \rho_B} \expval*{\dirac{\hat{y}_j}{A}{\rho_A}}_{B\vert \rho_B}  \ ,
\end{align}
and so on for moments of all orders. The right-hand side expressions in~\eqref{eq:effectiveExpValSwitching} and~\eqref{eq:effectiveMomentSwitching} need to be written in terms of the $B$-frame coordinates $z_i$\ and their moments. To that end, each expectation value needs to be expanded in the expectation values and moments of the kinematical generators $\{\hat{q}_A, \hat{p}_A,\hat{q}_B, \hat{p}_B;\hat{x}_i\} $\ as in~\eqref{eq:effExpand}, variables associated with $\hat{p}_B$ must be eliminated using the constraint conditions ($\langle \hat{a}\hat{p}_B \rangle = \langle \hat{a} (-\hat{p}_A - \hat{G}_S) \rangle$\ for all $\hat{a}$), while variables associated with $\hat{q}_B$\ are removed by $B$-gauge conditions~\eqref{eq:EffectiveGaugeCond}\ and $q_B=\rho_B$. We look at explicit examples of effective frame switching in Sect.~\ref{sect:translationalInvariant} and appendix~\ref{app:spin}.

As mentioned in the introduction and emphasized throughout the manuscript, the earlier work on (temporal) QRFs \cite{EffectivePoT1, EffectivePoT2, EffectivePoTChaos} already constructed explicit QRF transformations at the leading semiclassical order. To our knowledge it constitutes the first explicit construction of QRF transformations in the literature and it is worth briefly summarizing the method it utilized and comparing to the method of this section. This earlier work considered two-component canonical systems governed by constraints of the form\footnote{This form of the constraint is inspired by the Wheeler-DeWitt equation of quantized homogeneous cosmological models. The first clock changes in the PN framework were also formulated in two-component models of this kind (with $V=0$) \cite{Hohn:2018toe,Hohn:2018iwn}.}
\begin{equation} \label{eq:EffectivePoTConstraint}
    \hat{C} = \hat{p}_A^2-\hat{p}_B^2 - V(\hat{q}_A, \hat{q}_B) \ ,
\end{equation}
where $V(\hat{q}_A, \hat{q}_B)$\ is an arbitrary interaction potential between the two component systems. We note that this model lacks a separate ``observed system'' component: from the perspective of $A$\ the observed system is simply $B$\ and vice versa. While the constraint is not of the simple form discussed in Sect.~\ref{sect:trinity}, one may still use the ideal QRF orientation observables $\hat q_A,\hat q_B$ for the two frames (which, however, is not exactly covariant with respect to $\hat C$); this is exactly what the effective treatment in the above references invoked. To order $\hbar$\ this system possesses five independent constraint conditions~\eqref{eq:EffectiveConstraintFunctions} $C$, $C_{q_I}$, $C_{p_I}$, $I=A, B$ (analogous to~\eqref{eq:constraintTower}, but having a different form in terms of the classical and quantum variables due to the difference between constraints~\eqref{eq:EffectivePoTConstraint} and~\eqref{eq:idealConstraint}). Assuming the perspective of frame $I=A, B$ involves just three gauge-fixing conditions~\eqref{eq:EffectiveGaugeCond}
\begin{equation}\label{eq:EffectivePoTGauge}
\left( \Delta q_I\right)^2 = 0 \ , \quad \Delta(q_I q_J) = 0 \ , \quad \Delta( q_I p_J)=0 \ ,
\end{equation}
where $J\neq I$\ is the label for the other subsystem. One may then also completely fix the gauge by further setting $q_I=q_I^0=const$. The main idea for constructing an explicit transformation between the two QRF perspectives in those papers is the same as the one used here: given relational information from the perspective of $A$ the physical state corresponds to the entire gauge orbit generated by the constraint functions via the Poisson bracket~\eqref{eq:quantumPoisson}. Relational data from the perspective of $B$\ correspond to another point on the same gauge orbit---the one for which $B$-gauge conditions hold.

The method for actually computing the values assigned by the same physical state to relational data from the perspective of $B$\ used in~\cite{EffectivePoT1, EffectivePoT2, EffectivePoTChaos} is somewhat different from what was done earlier in this subsection. Instead of relying on the invariance of values of explicitly constructed relational Dirac observables, this older construction looked for a specific combination of constraint-generated flows that accomplish a suitable gauge transformation; namely, in the case of transforming from $A$\ to $B$\ we need\footnote{For convenience, in~\cite{EffectivePoT1, EffectivePoT2, EffectivePoTChaos} the last gauge condition listed in~\eqref{eq:EffectivePoTGauge} was replaced with $\Delta(q_A p_A)=-i\hbar/2$\ by using constraint functions and the first two gauge conditions. (See discussion leading up to~\eqref{eq:gaugeFixing}.)}
\begin{equation}
    \left\{ \begin{array}{l} \left(\Delta q_B \right) ^2 = w_1 \\ \Delta(q_B q_A) = 0 \\ \Delta( q_B p_A)=w_2 \end{array} \right. \longrightarrow \left\{ \begin{array}{l} \left(\Delta q_B \right) ^2 = 0 \\ \Delta(q_B q_A) = 0 \\ \Delta( q_B p_A)= 0 \end{array} \right. \ ,
\end{equation}
where $w_i$\ are the particular values taken by $B$'s moments in the $A$-gauge. If one wishes to completely fix the gauge, one will also need to implement $q_{B|A}\to q_B^0$, where $q_{B|A}$ is the value of $q_B$ in $A$-gauge. One then uses this specific combination of gauge flows to compute their effects on the classical and quantum variables of interest, in this way constructing a transformation from the full set of $A$-frame coordinates to the full set of $B$-frame coordinates, and, by construction, this must coincide with~\eqref{eq:truncatedSwitchingProduct}--\eqref{eq:effectiveMomentSwitching} if one applied it to constraints of the form discussed here. Indeed, both methods simply compare expectation values of the basic observables on the same two gauge fixing surfaces.

While the explicit method for constructing QRF transformations used in~\cite{EffectivePoT1, EffectivePoT2, EffectivePoTChaos} is therefore different, the shared treatment of the physical state as a gauge orbit in the quantum phase space of the kinematical system ensures that the transformations derived using either method carry identical information. Thus, at semiclassical order, the QRF transformations of the PN framework agree with those of the effective approach (constructed with either method summarized here).

\subsection{Changes of degenerate QRFs}

So far we assumed the QRFs involved in the transformations to be non-degenerate. Let us briefly discuss -- at a purely qualitative level -- what happens when the QRFs admit degenerate reorientation generators. For definiteness, let us assume we are given a constraint of the form 
 \begin{equation}
     \hat{C} =  \hat{p}_A^2 - \hat{p}_B^2 + \hat{G}_S.
 \end{equation}
 We may then factorize it, similarly to~\eqref{eq:DegenerateC}, either with respect to $A$ or $B$ as $\hat C=\hat{C}^A_+\,\hat{C}^A_-=\hat{C}^B_+\,\hat{C}_-^B$. The corresponding projectors $\Pi_\pm^A,\Pi_\pm^B$ commute with one another, as well as with $\hat{C}$. This means that the kinematical and physical Hilbert spaces decompose into four distinct sectors \cite{Hohn:2018iwn,DeVuyst:2024uvd,TrinityRel} 
 \begin{equation}
 \begin{split}
     \mathcal{H}_{\rm kin}&=\bigoplus_{\lambda_A,\lambda_B=+,-}\,\mathcal{H}_{\lambda_A\lambda_B}^{\rm kin}\,, \\ \mathcal{H}_{\rm phys} &=\bigoplus_{\lambda_A,\lambda_B=+,-}\,\mathcal{H}_{\lambda_A\lambda_B}\,,
    \end{split}
 \end{equation}
 where the $\hat{C}$-generated gauge transformations preserve each kinematical sector $\mathcal{H}_{\lambda_A\lambda_B}^{\rm kin}$. (Note that gauge-invariant observables are not restricted to preserve these sectors, cf.~\eqref{eq:degenerateInvAlg}.) This is the quantum incarnation of the fact that classically the constraint surface is comprised of four disjoint pieces not linked by gauge orbits \cite{Hohn:2018iwn} (see also \cite{Hohn:2018toe}). Since all of our QRF transformations are based on gauge transformations,\footnote{This is manifest in the algebraic and effective approaches but also true for the QRF transformations of the PN framework that we reviewed. Indeed, recall that the PW reduction $\mathcal{R}_R(\rho)$ is a gauge fixing. Thus,~\eqref{eq:temporalChange} is a change of gauge and thereby corresponds to a gauge transformation. In the quantum theory, this was made manifest in \cite{Krumm:2020fws} and in the classical setting in \cite{Vanrietvelde:2018pgb,Vanrietvelde:2018dit,Hohn:2018iwn,Hohn:2018toe}. Note that the PN framework admits a second kind of QRF transformation, purely at the relational observable level, that switches the QRF dressing and is not related to gauge transformations \cite{delaHamette:2021oex,Hoehn:2023ehz,Carrozza:2021gju,Goeller:2022rsx}.} we have that we can only change QRF \emph{per $\lambda_A\lambda_B$ sector}. 
 
 Indeed, for each QRF $R$ the PW reduction map is now defined per $\lambda_R$-sector, and so composing it with the inverse reduction of the other QRF as in~\eqref{eq:temporalChange} leads to QRF transformations per $\lambda_A\lambda_B$-sector \cite{TrinityRel,DeVuyst:2024uvd,Hohn:2018iwn,Hohn:2018toe}. Of course, one may take the direct sum of these maps to cover the entire physical Hilbert space and encompass observables that map between sectors. Similarly, in the algebraic and effective approaches, the expectation values $\omega(\hat{C}^i_\pm),\langle\hat C^i_{\pm}\rangle$, $i=A,B$, are gauge-invariant Dirac observables, so that the gauge flows leave the $\lambda_A\lambda_B$-sectors invariant and we can only switch QRFs sector-wise. Given the sector-wise equivalence of the three approaches discussed in Sect.~\ref{sssec_doubledeg}, as well as the agreement of their QRF changes for ideal QRFs discussed above, the sector-wise QRF changes in the degenerate case will likewise encode the same information for expectation values across the three approaches.

\section{Frame dependence of uncertainties and covariances}
\label{sect:rel}

Let us now consider a physical application of the QRF transformations formulated in Sect.~\ref{sect:switchingIdeal}. We will show that moments of observables, such as uncertainties $(\Delta a_i)^2$ and covariances $\Delta(a_ia_j)$ take different values in different frame perspectives -- these quantities are thus frame-dependent. It is natural to consider their transformations within the effective approach, as they are the fundamental quantum variables of interest. Of course, as we have shown above, it does not matter which formalism is used to transform them.
Some frame transformation properties of uncertainties and covariances were explored within the effective approach in~\cite{EffectivePoT1, EffectivePoT2, EffectivePoTChaos}. One important feature of the models treated in this older work is that they all consisted of a pair of subsystems. In this case, the frame transformation involves a complete switch between which kinematical subsystem is observed and which serves as reference, so that the two frames did not both describe the same third kinematical subsystem $S$ as here. As a result, variables such as $(\Delta f_S)^2$, that could be compared in both perspectives did not arise. Instead, the most interesting effects caused by the frame transformations in~\cite{EffectivePoT1, EffectivePoT2, EffectivePoTChaos} were due to the interaction between the two subsystems. Frame dependence of uncertainties of the type recently explored in~\cite{Suleymanov:2025nrr} does not directly appear in those older transformations. While~\cite{Suleymanov:2025nrr} makes use of the PN framework, here we show how the use of the (equivalent) semiclassical effective expansion streamlines the process, leading to the same conclusions in an efficient way, while slightly generalizing the argument.\footnote{An interesting exploration of uncertainties in the context of the distinct operational approach to QRFs has recently appeared in \cite{Riera:2024ehk}, also using quantum phase space techniques (albeit distinct ones). frame dependence was however not analyzed there.}

 We consider a system with two ideal frames governed by a constraint of the form \eqref{eq:constraintIdeal}. As in Sect.~\ref{sec:effectiveSwitching}, we assume that the system algebra $\mathcal{A}_S$ is generated by a set of elements $\hat{x}_i$, that form a Lie algebra with respect to the commutator, and that $\hat{G}_S \equiv \hat{G}_S(\hat{x}_i)$. Let us apply the results of Sect.~\ref{sect:switchingIdeal} to look at the transformation of the uncertainty of an arbitrary system observable $\hat{f}_S \equiv \hat{f}_S(\hat{x}_i) \in \mathcal{A}_S$.  Starting from the definition of uncertainty and applying \eqref{eq:truncatedSwitching} and \eqref{eq:idealDirac} we find 
 \begin{widetext}
\begin{align}
    \label{eq:covarianceTransform}
    \mathscr{T}_M(\Delta f_S)^2_{A | \rho_A} 
    &= \mathscr{T}_M \left[ (\Delta f_S)^2_{B | \rho_B} + \left\langle e^{-i(\hat{q}_A - \rho_A) \hat{G}_S/\hbar}
    \comm*{\hat{f}_S^2}{e^{i(\hat{q}_A - \rho_A) \hat{G}_S/\hbar}}\right\rangle_{B | \rho_B} \right. \nonumber\\
     & \hspace{10pt}-\left\langle e^{-i(\hat{q}_A - \rho_A) \hat{G}_S/\hbar}
    \comm*{\hat{f}_S}{e^{i(\hat{q}_A - \rho_A) \hat{G}_S/\hbar}}\right\rangle_{B | \rho_B}^2  \\
    &\left. \hspace{10pt} -2 \left\langle \hat{f}_S \right\rangle_{B|\rho_B} \left\langle e^{-i(\hat{q}_A - \rho_A) \hat{G}_S/\hbar}
    \comm*{\hat{f}_S}{e^{i(\hat{q}_A - \rho_A) \hat{G}_S/\hbar}} \right\rangle_{B | \rho_B} \right].\nonumber
\end{align}
\end{widetext}
In particular, we observe that whenever $\comm*{\hat{f}_S}{\hat{G}_S} \neq 0$ (and so $\hat O_{i}^\rho(\hat f_S)$, $i=A,B$ does \emph{not} reside in the intersection~\eqref{eq:alginter} of the relational system algebras), the uncertainty in the value of $f_S$, as viewed from $B$'s perspective, will differ from the uncertainty from $A$'s perspective by terms that depend on the covariances and uncertainties of the other operators including $\hat{q}_A$, as seen relative to frame $B$.\footnote{Making use of the nested commutator formula \cite[Eq.~(27)]{Trinity} for ideal frames 
\begin{align}
    \dirac{\hat{f}_S}{A}{\rho_A} &= \sum_{n=0} \frac{(i \widehat{\Delta q}_A)^n}{\hbar^n n!}\comm*{\hat{f}_S}{\hat{G}_S}_n \nonumber \\
    &= \hat{f}_S + \sum_{n=1} \frac{(i \widehat{\Delta q}_A)^n}{\hbar^n n!} \comm*{\hat{f}_S}{\hat{G}_S}_n,
\end{align}
equation~\eqref{eq:covarianceTransform} can also be written in a more suggestive form as 
\begin{align} 
    (\Delta f_S)^2_{A | \rho_A} &= (\Delta f_S)^2_{B | \rho_B} + \sum_{n=1} \omega_{B|\rho_B}(\frac{(i \widehat{\Delta q}_A)^n}{\hbar^n n!} \comm*{\hat{f}_S^2}{\hat{G}_S}_n) \nonumber\\ &- \omega_{B|\rho_B}^2(\sum_{n=1} \frac{(i \widehat{\Delta q}_A)^n}{\hbar^n n!}\comm*{\hat{f}_S}{\hat{G}_S}_n)  \\
    &- 2 \omega_{B|\rho_B}(\hat{f}_S) \omega_{B|\rho_B}(\sum_{n=1} \frac{(i \widehat{\Delta q}_A)^n}{\hbar^n n!}\comm*{\hat{f}_S}{\hat{G}_S}_n). \nonumber
\end{align}
}

In the following subsection we look at explicit examples of uncertainty and covariance transformations in the context of the translationally invariant system introduced in Sect.~\ref{sect:example}. We encourage the interested reader to also look at the example of effective frame transformations for a system with algebra generated by $\mathfrak{su}(2)$  discussed in appendix~\ref{app:spin}.

\subsection{Back to the translationally invariant system}
\label{sect:translationalInvariant}

Consider again the translationally invariant system of particles from Sect.~\ref{sect:example}, where gauge symmetry is generated by constraint~\eqref{eq:momcon}. Let us begin by deriving the transformation for the uncertainty in the position of particle $C$\ as we go from frame $B$\ to frame $A$. In frame $A$, this uncertainty can be computed via
\begin{equation} \label{eq:CSpread} 
    (\Delta q_C)^2_{A | \rho_A} = \left\langle \hat{q}_C^2 \right\rangle_{A | \rho_A} - \left\langle \hat{q}_C \right\rangle_{A | \rho_A}^2 \ .
\end{equation}
Using~\eqref{eq:truncatedSwitching}, we find
\begin{align}
    \mathscr{T}_M \left\langle \hat{q}_C\right\rangle_{A \vert \rho_A} &= \mathscr{T}_M \left\langle \dirac{\hat{q}_C}{A}{\rho_A} \right\rangle_{B \vert \rho_B}  \\ &= \mathscr{T}_M \left\langle\hat{q}_C - \hat{q}_A + \rho_A  \right\rangle_{B \vert \rho_B}. \nonumber
\end{align}
By using the Dirac observable $\dirac{\hat{q}_C^2}{A}{\rho_A} = (\dirac{\hat{q}_C}{A}{\rho_A})^2$, we obtain the analogous expression for $\mathscr{T}_M \left\langle \hat{q}_C^2\right\rangle_{A \vert \rho_A}$.  Expanding expectation values through~\eqref{eq:effExpand}, now allows us to write the right-hand-side of~\eqref{eq:CSpread} in terms of the quantum variables evaluated in frame $B$, yielding
\begin{align}
    \label{eq:transformVariance}
    &(\Delta q_C)^2_{A | \rho_A} \\
    &\quad = (\Delta q_C)^2_{B | \rho_B} + (\Delta q_A)^2_{B | \rho_B} - 2 \Delta(q_A q_C)_{B | \rho_B}. \nonumber
\end{align}
Since the expression for $\left\langle \dirac{\hat{q}_C^2}{A}{\rho_A} \right\rangle$\ only involves moments up to order two, the relation~\eqref{eq:transformVariance} is valid for all orders of truncation $M\geq2$. Suppose particle $C$\ is well localized when seen in frame $B$. According to~\eqref{eq:transformVariance}, that same particle may well have a high uncertainty in position when described relative to frame $A$. For example, if in frame $B$ the correlation between the positions of $A$\ and $C$\ are small, while the uncertainty in the position of $A$\ is large, we get $(\Delta q_C)^2_{A | \rho_A} \approx (\Delta q_A)_{B | \rho_B}^2$, which can be large independently from $(\Delta q_C)^2_{B | \rho_B}$.

Similarly, by considering the Dirac observable $\dirac{\hat{q}_B \hat{q}_C}{A}{\rho_A} = \dirac{\hat{q}_B}{A}{\rho_A} \dirac{\hat{q}_C}{A}{\rho_A}$ we obtain the transformation of covariance between the positions of $B$\ and $C$, when we switch between frames $D$ and $A$
\begin{align}
    \label{eq:covarianceTraGeneral}
    \Delta(q_B &q_C)_{A | \rho_A} = \Delta(q_B q_C)_{D | \rho_D} + (\Delta q_A)^2_{D | \rho_D} \nonumber \\ &- \Delta(q_A q_B)_{D | \rho_D} - \Delta(q_A q_C)_{D | \rho_D}.
\end{align}
Specializing the above result for the case where $B=D$ we obtain, after imposing the $B$-gauge conditions on the right hand side,
\begin{equation}
    \label{eq:covarianceTra}
    \Delta(q_B q_C)_{A | \rho_A} = (\Delta q_A)^2_{B | \rho_B}  - \Delta(q_A q_C)_{B | \rho_B}.
\end{equation}
Again, if the correlations between particle $C$ and $A$ are small from $B$'s perspective, but particle $A$ is not localized in that frame; then there are nonzero 
correlations between particles $C$ and $B$ from $A$'s perspective: $\Delta(q_B q_C)_{A | \rho_A} \approx (\Delta q_A)^2_{B | \rho_B}$. In particular, this could mean that the positions of particles $C$ and $B$ are entangled when described in frame $A$ while the positions of $C$ and $A$ are uncorrelated when viewed from $B$.

We could derive similar transformations for the variances and covariances of the momenta and of mixtures of position and momenta  by selecting appropriate relational Dirac observables and  performing the effective expansion. This procedure reproduces the transformations reported in~\cite{Suleymanov:2025nrr}. In particular, Eqs.~(\ref{eq:transformVariance}, \ref{eq:covarianceTraGeneral}, \ref{eq:covarianceTra}) are the same as
\cite[Eqs.~(16-18)]{Suleymanov:2025nrr}, which invoked the PN framework machinery, that we showed is equivalent to the effective approach.\footnote{The authors of~\cite{Suleymanov:2025nrr} followed the procedure in~\cite{Vanrietvelde:2018pgb} for constructing the physical Hilbert space, but did not directly invoke the frame change maps~\eqref{eq:temporalChange}. Instead, they expressed the physical momentum space wave function in different ways by solving the momentum constraint~\eqref{eq:momcon} for different momenta, circumventing the need for an explicit transformation from one frame to another. In fact, this is precisely the information contained in \cite[Eqs.~(24, 25, 35)]{Vanrietvelde:2018pgb}, and  writing the physical momentum wave function in different ways by solving for the momenta of different choices of QRF is exactly equivalent to explicitly using the QRF changes~\eqref{eq:temporalChange}.} In our treatment, on the other hand, we emphasize that all these relations stem from \eqref{eq:algFrameChange} which also works for more general systems, such as, for example, the one discussed in appendix~\ref{app:spin}, where the system algebra is generated by $\mathfrak{su}(2)$. 

Finally, as a sanity check, let us show how the same equations follow directly from the equivalent PN framework using the QRF transformations \eqref{eq:temporalChange}. As an example, we will re-derive \eqref{eq:transformVariance}. We start by considering the reduced algebras \eqref{eq:redalgAB}, stemming from jumping into the perspectives of particle $A$ and $B$, respectively
\begin{equation}
    \mathcal{A}_{I_{\bar{A}}|A} = \{ \hat{q}_i, \hat{p}_i | i \in I_{\bar{A}} \}, \quad \mathcal{A}_{I_{\bar{B}}|B} = \{ \hat{q}_i, \hat{p}_i | i \in I_{\bar{B}} \},
\end{equation}
where $I_{\bar{i}} := I \setminus \{i\}$, and
which can be transformed into each other through the QRF transformation \eqref{eq:temporalChange}. Applying this map to $\hat{q}_C$, or more precisely invoking~\eqref{eq:PNobstrans} for $\hat f_S=\hat q_C$ and setting $\hat f_B=\mathds1_B$ and recalling~\eqref{eq:idQRFposrel}, results in 
\begin{align}
    &V_{A \to B}(0,0)\, \hat{q}_C^2\, V^\dagger_{A \to B} (0,0) \nonumber\\ &\quad = (V_{A \to B}(0,0) \,\hat{q}_C \,V^\dagger_{A \to B} (0,0))^2  \\
    &\quad = (\hat{q}_C - \hat{q}_A)^2 = \hat{q}_C^2 + \hat{q}_A^2 - 2 \hat{q}_C \hat{q}_A. \nonumber
\end{align}
Taking the expectation value of the right hand side in a relational state corresponding to the $B$-frame $\ket*{\psi_{I_{\bar{B}} | B}(0)}$, is then equivalent to evaluating $\hat{q}_C^2$ in the corresponding relational state in $A$-gauge $\ket*{\psi_{I_{\bar{A}} | A}(0)}$, cf.~\eqref{eq:idealexpectransf}. Employing an expansion as in~\eqref{eq:effExpand} then  leads to the same expression as \eqref{eq:transformVariance}.

We note that this QRF dependence of uncertainties and covariances is entirely due to subsystem relativity manifested in~\eqref{eq:alginter} and discussed below~\eqref{eq:idealexpectransf}.

\section{From physical states to algebraic states and gauge flows}
\label{sect:generalGaugeFreedom}

In the preceding sections, we have demonstrated that the Hilbert-space based methods of the PN framework, the algebraic, and the related semiclassical effective constructions (at appropriate order) encode equivalent relational information when applied to an ideal non-interacting QRF. As we have seen, the connection to the algebraic states is most direct through the Page-Wootters conditional states~\eqref{eq:stateReal}, which can be viewed as a form of gauge-fixing. The Hilbert-space version of this gauge-fixing is accomplished via a projection to a ``Schr\"odinger'' reduced Hilbert space (see equation~(\ref{eq:PWreduction})), which, unlike the projection to the physical Hilbert space, depends on the choice of gauge provided by a particular configuration of the chosen reference system. The algebraic version of this gauge is imposed on algebraic kinematical states---linear functionals on an algebra of kinematical observables---by way of restricting to states that are left-multiplicative with respect to the reference configuration observable (see equation~(\ref{eq:AlgebraicGauge})).

In this section, we reconcile two formal differences between these two approaches to fixing gauge. First, the Hilbert-space-based approach involves constructions and maps between several Hilbert spaces beyond the original kinematical Hilbert space, whereas the algebraic approach only deals with states on the kinematical observable algebra. Second, reference gauge~(\ref{eq:AlgebraicGauge}) is explicitly designed to fix a large class of gauge flows~(\ref{eq:AlgebraicGaugeFlow}) on algebraic solution states of the constraint, whereas there are no gauge flows on Hilbert space solutions to the constraint, since its action on solutions is trivial. We will show that these differences ``melt away" once one focuses on generalized algebraic states constructed by extending the physical inner product~(\ref{eq:PhysicalIP}) so that it can be used to compute (generalized) matrix elements of kinematical observables through the choice of a gauge. Gauge flows manifest themselves as gauge-choice-dependence of the values assigned by extended states to non-invariant observables.

Sect.~\ref{sec:physIPextension} and~\ref{sec:GaugeFlowsExplained} discuss the above gauge choice in a very general way, making very few assumptions about the constrained system. Sect.~\ref{sec:ReferenceGauge} and~\ref{sec:ClockGaugeProjection} focus on the gauge associated with orientation of a reference system, assuming that the orientation space is the real line but without assuming that the reference system is ideal.

\subsection{Extensions of the physical inner product to non-invariant observables} \label{sec:physIPextension}

The key Hilbert space operation for constructing solutions to the constraint from kinematical states is the physical state map $\Pi$\ defined in~\eqref{eq:projector}.\footnote{To make comparison with QRF literature easier, we note that $\Pi$\ is often denoted as $\delta(\hat{C})$ (for example in~\cite{Trinity}).} The result of applying $\Pi$\ to a state in $\mathcal{H}$ is a generalized state in $\mathcal{H}^*$, which can be evaluated on ``nicely-behaved'' elements of $\mathcal{H}$, formally this is done via the kinematical inner product.  As a functional on $\mathcal{H}$\ the result satisfies
\begin{equation}
\hat{C} \, \Pi | \psi \rangle = 0 \ ,
\end{equation}
in the sense that $\langle \phi | \hat{C} \Pi | \psi \rangle = 0$\ for all $|\phi \rangle, | \psi \rangle \in \mathcal{H}$\ that are in the domain of the Abelian algebra generated by $\hat{C}$.  From now on, we will shorten the notation: for equality of a pair of sequences of operations $A$\ and $B$\ we will simply write $A=B$, to mean that there are dense subsets $D_1, D_2 \subset \mathcal{H}$, such that $\langle \phi| A |\psi \rangle =\langle \phi| B |\psi \rangle$\ for all $|\phi\rangle \in D_1$\ and $|\psi \rangle \in D_2$. For example, it follows immediately that
\begin{equation}\label{eq:ExpCPiPhys}
e^{is\hat{C}/\hbar} \Pi = \Pi \ .
\end{equation}

In the Hilbert space methods of the PN framework $\Pi$\ forms the basis for both constructing solutions to the constraint~(\ref{eq:projector}) and defining the physical inner product~(\ref{eq:PhysicalIP}) between a pair of solution states $|\psi_\text{phys} \rangle$\ and $|\phi_\text{phys}\rangle$.\footnote{As we have done for the PN construction in Sect.~\ref{sect:trinity}, throughout Sect.~\ref{sect:generalGaugeFreedom} we assume that~\eqref{eq:PhysicalIP} defines a positive definite inner product.} We can use the physical inner product to compute expectation values and matrix elements of strong Dirac observables $\hat{O}$ relative to physical states
\begin{equation} \label{eq:PhysIPMatrix}
    \left( \psi_\text{phys} \left| \hat{O} \right| \phi_\text{phys} \right) = \langle \psi | \hat{O} \Pi | \phi\rangle \ ,
\end{equation}
where $| \psi \rangle$\ and $| \phi \rangle$\ are any kinematical states such that $\Pi|\psi\rangle =|\psi_\text{phys}\rangle$\ and $\Pi|\phi\rangle =|\phi_\text{phys}\rangle$.
Since the final expression on the right involves states on the kinematical Hilbert space, it can be used to assign values also to kinematical observables that do not commute with $\hat{C}$. Of course, $\Pi$\ is many-to-one and there is an equivalence class of (generalized) states $| \psi \rangle$\ and $| \phi \rangle$ that give identical physical expectation values, but the choice of representative states does matter for non-Dirac observables. We can view a particular choice of such states as a gauge choice that allows us to extend evaluation of a physical matrix element in~\eqref{eq:PhysIPMatrix} to kinematical observables $\hat A$
\begin{equation}
\left( \psi_\text{phys} \left| \hat{A} \right| \phi_\text{phys} \right)_\Phi := \langle \psi_\text{phys} | \Phi \hat{A}  | \phi_\text{phys}\rangle \ ,
\end{equation}
where the gauge $\Phi$\ represents a particular choice of a representative state for each solution state, so that $|\psi \rangle = \Phi | \psi_\text{phys}\rangle$\ with $\Pi|\psi\rangle =|\psi_\text{phys}\rangle$, and the notation on the right hand side means that $\Phi\hat A$\ is to be evaluated in \emph{physical states} using the \emph{kinematical inner product}. In short, a gauge is a particular choice of representing the physical inner product via the kinematical inner product.

In situations of primary interest to us, the states $\Phi | \psi_\text{phys} \rangle$\ will be constructed using fixed orientation states of the reference system~(\ref{eq:clock states}), which are typically generalized states in $\mathcal{H}^*_{R}$, rather than proper states in $\mathcal{H}_R$. For our purposes, then, a gauge is a linear map $\Phi$\ from the space of solutions to the constraint $\mathcal{H}_\text{phys}$\ to sufficiently nicely behaved elements of $\mathcal{H}^*$, the dual of the kinematical Hilbert space, such that for any pair of solutions we can use the kinematical inner product to evaluate
\begin{equation}
\langle \psi_\text{phys} | \Phi | \phi_\text{phys} \rangle = \overline{\langle \phi_\text{phys} | \Phi | \psi_\text{phys} \rangle} < \infty \ ,
\end{equation}
and $\Pi (\Phi |\psi_\text{phys} \rangle ) = |\psi_\text{phys} \rangle$, so that, on $\mathcal{H}$\ we have
\begin{equation} \label{eq:GaugeCondition}
\Pi\, \Phi\, \Pi = \Pi \ ,
\end{equation}
and,  similarly,  on $\mathcal{H}_\text{phys}$,  we get $\Phi\, \Pi\, \Phi = \Phi$. Note that~(\ref{eq:GaugeCondition}) by itself already implies that $\langle . |\Phi| . \rangle$\ is a bounded complex-sesquilinear form on $\mathcal{H}_\text{phys}$, \emph{we therefore take~(\ref{eq:GaugeCondition}) to be the defining property of a gauge}.\footnote{Relation~\eqref{eq:GaugeCondition} features in the Hilbert space-based constructions of the PN framework in the form of the property that maps~\eqref{eq:PWreduction} and~\eqref{eq:PWinverse} from $\mathcal{H}_\text{phys}$\ to the Schr\"odinger picture invert each other \cite{Trinity,delaHamette:2021oex}.} In particular, it ensures that
\begin{equation} \label{eq:GaugePhysIP}
\langle \psi_\text{phys} | \Phi | \phi_\text{phys} \rangle = \left( \psi_\text{phys}| \phi_\text{phys} \right) \ ,
\end{equation}
and, more generally, for any Dirac observable $\hat{O}$, 
\begin{equation}\label{eq:RelOMatrix}
\langle \psi_\text{phys}| \Phi\, \hat{O} | \phi_\text{phys} \rangle = \left( \psi_\text{phys} \left| \hat{O}\, \right| \phi_\text{phys} \right) \ ,
\end{equation}
To show this explicitly, we note that there are $| \psi \rangle$\ and $| \phi \rangle$\ in $\mathcal{H}$, such that $\Pi|\psi\rangle =|\psi_\text{phys}\rangle$\ and $\Pi|\phi\rangle =|\phi_\text{phys}\rangle$. Furthermore, $\hat{O}\Pi = \Pi \hat{O}$\ and we quickly get
\begin{equation}
\begin{split}
\langle \psi_\text{phys}| \Phi \, \hat{O} | \phi_\text{phys} \rangle &= \langle \psi |  \Pi\, \Phi \,\hat{O}\, \Pi |\phi\rangle 
\\ 
&= \langle \psi |  \Pi \, \Phi \, \Pi\,\hat{O} |\phi\rangle 
\\
&= \langle \psi |  \Pi\,\hat{O}  |\phi \rangle \\
&= \left( \psi_\text{phys} \left| \hat{O}\, \right| \phi_\text{phys} \right) \ .
\end{split}
\end{equation}
Given a gauge $\Phi$, there will generally be some kinematical observables that do not commute with $\hat{C}$\ for which the generalized matrix elements $\langle \psi_\text{phys}| \Phi\, \hat{A} | \phi_\text{phys} \rangle$\ can also be computed, where the value
does depend on the choice of $\Phi$\ and the gauge choice matters.
Our main example of a gauge is the reference frame gauge $\Theta(\rho)$\ defined in Sect.~\ref{sec:ReferenceGauge} by direct analogy with the reduction map~\eqref{eq:PWreduction}. Before we discuss this example in detail, we first consider the general properties of gauge fixing applied to kinematical algebraic states.

\subsection{Gauge freedom of algebraic states} \label{sec:GaugeFlowsExplained}

The previous subsection introduced gauge freedom associated with evaluating expectation values and generalized matrix elements of kinematical observables with respect to physical states.  In order to better understand this gauge freedom,  it will be convenient to focus on \emph{algebraic} states---linear assignments of complex values to a subset of kinematical operators. Since we will not specify the precise algebra of kinematical operators on which these states ought to be well-defined, we will refer to them as ``algebraic'' only loosely, in the sense that they are not Hilbert-space states, but rather prescriptions for computing generalizations of matrix elements of operators. Specifically, any pair of states $|\psi \rangle$, $|\phi \rangle$ in $\mathcal{H}$ can be used to define an algebraic state via the inner product
\begin{equation}
\tilde{\omega} (.) := \langle \psi | . | \phi \rangle \ .
\end{equation}
Assuming appropriate care is taken with the domains of evaluated operators, both $|\psi \rangle$\ and $|\phi \rangle$\ could also be elements of the dual space $\mathcal{H}^*$. Since the above state involves a single use of the (kinematical) inner product, by loose analogy, we will call such states \emph{pure}. More generally, such states can be combined into generalizations of \emph{mixed} states via linear combinations $\tilde{\omega}(.) = a\tilde{\omega}_1(.)+b \tilde{\omega}_2(.)$. (If we want to preserve normalization and positivity, this sum needs to be convex.) Any solution to the constraint $|\phi_\text{phys} \rangle$\ can be used to generate an algebraic \emph{solution state} by selecting an appropriate ``partner'' state $| \psi \rangle$
\begin{equation}
\omega (.) := \langle \psi| . | \phi_\text{phys} \rangle = \langle \psi | (.) \Pi | \phi \rangle \ ,
\end{equation}
for some $|\phi \rangle$\ such that $\Pi |\phi \rangle= | \phi_\text{phys} \rangle$. Dropping the explicit reference to the ``bra'' and ``ket'' components of algebraic states, we will assume that a general solution state (not necessarily pure) has the form
\begin{equation}\label{eq:GeneralSolutionState}
\omega(.) = \tilde{\omega} ((.)\Pi) \ ,
\end{equation}
where $\tilde{\omega}$\ is some algebraic state within which the group averaging operation $\Pi$\ is well-defined. By analogy with the purely algebraic approach of Sect.~\ref{sec:AlgebraicMethod} we then have $\omega(\hat{A} \hat{C})=0$\ for all operators $\hat{A}$\ for which $\omega(\hat{A} \hat{C})$\ can be evaluated.\footnote{The symmetric alternative choice of $\omega(\hat{C}\hat{A})=0$\ for all $\hat{A}$, corresponds to $\omega(.):=\tilde{\omega}(\Pi(.))$.} 

Suppose we wanted to compute the matrix element of some kinematical observable $\hat{A}$\ relative to a pair of solution states $| \psi_\text{phys} \rangle$\ and $| \phi_\text{phys} \rangle$\ in gauge $\Phi$. The corresponding expression is
\begin{equation}
\langle \psi_\text{phys}| \Phi \, \hat{A} | \phi_\text{phys} \rangle  = \langle \psi |  \Pi\, \Phi \,\hat{A}\, \Pi |\phi\rangle \ .
\end{equation}
This suggests that the general form of a $\Phi$-gauge solution state is
\begin{equation} \label{eq:GaugeFixedSolution}
\omega(.) = \tilde{\omega} \left( \Pi \,\Phi \, (.)\, \Pi\right) \ ,
\end{equation}
for some appropriately well-behaved $\tilde{\omega}$. If $|\psi \rangle$\ is in the domain of $\Pi$, the pure solution state $\omega(.) = \langle \psi | (.) \Pi| \phi \rangle$\ can always be viewed as having the gauge-fixed form~(\ref{eq:GaugeFixedSolution}) 
\begin{equation} \label{eq:PureSolutionState}
\omega(.)= \langle \psi | \Pi\, \Phi (.) \Pi | \phi\rangle \ ,
\end{equation}
where $\Phi$\ is any gauge such that $\Phi (\Pi |\psi \rangle) = |\psi \rangle$.

Suppose we have a pair of algebraic states that solve the constraint $\omega_{i}(.) = \tilde{\omega}_i ((.) \Pi)$, $i=1, 2$. We say that the two states are physically equivalent if
\begin{equation} \label{eq:EquivalentStates}
\omega_1(\Pi\, \Phi (.)) = \omega_2(\Pi\, \Phi (.))
\end{equation}
for any gauge $\Phi$. As expected, for a pair of physically equivalent states we have
\begin{equation}
    \omega_1( \hat{O}) = \omega_2( \hat{O}) \ ,
\end{equation}
for any Dirac observable $\hat{O}$. To see this explicitly we use~(\ref{eq:GaugeCondition}) and the fact that $\hat{O}\Pi=\Pi \hat{O}$
\begin{equation}
\label{eq:ObservableInEquivStates}
\begin{split}
    \omega_1( \hat{O}) &= \tilde{\omega}_1 (\hat{O} \Pi )
    \\
    &= \tilde{\omega}_1 (\Pi\hat{O}  )
    \\
    &= \tilde{\omega}_1 (\Pi \Phi\hat{O} \Pi)
    \\
    &=\omega_1 (\Pi \Phi\hat{O} )
    \\
    &=\omega_2 (\Pi \Phi\hat{O} ) =\omega_2( \hat{O}) \ ,
\end{split}
\end{equation}
where $\Phi$\ is an arbitrary gauge and the last equality is obtained by following the preceding set of transformations in reverse. Definition~(\ref{eq:EquivalentStates}) together with property~(\ref{eq:GaugeCondition}) strongly suggest the form of the gauge transformation from $\Phi_A$-gauge to $\Phi_B$-gauge which (it is straightforward to check) preserves physical equivalence
\begin{equation} \label{eq:FormalGaugeTransform}
    \omega_B (.):= \omega_A(\Pi\, \Phi_B (.)) \ .
\end{equation}
This makes intuitive sense if we think of the pure algebraic states of the form~(\ref{eq:PureSolutionState}): we can change the gauge by first taking the state on the left to its invariant equivalence class via $\Pi$\ and then fixing the new gauge $\Phi_B$. The key objects guiding gauge transformations are then products of the form $\Phi_A \, \Pi\, \Phi_B$.\footnote{The transformations~\eqref{eq:temporalChange} between a pair of QRFs indeed has an equivalent form. Explicitly, using a pair of reference gauges $\Theta_{A, B}(\rho)$\ defined in~\eqref{eq:ThetaDef}, we have $\Theta_A(\rho_A) \Pi \Theta_B(\rho_B) = | \rho_A \rangle \left( \mathcal{R}_A^{\dag} (\rho_A)\circ \mathcal{R}_B(\rho_B) \right) \langle \rho_B |$, where here the order of operations is from left to right because we are applying gauge changes on the ``bra'' side of the algebraic state.}

These formal gauge transformations are in general discrete---not part of any one--parameter family. In order to understand the local continuous gauge freedom implied by~(\ref{eq:EquivalentStates}), it is interesting to look at gauge transformations induced by unitary operators on the kinematical Hilbert space. Given a gauge $\Phi$\ and a pair of self-adjoint Dirac observables $\hat{O}_1$\ and $\hat{O}_2$\ we can construct another gauge $\Phi'=e^{i\hat{O}_1\hat{C}}\Phi e^{i\hat{O}_2\hat{C}}$. Since $e^{i\hat{O}_i\hat{C}}$\ map eigenstates of $\hat{C}$\ to other eigenstates of $\hat{C}$\ with the same eigenvalues, it follows that  $\langle \phi_\text{phys}| \Phi' |\psi_\text{phys} \rangle<\infty$. It is also straightforward to verify condition~(\ref{eq:GaugeCondition}). Now suppose we would like to transform a state $\omega$\ in $\Phi$\ gauge to its physically equivalent counterpart in the $\Phi'$\ gauge. Then, using~(\ref{eq:FormalGaugeTransform})
\begin{align}
\omega'(.) &= \omega \left( \Pi \Phi'(.) \right) = \omega \left( \Pi \, e^{i\hat{O}_1\hat{C}}\Phi e^{i\hat{O}_2\hat{C}} (.) \right)  \nonumber \\ &= \omega \left( \Pi \, \Phi e^{i\hat{O}_2\hat{C}} (.) \right) \ .
\end{align}
Furthermore, using the form~(\ref{eq:GaugeFixedSolution}) for $\omega$, we see that the transformation is simply accomplished by the unitary operator on the right
\begin{align}\label{eq:UnitaryGaugeTransform}
\omega'(.)&= \tilde{\omega} \left( \Pi \,  \Phi\, \Pi \, \Phi e^{i\hat{O}_2\hat{C}}\, (.) \,\Pi \right) 
\\
&= \tilde{\omega} \left( \Pi\, \,  \Phi\, e^{i\hat{O}_2\hat{C}}\, (.) \Pi \right) = \omega \left( e^{i\hat{O}_2\hat{C}} (.) \right) \ . \nonumber
\end{align}

Conversely, we can see that a broader class of invertible transformations preserves the physical equivalence class of solution states. Suppose that for some invertible operator $\hat{L}$\ we have $\omega(.)$\ physically equivalent to $\omega(\hat{L}(.)\hat{L}^{-1})$\ for any solution state $\omega$\ of the form~(\ref{eq:GeneralSolutionState}), then
\begin{equation}
\hat{L}\, \Pi\, \Phi (.)\, \hat{L}^{-1}\, \Pi = \Pi\, \Phi (.)\, \Pi \, 
\end{equation}
which can be satisfied by demanding $\hat{L}\, \Pi = \hat{L}^{-1}\, \Pi= \Pi$. A general class of such invertible operations is given by
\begin{equation}
\hat{L}=e^{i\lambda\hat{A}\hat{C}} \ ,
\end{equation}
which formally matches the gauge  flows $S_{\hat{A}\hat{C}}(\lambda)$ of the algebraic approach in~(\ref{eq:AlgebraicGaugeFlow}). We note that, if $\hat{A}$\ is not self-adjoint or does not commute with $\hat{C}$, the exponential $\hat{L}=e^{i\lambda\hat{A}\hat{C}}$\ will not in general be a densely defined operator, however, as a formal power series, it will satisfy the above condition. As in the case of the gauge transformation~(\ref{eq:UnitaryGaugeTransform}), only one of the factors matters here, since
\begin{align}\label{eq:GeneralGaugeFlow}
\omega \left( e^{i\lambda\hat{A}\hat{C}}(.)e^{-i\lambda\hat{A}\hat{C}} \right) &= \tilde{\omega} \left( e^{i\lambda\hat{A}\hat{C}}(.)e^{-i\lambda\hat{A}\hat{C}}\Pi \right) \nonumber\\ 
&= \tilde{\omega} \left( e^{i\lambda\hat{A}\hat{C}}(.)\Pi \right) \\
&= \omega \left( e^{i\lambda\hat{A}\hat{C}}(.) \right) \ .\nonumber
\end{align}

We note that a non-pure algebraic solution state~(\ref{eq:GeneralSolutionState}) cannot in general be viewed as gauge-fixed relative to some gauge of the form~(\ref{eq:GaugeFixedSolution}). This makes intuitive sense if we think of the example of a solution state that is a linear combination of a pair of states fixed with respect to different gauges
\begin{equation}
\begin{split}
\omega(.) &= a \omega_1(\Pi\,\Phi_1(.) \Pi) + b\omega_2(\Pi\,\Phi_2(.) \Pi) \\
&\neq \tilde{\omega}(\Pi \, \tilde{\Phi} (.) \Pi) \ .
\end{split}
\end{equation}
Similarly, while the transformation~(\ref{eq:GeneralGaugeFlow}), where $\hat{A}$\ is not a Dirac observable, does preserve gauge equivalence in the sense of~(\ref{eq:EquivalentStates}), it does not, in general, take the state to a state in a new well-defined gauge of the form~(\ref{eq:GaugeFixedSolution}). In other words, gauge-fixed algebraic states are special: a general algebraic state that solves the constraint, does not correspond to any particular coherent choice of a gauge.

We conclude the discussion of the general properties of gauges by noting that any valid gauge can be extended to a covariant family that resolves the identity.  Suppose that we have a gauge $\Phi(0)$, such that it satisfies~(\ref{eq:GaugeCondition}) $\Pi \Phi(0) \Pi = \Pi$. Define a one-parameter family of gauges
\begin{equation}
\Phi(s) := e^{-is\hat{C}/\hbar} \Phi(0) e^{is\hat{C}/\hbar}
\end{equation}
This family is covariant in the sense of
\begin{equation} \label{eq:GaugeCovariance}
\Phi(s') = e^{-i\hat{C} (s'-s)/\hbar} \Phi(s) e^{i\hat{C}(s'-s)/\hbar} \ .
\end{equation}
The resolution of identity when restricted to $\mathcal{H}_\text{phys}$\ follows since
\begin{equation}
\begin{split}
\Pi &= \Pi\, \Phi(0)\, \Pi 
\\ 
&= \frac{1}{2\pi \hbar} \int_{\mathbb{R}} ds\, e^{ is\hat{C}/\hbar}\, \Phi(0)\, \Pi 
\\
&= \frac{1}{2\pi \hbar} \int_{\mathbb{R}} ds\, e^{ is\hat{C}/\hbar}\, \Phi(0)\, e^{ -is\hat{C}/\hbar}\, \Pi \\
&= \left( \frac{1}{2\pi \hbar} \int_{\mathbb{R}} ds\,  \Phi(s) \right) \Pi .\label{eq:PhiGaugeIdResolution}
\end{split}
\end{equation}
Thus $\left( \frac{1}{2\pi \hbar} \int_{\mathbb{R}} ds\,  \Phi(s) \right)$\ is an identity on $\mathcal{H}_\text{phys}$.

\subsection{The reference system gauge} \label{sec:ReferenceGauge}

In this subsection, we construct the gauge $\Theta(\rho)$\ associated with setting the reference system to a particular orientation and look at its basic properties. We define this gauge as the projection to fixed orientation reference states $|\rho\rangle$ (see equation~(\ref{eq:clock states})),  which,  in the language of \cite{Trinity},  can be written as
\begin{equation} \label{eq:ThetaDef}
 \Theta(\rho) := |\rho \rangle \langle \rho |_R\otimes \mathds{1}_S \ .
 \end{equation}
If for each $\rho$\ there is some self-adjoint operator $\hat{Z}_\rho$\ such that $\hat{Z}_\rho |\rho \rangle_R \otimes |\psi \rangle_S = \rho | \rho \rangle \otimes |\psi \rangle_S$\ then we can also construct the `projection' (effect density) via group-averaging
 \begin{align}
 \Theta(\rho) &= \int_{\mathbb{R}} ds\, e^{is(\hat{Z}_\rho - \rho ) / \hbar}
 \\&=  \int_{\mathbb{R}} ds\, e^{-is(\hat{Z}_\rho - \rho ) / \hbar} \ . \nonumber
\end{align}
Whichever way this projection is constructed, we will assume that it is covariant as defined in~\eqref{eq:GaugeCovariance}\footnote{Note that for constraints~\eqref{eq:Constraint} that have no interaction between the $S$\ and $R$\ subsystems, $e^{- is\hat{C}/\hbar} \, \Theta(\rho)\, e^{is\hat{C}/\hbar} = e^{- is\hat{G}_R/\hbar} \, \Theta(\rho)\, e^{is\hat{G}_R/\hbar}$. Here we will not be using this stronger property.} and resolves the identity
\begin{equation}\label{eq:ThetaIdResolution}
\frac{1}{2\pi\hbar} \int_{\mathbb{R}} d\rho\,  \Theta(\rho) = \mathds{1} \ .
\end{equation}
Note that the above condition holds on $\mathcal{H}$\ and is therefore stronger than~\eqref{eq:PhiGaugeIdResolution}.

We can now show that $\Theta(\rho)$\ indeed satisfies~(\ref{eq:GaugeCondition}) for arbitrary $\rho$,\footnote{As noted earlier, this type of identity appears within the PN framework when one considers the map from physical states to gauge-dependent Schr\"odinger states. In particular, using~\eqref{eq:PWreduction} and~\eqref{eq:PWinverse}, one finds $ \mathcal{R}_R^{\dag}(\rho) \circ \mathcal{R}_R (\rho)= \Pi \Theta(\rho)$. Thus, the inversion relation $ \mathcal{R}_R^{\dag}(\rho) \circ \mathcal{R}_R (\rho)= \mathds{1}_\text{phys}$\ (equation (42) in \cite{Trinity}) immediately implies~\eqref{eq:PiThetaPi}.}
\begin{equation} \label{eq:PiThetaPi}
\Pi\, \Theta(\rho) \, \Pi = \Pi \ .
\end{equation}
To verify it step-by-step
\begin{align}
\Pi\, \Theta(\rho) \,\Pi &=  \frac{1}{2\pi \hbar} \int_{\mathbb{R}} ds\, e^{is\hat{C}/\hbar} \, \Theta(\rho) \,\Pi
\nonumber\\
&= \frac{1}{2\pi \hbar} \int_{\mathbb{R}} ds\, e^{is\hat{C}/\hbar}\,  \Theta(\rho) e^{-is\hat{C}/\hbar}\,\Pi
\nonumber\\
&= \frac{1}{2\pi \hbar} \int_{\mathbb{R}} ds\,  \Theta(\rho-s)\, \Pi
\\
&= \frac{1}{2\pi \hbar} \int_{\mathbb{R}} ds'\,  \Theta(-s')\, \Pi
\nonumber\\
&= \frac{1}{2\pi \hbar} \int_{\mathbb{R}} ds'\,  \Theta(s') \, \Pi = \Pi \ , \nonumber
\end{align}
where we have used~(\ref{eq:ExpCPiPhys}),  (\ref{eq:GaugeCovariance}), and (\ref{eq:ThetaIdResolution}) in that order. 

Following~(\ref{eq:GaugeFixedSolution}), we define an algebraic gauge-fixed state, corresponding to the reference subsystem having a fixed orientation $\rho$, as
\begin{align}\label{eq:PWGeneralGaugeState}
\omega_{R|\rho}(.) := \tilde{\omega} \left( \Pi\, \Theta(\rho) (.) \Pi\, \right) \ ,
\end{align}
where $\tilde{\omega}(.)$\ is an arbitrary algebraic kinematical state on which the above projective operations are well-defined. The core example is $\tilde{\omega} (.)=\langle \phi | . |\psi\rangle$ for some $|\phi \rangle, |\psi \rangle \in \mathcal{H}$\ that can be represented as Schwartz wave functions on the spectrum of $\hat{C}$.  Setting $|\phi \rangle =|\psi \rangle$\ matches our earlier definition of fixed reference states in the case of an ideal reference frame given in equation~(\ref{eq:stateReal}).

For the reference gauge, the covariance condition~(\ref{eq:GaugeCovariance}) also provides a simple example of a gauge transformation for an algebraic state of the type~(\ref{eq:UnitaryGaugeTransform}).  Using~(\ref{eq:GaugeCovariance}), gauges corresponding to two different reference orientations $\rho_1$\ and $\rho_2$\ are related by the transformation
\begin{equation}
\Theta(\rho_2)=e^{-i(\rho_2-\rho_1)\hat{C}/\hbar}\Theta(\rho_1)e^{i(\rho_2-\rho_1)\hat{C}/\hbar} \ .
\end{equation}
Given a solution state $\tilde{\omega}((.)\Pi)$, the two gauge-fixed algebraic states that one obtains $\omega_{R|\rho_i}(.)=\tilde{\omega}(\Pi\Theta(\rho_i)(.)\Pi)$\ are related by a unitary transformation $\omega_{R|\rho_2}(.) = \omega_{R|\rho_1}\left( e^{i(\rho_2-\rho_1)\hat{C}/\hbar}(.) \right)$\ and are physically equivalent in the sense of equation~(\ref{eq:EquivalentStates}).

Suppose we have a system with two valid reference subsystems $A$\ and $B$. Given a state $\omega_{A|\rho_A}$\ gauge-fixed by fixing the orientation of $A$\ to $\rho_A$\ of the form~(\ref{eq:PWGeneralGaugeState})\footnote{This also works for the weaker reference gauge-fixed states of the form~(\ref{eq:QDSRGeneralGaugeState}) considered in Sect.~\ref{sec:WeakRefGauge}.}, we would like to construct the corresponding state $\omega_{B|\rho_B}$\ gauge fixed by fixing the orientation of the quantum reference frame $B$\ to $\rho_B$. We follow the transformation~(\ref{eq:FormalGaugeTransform}) to define
\begin{align}\label{eq:GenericClockTransform}
    \omega_{B|\rho_B} (.) := \omega_{A|\rho_A} \left( \Pi \Theta_B (\rho_B) (.) \right)
\end{align}
(Compare this transformation to \cite[Eq.~(84)]{Trinity} and \cite[Thm.~4]{delaHamette:2021oex}). It is not difficult to check that $\omega_{B|\rho_B} (.)$\ and $\omega_{B|\rho_B} (.)$\ are physically equivalent in the sense of~(\ref{eq:EquivalentStates}) and therefore, by the same argument as in~(\ref{eq:ObservableInEquivStates})
\begin{equation}
\omega_{B|\rho_B} (\hat{O}) := \omega_{A|\rho_A} (\hat{O}) \ ,
\end{equation}
for any Dirac observable $\hat{O}$.

\subsection{Evaluating kinematical and relational observables}\label{sec:ClockGaugeProjection}

In this section, we show that the values assigned by our formal gauge-fixed algebraic states to relational Dirac observables of~\cite{Trinity} match those assigned by these same states to kinematical observables which are dressed by the same quantum frame that was used to build the relational observables. This mimics our discussion in Sect.~\ref{sect:equivalence} and matches the main results of~\cite{Trinity}. However, our present discussion elucidates these results from the novel vantage point of algebraic states. 

We first establish a few identities that will be useful for connecting our construction to results in \cite{Trinity}. The projection onto the physically accessible subspace of the observed system~\eqref{eq:PWprojector} can be trivially extended to the entire kinematical Hilbert space $\mathcal{H}$
\begin{equation} \label{eq:PiSys}
\hat{\pi}:=\mathds{1}_R\otimes \Pi_{|R}
\end{equation}
We use the ``hat'' to emphasize that this definition gives us a genuine (bounded) projection operator that is well-defined on the entire kinematical Hilbert space. Recall that, in this section, we are not assuming the reference frame to be ideal as was done in Sect.~\ref{sect:equivalence} through~\ref{sect:rel}. For any reference system where $\sigma_S\subseteq (-\sigma_R)$, which is always true for an ideal reference frame but also in some non-ideal cases, $\Pi_{|R}=\mathds{1}_S$~\cite{Trinity}, so that $\hat{\pi}=\mathds{1}$. 

Using the Hilbert space form of $\Theta(\rho)$\ in~(\ref{eq:ThetaDef}), it is immediately clear that
\begin{equation}\label{eq:PiThetaCommute}
\Theta(\rho) \, \hat{\pi} =  \hat{\pi} \, \Theta(\rho) \ .
\end{equation}
The defining property~\eqref{eq:PWprojector} of $\Pi_{|R}$\ can be written using operations on the entire kinematical Hilbert space
\begin{equation} \label{eq:SystemProjectorIdentity2}
\Theta(\rho) \,\Pi\, \Theta(\rho)  = \Theta(\rho) \,\hat{\pi} \ .
\end{equation}
In the special case where $\sigma_S\subseteq (-\sigma_R)$\ this simplifies to $
\Theta(\rho)\, \Pi \, \Theta(\rho)  = \Theta(\rho)$. Finally, we establish that
\begin{equation} \label{eq:PiPiCommute}
\hat{\pi}\, \Pi = \Pi\, \hat{\pi} =\Pi \ .
\end{equation}
Explicitly,
\begin{align}
\hat{\pi}\, \Pi &= \frac{1}{2\pi \hbar} \int_{\mathbb{R}} ds\,  \Theta(s)\, \hat{\pi}\, \Pi
\nonumber\\
&=  \frac{1}{2\pi \hbar} \int_{\mathbb{R}} ds\,  \Theta(s)\, \Pi\, \Theta(s)\, \Pi
\\
&= \frac{1}{2\pi \hbar} \int_{\mathbb{R}} ds\,  \Theta(s)\, \Pi = \Pi \ , \nonumber
\end{align}
where we have used~(\ref{eq:ThetaIdResolution}), (\ref{eq:SystemProjectorIdentity2}), (\ref{eq:PiThetaPi}), and, finally,~(\ref{eq:ThetaIdResolution}) again. The identity for the product in reverse order can be verified in an identical manner.

\subsubsection{Observables in gauge-fixed states}

Relational observables in \cite{Trinity} are constructed via G-twirl, which is essentially an incoherent group-averaging of kinematical operators
\begin{equation}\label{eq:GTwirl}
\mathcal{G}(\hat{A}) := \frac{1}{2\pi \hbar} \int_{\mathbb{R}} ds\, e^{ is\hat{C}/\hbar} \,\hat{A}\, e^{-is\hat{C}/\hbar} \ .
\end{equation}
Formally, $[\mathcal{G}(\hat{A}), \hat{C}]=0$, although, in general, the integral does not converge to a densely-defined operator (for example, the twirl is not defined if $[ \hat{A}, \hat{C} ] = 0$). Nevertheless, this operation successfully defines a class of relational Dirac observables via~\eqref{eq:diracObservable}, which we rewrite in the language of this section. For any $\hat{f}_S \in \mathcal{A}_S$ 
\begin{equation}\label{eq:RelationalObservable}
\hat{O}^\rho_R(\hat f_S) := \mathcal{G}\left( \Theta(\rho) \hat{f}_S \right)
\end{equation}
is the relational observable for measuring the system variable $f_S$\ when the quantum reference frame $R$ is in orientation $\rho$. While~\eqref{eq:GTwirl} is not an algebra homomorphism on its domain within $\mathcal{A}$\ (that is, $\mathcal{G}(\hat{A}\hat{B}) \neq \mathcal{G}(\hat{A})\mathcal{G}(\hat{B})$, in general), \eqref{eq:RelationalObservable} does define a unital $*$-homomorphism when restricted to $\mathcal{A}_{S|R}$ in~\eqref{eq:reducedAlgebra} and acting on $\mathcal{H}_{\rm phys}$~\cite{Trinity,delaHamette:2021oex,Bartlett:2006tzx} (for ideal QRFs it is also kinematically a homomorphism).

Let $\omega_{R|\rho}(.)$\ be a gauge-fixed state corresponding to a particular orientation $\rho$ of the reference system as in equation~(\ref{eq:PWGeneralGaugeState}). We find that for any system observable $\hat{f}_S\in \mathcal{A}_S$, the value of $\hat{O}^\rho_R(\hat f_S)$\ in such a state coincides with the value assigned by this state to the original kinematical system observable $\hat{f}_S$. That is
\begin{equation}\label{eq:RelOExpValue}
\omega_{R|\rho}\left( \hat{O}^\rho_R(\hat f_S) \right) =\omega_{R|\rho}\left( \hat{f}_S \right) \ .
\end{equation}
This is straightforward to verify explicitly using~(\ref{eq:ExpCPiPhys}) and~(\ref{eq:PiThetaPi}):
\begin{align}
 &\Pi\, \Theta(\rho)\,\hat{O}^\rho_R(\hat f_S)\,\Pi = \Pi\, \Theta(\rho)\,\mathcal{G}\left( \Theta(\rho)\, \hat{f}_S \right) \,\Pi
\nonumber\\
&\quad=\Pi\, \Theta(\rho)\,\frac{1}{2\pi \hbar} \int_{\mathbb{R}} ds\, e^{ is\hat{C}/\hbar}\, \Theta(\rho)\, \hat{f}_S\, e^{-is\hat{C}/\hbar}\, \Pi 
\nonumber\\
&\quad= \Pi\, \Theta(\rho)\, \frac{1}{2\pi \hbar} \int_{\mathbb{R}} ds\, e^{ is\hat{C}/\hbar}\, \Theta(\rho)\, \hat{f}_S\,  \Pi 
\\
&\quad= \Pi\, \Theta(\rho)\, \Pi\, \Theta(\rho)\, \hat{f}_S\,  \Pi
\nonumber\\
&\quad= \Pi\, \Theta(\rho)\, \hat{f}_S\,  \Pi \ . \nonumber
\end{align}
This establishes \eqref{eq:RelOExpValue}, matching~\eqref{eq:equivalenceAlgebraic}. Comparing the two results, we note that the derivation of~\eqref{eq:RelOExpValue} uses a stronger assumption about the form~\eqref{eq:PWGeneralGaugeState} of the algebraic state $\omega_{R|\rho}$, while, on the other hand, it does not rely on the QRF being ideal.

 Let us now construct the analogue of relation~\eqref{eq:Phys-to-PWexp} derived using~\eqref{eq:PWExpVal} (\cite[Thm.~4]{Trinity}). We start by writing projected system observables in the language of this section. Using definition (43) of \cite{Trinity}\footnote{Within our notation, $\hat{f}_S^\text{phys}$\ and $\mathcal{E}_S^{\rho} (\hat{f}_S^\text{phys})$\ of~\cite{Trinity}, correspond to $\hat{f}_{S|R}$\ (see~\eqref{eq:fSR}) and $\hat{\mathcal{O}}^\rho_R (\hat{f}_S)$ (see~\eqref{eq:relobsphys}) respectively.} as well as our results~(\ref{eq:PiThetaCommute}) and~(\ref{eq:PiPiCommute}) 
\begin{align} \label{eq:ProjectedSObs}
\hat{\mathcal{O}}^\rho_R (\hat{f}_S) &:= \Pi \,\Theta(\rho) \,\hat{\pi} \hat{f}_S \,\hat{\pi}
\nonumber \\
&= \Pi \,\hat{\pi}\,\Theta(\rho)  \, \hat{f}_S\, \hat{\pi}
\\
&=  \Pi\, \Theta(\rho) \, \hat{f}_S \,\hat{\pi} \ . \nonumber
\end{align}
(This is a generalization of~\eqref{eq:fSR} to non-ideal QRFs.) Equation~(38) of \cite{Trinity} defines the Page-Wootters inner product on the space of constraint solutions, and we immediately get
\begin{align}\label{eq:PW=Phys}
\langle \phi_\text{phys} | \psi_\text{phys} \rangle_\text{PW} &:= \langle \phi_\text{phys} |\Theta(\rho)| \psi_\text{phys} \rangle \nonumber
\\
&=( \phi_\text{phys} | \psi_\text{phys} ) \ , 
\end{align}
where the second equality is the basic property of a gauge~\eqref{eq:GaugePhysIP}. Using a pair of representative kinematical states satisfying $\Pi |\psi \rangle = | \psi_\text{phys} \rangle$ and $\Pi |\phi \rangle = | \phi_\text{phys} \rangle$, we can express~\eqref{eq:PW=Phys} via the kinematical inner product, allowing us to relate physical matrix elements of projected relational observables of the system to the values assigned to kinematical system observables by algebraic solution states. Using~\eqref{eq:ProjectedSObs}, (\ref{eq:PiThetaPi}) and~(\ref{eq:PiPiCommute}),
\begin{align}
( \phi_\text{phys} |\hat{\mathcal{O}}^\rho_R (\hat{f}_S)| \psi_\text{phys} )  &\equiv \langle \phi_\text{phys} |\hat{\mathcal{O}}^\rho_R (\hat{f}_S)| \psi_\text{phys} \rangle_\text{PW} \nonumber
\\
&=\langle \phi_\text{phys} |\Theta(\rho)\,\hat{\mathcal{O}}^\rho_R (\hat{f}_S)| \psi_\text{phys} \rangle \nonumber
\\
&= \langle \phi| \Pi\, \Theta(\rho)\, \hat{\mathcal{O}}^\rho_R (\hat{f}_S)\,\Pi|\psi \rangle \nonumber
\\
&=  \langle \phi|  \Pi\, \Theta(\rho)\, \Pi\, \Theta(\rho) \, \hat{f}_S\, \hat{\pi}\,  \Pi | \psi \rangle \nonumber
\\
&= \langle \phi| \Pi\, \Theta(\rho)\,  \hat{f}_S  \, \Pi | \psi \rangle \nonumber
\\
&= \omega_{R|\rho} \left( \hat{f}_S \right) \ ,
\end{align}
where $\omega_{R|\rho}$\ is of the form~\eqref{eq:PWGeneralGaugeState}, with $\tilde{\omega} (.)=\langle \phi | . |\psi\rangle$. We have essentially re-derived a version of Theorem 4 in \cite{Trinity}.

\subsubsection{Weak reference gauge}
\label{sec:WeakRefGauge}

Almost-positive algebraic states introduced in Sect.~\ref{sect:algebraicClock} are right-null states of the constraint and left-multiplicative states of the reference observable. This suggests defining gauge-fixed states in a slightly weaker form than~(\ref{eq:GaugeFixedSolution}) and, correspondingly, (\ref{eq:PWGeneralGaugeState}), by dropping the $\Pi$\ projector on the left
\begin{align}\label{eq:QDSRGeneralGaugeState}
\omega_{R|\rho}(.) := \tilde{\omega} \left( \Theta(\rho) (.) \Pi \right) \ ,
\end{align}
where, again, $\tilde{\omega}$\ is an arbitrary kinematical state on which the above projective operations are well-defined.

With this weaker condition on the gauge-fixed state, we find that the analogue of~(\ref{eq:RelOExpValue}) holds only up to commutators of the relevant system observable with the projection operator $\hat{\pi}$. Following the same steps as in the proof of~(\ref{eq:RelOExpValue}) but, this time, using identities~(\ref{eq:SystemProjectorIdentity2})\ and~(\ref{eq:PiPiCommute}) at the end, we get
\begin{align}
  \Theta(\rho)\,\hat{O}^\rho_R(\hat f_S)\,\Pi &= \Theta(\rho)\,\mathcal{G}\left( \Theta(\rho)\, \hat{f}_S \right) \,\Pi
\nonumber\\
&=  \Theta(\rho)\, \Pi\, \Theta(\rho)\, \hat{f}_S\,  \Pi
\nonumber\\
&= \Theta(\rho)\, \hat{\pi} \hat{f}_S\,  \Pi
\\
&= \Theta(\rho)\, \hat{f}_S\,  \Pi + \Theta(\rho)\, [\hat{\pi},\hat{f}_S]\,  \Pi \ .\nonumber
\end{align}
Thus, for a gauge-fixed state~(\ref{eq:QDSRGeneralGaugeState}) we have
\begin{equation}
\omega_{R|\rho}\left( \hat{O}^\rho_R(\hat f_S) \right) =\omega_{R|\rho}\left( \hat{f}_S \right) +
\omega_{R|\rho}\left( [\hat{\pi},\hat{f}_S] \right) \ .
\end{equation}
Of course, for an ideal reference frame, we have $\hat{\pi}=\mathds{1}$, and we recover~(\ref{eq:RelOExpValue}) for all system observables even with the weaker condition~(\ref{eq:QDSRGeneralGaugeState}).

\section{Discussion and outlook}
\label{sect:discussion}

The main purpose of this report is to bring into focus the strong relation between three different approaches to implementing quantum reference frames (QRFs) associated with gauge symmetries: the perspective-neutral (PN) approach \cite{delaHamette:2021oex,Hoehn:2023ehz,Trinity,TrinityRel,Chataignier:2024eil,Hohn:2018toe,Hohn:2018iwn,AliAhmad:2021adn,Hoehn:2021flk,Carrozza:2024smc,Giacomini:2021gei,Castro-Ruiz:2019nnl,delaHamette:2021piz,Vanrietvelde:2018dit,Vanrietvelde:2018pgb,Suleymanov:2023wio,DeVuyst:2024pop,DeVuyst:2024uvd,Hoehn:2023axh,Araujo-Regado:2025ejs,Suleymanov:2025nrr}, 
the effective approach \cite{EffectiveConstr, EffectiveConstrRel, Bojowald:2009jk,EffectivePoT1,EffectivePoT2,EffectivePoTChaos}, 
and the algebraic approach \cite{QDSR,AlgebraicPoT,AlgebraicClocks}. Each of the three approaches has its own method for `jumping into a perspective' that corresponds to a specific orientation of a chosen reference frame.  Within the PN framework this is accomplished through a (gauge-fixing) reduction map, while for the effective and algebraic approaches a perspective is set through imposing gauge-fixing conditions, which amounts to making the reference observable act like a multiple of the identity in the algebraic approach, corresponding to setting the frame fluctuations and covariances to zero within the effective approach (Sect.~\ref{sect:algebraicClock} and \ref{sect:clockGauge}). 

Despite their a priori differences, we explicitly find that the QRFs constructed using these three approaches carry equivalent physical information when the QRFs involved are ideal. This equivalence has two layers. First, evaluating the relational Dirac observables of the PN framework~\eqref{eq:diracObservable} using a relational algebraic state (i.e.\ gauge-fixed as in~\eqref{eq:algebraicFrame}) yields the same result~\eqref{eq:equivalenceAlgebraic} as the expectation value for the corresponding non-invariant system observable in the same relational state. Second, a concrete Hilbert-space realization~\eqref{eq:stateReal} of a gauge-fixed algebraic state is accomplished through the use of a  Page-Wootters-type projection that matches the algebraic reference gauge. The values assigned by this constructed algebraic state to non-invariant observables can be expressed both as physical expectation values~\eqref{eq:algDirac} of the relational Dirac observable and as a Page-Wootters-state expectation value~\eqref{eq:Phys-to-PWexp}. Both of these results tell us that algebraic gauge-fixing captures the same relational data as relational observables and relational states of the PN framework. The same result linking the effective approach with the PN formalism also holds at the appropriate semiclassical order due to the close relation between the algebraic and the effective QRF constructions.

The connection between the approaches established here immediately allowed us to use the PN framework machinery to develop explicit QRF transformations within the other two approaches in Sect.~\ref{sect:switchingIdeal}, in the case of the effective approach being equivalent in content to the older transformations in \cite{EffectivePoT1,EffectivePoT2,EffectivePoTChaos}.  We further believe that the links with algebraic and effective methods add valuable techniques to the perspective-neutral toolkit.  Since the algebraic approach works with the space of linear functionals on the kinematical observable algebra,  it circumvents the need for constructing an explicit physical Hilbert space.  It thus provides a promising venue for developing approximate methods that work ``locally'' - that is, for a subset of states of the reference systems. The effective semiclassical approach can be viewed as one such approximation scheme that applies as long as the expectation values of the basic generators and an (infinite) set of covariances and higher-order moments satisfy the semiclassical hierarchy.  Furthermore, the effective approach provides a natural starting point for perturbative expansions in $\hbar$ to obtain quantum corrections, even when interactions between system and frame are present, in terms of experimentally interesting quantities -- the expectation values and moments of observables. Both the algebraic and the effective approaches naturally encompass couplings in the constraint between the QRF and observed systems, which is an under-developed feature within the PN framework.

We already see the usefulness of QRF transformations that utilize the effective approach in Sect.~\ref{sect:rel} where we look at the frame dependence of uncertainties and covariances. Given any observed system's observable $\hat{f}_S$\ that is not an invariant, so that $[\hat{f}_S, \hat{C}] \neq 0$, we find that the transformation of the associated uncertainty $(\Delta f_S)^2$ from one ideal frame,  say Alice's,  into another ideal frame, say Bob's, depends on covariances between the system degrees of freedom and the orientation of Alice's frame~\eqref{eq:covarianceTransform}.  Applying our results to a translationally invariant system of interacting particles in one dimension (Sect.~\ref{sect:translationalInvariant}), we reproduce the corresponding QRF transformations of uncertainties and covariances reported in~\cite{Suleymanov:2025nrr} without using any explicit Hilbert space states.

Of course, as already emphasized, our strongest results linking the three approaches assume ideal reference frames. However, the PN framework extends to a general class of non-ideal reference frames that form a positive operator-valued measure  rather than a self-adjoint QRF orientation observable. The semiclassical effective approach has also been developed, at least to order $\hbar$, for a class of non-ideal reference frame models \cite{EffectivePoTChaos,EffectivePoT1,EffectivePoT2}, though this treatment still explicitly relies on using one of the ``ideal'' observables of the reference system to fix gauge freedom. On the other hand, extending the algebraic approach to non-ideal reference frames would involve a non-trivial modification of the procedure described in Sect.~\ref{sec:AlgebraicMethod}. The main obstacle when the reference frame is non-ideal, is that the reference operator $\hat{R}$ one obtains from \eqref{eq:time operator} is neither self-adjoint, nor does it have the reference states $\ket{\rho}$ \eqref{eq:clock states} as its eigenstates. This means, that in the algebraic approach, there is no $*$-invariant reference variable $\hat{R}$  to define left-multiplicative algebraic states \eqref{eq:algebraicFrame}. The most promising remedy is to follow our insights from Sect.~\ref{sect:generalGaugeFreedom} and work with generalized projections such as $\Pi$, $\Phi$, and $\Theta(\rho)$. Adding such operations either as elements of the kinematical algebra $\mathcal{A}$\ or its homomorphisms requires us to go beyond the simple polynomial $*$-algebra that have been used within the algebraic approach up to this point. 

Comparison becomes even more complicated when interactions between the observed system and the reference frame are introduced. Although progress has been made for interacting systems within the PN framework  \cite{Smith2019quantizingtime,Hoehn:2023axh,Castro-Ruiz:2019nnl}, the effective approach \cite{EffectivePoTChaos}, and in the algebraic approach \cite{AlgebraicPoT}, the cases they can handle are quite restricted and different from each other. Some form of gauge-fixing of the type explored in the previous section is the most promising direction for applying quantum reference frame methods to interacting reference frames, where group averaging and the related G-twirling may be hard to perform explicitly. The idea would be to substitute these operations by constructing ``local'' transformations between gauges, for example where the formal transformation $\Phi_A \Pi \Phi_B$\ collapses into a collection of finite shifts of the form~\eqref{eq:UnitaryGaugeTransform}.

\subsection*{Acknowledgements}
A large portion of this research was conducted while AT was visiting the Okinawa Institute of Science and Technology (OIST) through the Theoretical Sciences Visiting Program (TSVP). This work was supported by funding from Okinawa Institute of Science and Technology Graduate University and also made possible through the support of the ID\# 62312 grant from the John Templeton Foundation, as part of the \href{https://www.templeton.org/grant/the-quantum-information-structure-of-spacetime-qiss-second-phase}{\textit{`The Quantum Information Structure of Spacetime'} Project (QISS)}.~The opinions expressed in this project are those of the authors and do not necessarily reflect the views of the John Templeton Foundation. 

\subsection*{Contribution statement}

All authors contributed equally to the development and writing of this paper. No AI tools were used in its creation.

\bibliographystyle{quantum}
\bibliography{quinternity}

\onecolumn \newpage
\appendix

\section{Details of derivations}
\label{app:derivations}

This appendix contains the details of some of the derivations outlined in Sect.~\ref{sect:equivalence} and \ref{sect:switchingIdeal}.

\subsection{Expectation values in gauge-fixed algebraic states}
\label{app:equivalenceAlgebraic}

Here, we will explicitly derive equation~\eqref{eq:equivalenceAlgebraic}.
We make use of the definition of the Dirac observable with respect to an ideal frame (\ref{eq:idealDirac}) to write the left hand side for general orientation $\rho$ as
\begin{align}
    \omega_{R \vert \rho'}(\hat{O}_R^{\rho}(\hat{f}_S)) &= 
    \omega_{R \vert \rho'} \left( e^{-i(\hat{q}_R - \rho ) \hat{G}_S / \hbar} \hat{f}_S e^{i(\hat{q}_R - \rho ) \hat{G}_S / \hbar} \right).
\end{align}
We now make use of \eqref{eq:algebraicFrame}. For the exponential of two commuting operators to the left of $\hat{f}_S$ this is trivial: we expand it as an infinite series, use the frame gauge-fixing condition,  \eqref{eq:algebraicFrame} and write the resulting series as an exponential again. 
However, care has to be taken for the exponential on the right, since generally $\comm*{\hat{f}_S}{\hat{G}_S} \neq 0$. In this case, we similarly expand the exponential since $\comm*{\hat{q}_R}{\hat{G}_S} = 0$, and make use of the algebraic constraint property \eqref{eq:algebraicFrame} to write
\begin{align}
    \omega_{R \vert \rho'}(\hat{O}_R^{\rho}(\hat{f}_S)) &= \omega_{R \vert \rho'}( e^{-i(\rho' - \rho ) \hat{G}_S / \hbar} \hat{f}_S e^{i(\hat{q}_R - \rho ) \hat{G}_S / \hbar}) \nonumber \\
    &= \omega_{R \vert \rho'} \left( e^{-i(\rho' - \rho) \hat{G}_S / \hbar} \hat{f}_S 
    \left[ \sum_{n = 0}^{+ \infty} \frac{i^n}{n!}  \hbar^{-n} (\hat{q}_R - \rho )^n \hat{G}_S^n \right] \right)  \\
    &= \omega_{R \vert \rho'} \left(e^{-i(\rho' - \rho) \hat{G}_S / \hbar} \hat{f}_S 
    \left[ \sum_{n = 0}^{+ \infty} \frac{i^n}{n!} \hbar^{-n} (\hat{q}_R - \rho )^n (- \hat{p}_R)^n  \right] \right). \nonumber
\end{align}
Since $\hat{p}_R$ and $\hat{q}_R$ both commute with $\hat{f}_S$ and $\hat{G}_S$ we can commute the sum inside the square parentheses all the way to the left. Using the algebraic frame gauge condition again, we obtain
\begin{align}
    \omega_{R \vert \rho'}( e^{-i(\rho' - \rho) \hat{G}_S / \hbar} \hat{f}_S e^{i(\hat{q}_R - \rho ) \hat{G}_S / \hbar} ) =
    \omega_{R \vert \rho'} \left( \left[ \sum_{n = 0}^{+ \infty} \frac{i^n}{n!} \hbar^{-n} (\rho' - \rho)^n (- \hat{p}_R)^n  \right] e^{-i(\rho' - \rho) \hat{G}_S / \hbar} \hat{f}_S \right).
\end{align}
Finally, we reverse the steps by commuting this series to the right, use the algebraic constraint property, and finally rewrite the series as an exponential to get
\begin{equation}
    \omega_{R \vert \rho'}(\hat{O}_R^{\rho}(\hat{f}_S)) = \omega_{R \vert \rho'}( e^{-i(\rho' - \rho) \hat{G}_S / \hbar} \hat{f}_S e^{i(\rho' - \rho) \hat{G}_S / \hbar}).
\end{equation}
The equality (\ref{eq:equivalenceAlgebraic}) clearly follows after setting $\rho = \rho'$ as claimed.

Next, let us prove that, for the concrete realization of the gauge-fixed algebraic state provided by~\eqref{eq:stateReal}, the relation \eqref{eq:equivalenceAlgebraic} holds even for a non-ideal frame $R$ (still assumed to be non-interacting). We simply evaluate
\begin{align}
    \omega_{R \vert \rho}( \hat{f}_{S} ) 
    &= \expval{(\dyad{\rho}_R \otimes \mathds{1}_S)\hat{f}_{S}}{\psi_\text{phys}}_\text{kin} \nonumber \\
    &= \expval{(\dyad{\rho}_R \otimes \mathds{1}_S) \,\Pi
    \,(\dyad{\rho}_R \otimes \mathds{1}_S) \hat{f}_{S}}{\psi_\text{phys}}_\text{kin} \\
    &= \expval{(\dyad{\rho}_R \otimes \mathds{1}_S) \,\hat{O}_R^{\rho}(\hat{f}_S)\, }{\psi_\text{phys}}_\text{kin} = 
    \omega_{R \vert \rho}(\hat{O}_R^{\rho}(\hat{f}_S)) . \nonumber
\end{align}
To go to the second line, we made use of the identity $\Pi\left(\dyad{\rho}_R\otimes\mathds1_S\right)\Pi=\Pi$, which follows from the resolution of the identity~\eqref{eq:idResolution} (note that $\bra{\psi_{\rm phys}}$ contains a $\Pi$). To go to the last line, we made use of \eqref{eq:relobsphys} and~\eqref{eq:PWExpVal}.

\subsection{Frame transformation for algebraic states}
\label{sect:algebraicFrameSwitchting}

To prove \eqref{eq:algTransf} we start with the form taken by relational Dirac observables in an ideal reference frame~\eqref{eq:idealDirac}, and use the fact that $[\hat{f}_B, \hat{G}_S]=0$ and $[\hat{f}_S, \hat{p}_B] = 0$ 
\begin{align}
    \omega_{B \vert \rho_B}(\dirac{\hat{f}_B \hat{f}_S}{A}{\rho_A}) &= 
    \omega_{B \vert \rho_B}( e^{-i (\hat{q}_A - \rho_A) \hat{p}_B/\hbar} 
    \hat{f}_B e^{i (\hat{q}_A - \rho_A) \hat{p}_B/\hbar} 
    e^{-i (\hat{q}_A - \rho_A) \hat{G}_S/\hbar} 
    \hat{f}_S e^{i (\hat{q}_A - \rho_A) \hat{G}_S/\hbar}) \nonumber \\
    &= \omega_{B \vert \rho_B}(e^{-i (\hat{q}_A - \rho_A) \hat{p}_B/\hbar} 
    \hat{f}_B e^{i (\hat{q}_A - \rho_A) \hat{p}_B/\hbar} 
    \dirac{\hat{f}_S}{A}{\rho_A}).
\end{align}
Even though the definition of $\dirac{\hat{f}_S}{A}{\rho_A}$ requires there to be an extra $\hat{p}_B$ 
in the exponentials next to $\hat{f}_S$ since here the transformations of the observed system $BS$\ are generated by $(\hat{p}_B+\hat{G}_S)$, they simply annihilate to an identity operator 
because $\hat{p}_B$ and $\hat{f}_S$ commute. Next, we note that $[\hat{q}_B, e^{i (\hat{q}_A - \rho_A) \hat{p}_B/\hbar} ]=-(\hat{q}_A - \rho_A)  e^{i (\hat{q}_A - \rho_A) \hat{p}_B/\hbar} $, so that conjugation by this exponential gives
\[
e^{-i (\hat{q}_A - \rho_A) \hat{p}_B/\hbar} \hat{q}_B e^{i (\hat{q}_A - \rho_A) \hat{p}_B/\hbar} = \hat{q}_B- (\hat{q}_A - \rho_A) \ .
\]
Since $\hat{f}_B$\ is some polynomial in $\hat{p}_B$, and $\hat{q}_B$, conjugation has the effect of shifting every $\hat{q}_B$\ in the expression for $\hat{f}_B$\ in the above way giving us
\begin{equation}
\begin{split}
    \omega_{B \vert \rho_B}(\dirac{\hat{f}_B \hat{f}_S}{A}{\rho_A}) &= \omega_{B \vert \rho_B}(\hat{f}_B(\hat{q}_B - (\hat{q}_A - \rho_A), \hat{p}_B) 
    \dirac{\hat{f}_S}{A}{\rho_A})  \\
    &= \omega_{B \vert \rho_B}(\hat{f}_B(\rho_B - (\hat{q}_A - \rho_A), \hat{p}_B) 
    \dirac{\hat{f}_S}{A}{\rho_A}).
\end{split}
\end{equation}
In the last step we made use of the algebraic frame property \eqref{eq:algebraicFrame} and the fact that $\hat{f}_B$\ is ordered so that powers of its first argument appear on the left. 
The final step consists of commuting $\hat{f}_B$ to the right, $\dirac{\hat{f}_S}{A}{\rho_A}$ is made up of operators which commute with $\hat{q}_A$ and $\hat{p}_B$, 
and make use of the algebraic constraint property \eqref{eq:algebraicFrame} to obtain
\begin{equation}
    \omega_{B \vert \rho_B}(\dirac{\hat{f}_B \hat{f}_S}{A}{\rho_A}) = \omega_{B \vert \rho_B}(\dirac{\hat{f}_S}{A}{\rho_A} \hat{f}_B(\rho_B - (\hat{q}_A - \rho_A), -\hat{p}_A - \hat{G}_S) ).
\end{equation}
Together with \eqref{eq:algFrameChange} this leads to the desired result \eqref{eq:algTransf}.

\section{Additional example: System algebra generated by $\mathfrak{su}$(2)}
\label{app:spin}

In this appendix we explicitly work out some of the QRF transformations for a model where the generating observables of system $S$\ form an $\mathfrak{su}$(2) Lie algebra with respect to the commutator $\comm*{\hat{J}_i}{\hat{J}_j} = i \hbar \eps_{ijk} \hat{J}_k$, where the indices $i, j, k$\ take values $x$, $y$, or $z$. We consider the following simple constraint
\begin{equation}
    \hat{C} =  \hat{p}_A + \hat{p}_B - \beta \hat{J}_z,
\end{equation}
with $\beta$ a constant such that the three terms have the same dimension. The transformations for observables associated purely with the reference systems $A$\ and $B$\ are given directly by~\eqref{eq:algTransf} with $\hat{f}_S$\ set to $\mathds{1}$, and are not sensitive to the properties of $\hat{G}_S$. We therefore focus on the transformations associated with system $S$ variables, starting with the generators $\hat{J}_i$. Equation~(\ref{eq:changeGenerator}) immediately yields $\omega_{A \vert \rho_A}(\hat{J}_z) = \omega_{B \vert \rho_B}(\hat{J}_z)$:  $\hat{J_z}$ clearly commutes with the system generator and is therefore a Dirac observable.
For a non-trivial transformation, we look at $\hat{J}_x$. 
The Dirac observable associated to $\hat{J}_x$ can be found from writing (\ref{eq:idealDirac}) as an infinite sum over nested commutators \cite{Trinity} and subsequently splitting it in an even and odd part
\begin{align}
    \label{eq:spinDirac}
    \dirac{\hat{J}_x}{A}{\rho_A} &= \sum_{n=0} \frac{(-i \beta (\hat{q}_A-\rho_A))^n}{\hbar^n n!}  \comm*{\hat{J}_x}{\hat{J}_z}_n \nonumber \\
    &= \sum_{n=0} \frac{(i \beta (\hat{q}_A-\rho_A))^{2n}}{(2n)!} \hat{J}_x + i \sum_{n=0} \frac{(i \beta (\hat{q}_A-\rho_A))^{2n+1}}{(2n+1)!} \hat{J}_y \\
    &= \cos(\beta(\hat{q}_A - \rho_A)) \hat{J}_x - \sin(\beta(\hat{q}_A - \rho_A)) \hat{J}_y , \nonumber
\end{align}
with $\comm*{\hat{J}_x}{\hat{J}_z}_n$ being the $n$-th nested commutator of $\hat{J}_x$\ with $\hat{J}_z$.
The factor of $\hbar^{-n}$ gets canceled by the $\hbar^n$ stemming from the $n$-th nested commutator. A similar expression can be derived for $\hat{J}_y$.
Hence, the transformation \eqref{eq:algFrameChange} reads
\begin{equation}\label{eq:StateJxTransformation}
    \omega_{A \vert \rho_A}(\hat{J}_x)
    = \omega_{B \vert \rho_B} \left( 
     \cos(\beta(\hat{q}_A - \rho_A  )) \hat{J}_x - \sin(\beta(\hat{q}_A - \rho_A )) \hat{J}_y
     \right). 
\end{equation}

To write the effective transformation following~\eqref{eq:truncatedSwitching}, we create a truncated version of this expression. A direct way of accomplishing this is to expand the expression inside the state on the right-hand side of~\eqref{eq:StateJxTransformation} by writing generators as $\hat{y}_i = \widehat{\Delta y}_i + \expval*{\hat{y}_i}$\ and keeping terms up to 2nd order in products of $\widehat{\Delta y}_i$ (operators in products in~\eqref{eq:spinDirac} commute so there will not be explicit powers of $\hbar$\ in the expansion).\footnote{It is also possible to use the general procedure (\ref{eq:effExpand}), which assumes Weyl ordering of generators inside of the expression being expanded. This is easy to implement for an expression such as~\eqref{eq:spinDirac} where generators appearing in products commute with each other.}
However, note that we are expanding the argument of a function $\expval*{F(\hat{q}_A)}$ 
about $\expval*{\hat{q}_A} \equiv q_A$ which is not equal to $\rho_A \equiv \expval*{\hat{q}_A}_{A | \rho_A}$ when evaluated in a $B$-gauge state. We find to 2nd order
\begin{equation}\label{eq:EffectiveJxTransformation}
\begin{split}
    \mathscr{T}_2(J_x)_{A | \rho_A} &= (O^{\rho_A}_A(J_x))_{B | \rho_B}
    - \beta \sin(\beta (q_A - \rho_A  )_{B | \rho_B}) \Delta(q_A J_x)_{B | \rho_B} \\
    &\hspace{10pt}- \beta \cos(\beta(q_A - \rho_A  )_{B | \rho_B} ) \Delta(q_A J_y)_{B | \rho_B}
    - \frac{1}{2} \beta^2 (O^{\rho_A}_A(J_x))_{B | \rho_B} (\Delta q_A)^2_{B | \rho_B}
    ,
\end{split}
\end{equation}
where for compactness we use $(O^{\rho_A}_A(J_x))_{B | \rho_B}$\ to denote the ``classical'' expression for this relational Dirac observable evaluated in the $B$-frame. That is
\begin{equation}\label{eq:DiracObservableClassical}
    (O^{\rho_A}_A(J_x))_{B | \rho_B} := 
    \cos(\beta(q_A - \rho_A )_{B | \rho_B}) (J_x)_{B | \rho_B} - \sin(\beta(q_A - \rho_A)_{B | \rho_B}) (J_y)_{B | \rho_B},
\end{equation}
and recall that $(.)_{i | \rho_i}$ with $i = A,B$ means that expectation values of operators, such as $q_i$ and $J_i$, are  evaluated in that gauge. As a quick check, we see that setting $B=A$, with $\rho_B=\rho_A$\ on the right-hand side and imposing the corresponding gauge conditions $\Delta(q_A y_i)_{A | \rho_A} = 0$\ and $(q_A)_{A|\rho_A} =\rho_A$, recovers $(J_x)_{A | \rho_A}$.

A system algebra $\mathcal{A}_{BS}$\ corresponding to using $A$\ as the QRF has five generators $\{\hat{q}_B, \hat{p}_B; \hat{J}_i\}$. When expanded to semiclassical order $M=2$\ it should possess 20 degrees of freedom~\cite{SemiclassicalLie}, explicitly we have five expectation values, five uncertainties $(\Delta y_i)^2$, and ten covariances: $\Delta(q_Bp_B)$, three each of the forms $\Delta(q_BJ_i)$\ and $\Delta(p_BJ_i)$, and three of the form $\Delta (J_i J_j)$, $i\neq j$. The truncated degrees of freedom of the observed system in gauge $B$\ are instead generated by $\{\hat{q}_A, \hat{p}_A; \hat{J}_i\}$\ and can be accounted for in an identical fashion. Creating an explicit transformation from $B$\ to $A$\ therefore involves expressing all 20 classical and quantum variables of frame $A$\ in terms of those of frame $B$. We already see how this is accomplished for transforming expectation values in~\eqref{eq:EffectiveJxTransformation} let us now show how to create an explicit frame transformation for a second-order moment, using $\Delta(q_BJ_x )$\ as the example. 
Note that, in frame $B$, $\Delta(q_BJ_x )=0$\ is one of the gauge conditions, however, when we transform into frame $A$, $\Delta(q_BJ_x )$\ becomes an independent unrestricted degree of freedom.

To find the value this covariance takes upon transforming from $B$\ to $A$\ we make use of \eqref{eq:effectiveMomentSwitching}
\begin{equation}
    \Delta(q_B J_x)_{A | \rho_A} = \expval{\hat{O}_A^{\rho_A}(\hat{q}_B) \  \hat{O}_A^{\rho_A} (\hat{J}_x) }_{B | \rho_B} - \expval{ \hat{O}_A^{\rho_A}(\hat{q}_B)}_{B | \rho_B} \expval{\hat{O}_A^{\rho_A}(\hat{J}_x)}_{B | \rho_B} \ .
    \label{eq:qubitCovar}
\end{equation}
Expanding the first term on the right-hand side to order $M = 2$ in $B$-gauge by making use of \eqref{eq:idQRFposrel} and \eqref{eq:spinDirac}
leads to
\begin{equation}
\begin{split}
    \label{eq:trCovarIntermed}
    \mathscr{T}_2 &\expval{(\hat{q}_B - \hat{q}_A + \rho_A ) \hat{O}_A^{\rho_A}(\hat{J}_x) }_{B | \rho_B} = (\rho_B + \rho_A - q_A)_{B | \rho_B} (O_A^{\rho_A}(J_x))_{B | \rho_B}
    \\
    &\hspace{10pt} -[\cos(\beta(q_A - \rho_A )_{B | \rho_B})+ \beta (\rho_B + \rho_A - q_A)_{B | \rho_B} \sin(\beta(q_A - \rho_A )_{B | \rho_B})] \Delta(q_A J_x)_{B | \rho_B}  \\
    &\hspace{10pt}+[\sin(\beta (q_A - \rho_A)_{B | \rho_B}) - \beta (\rho_B + \rho_A - q_A)_{B | \rho_B} \cos(\beta(q_A-\rho_A )_{B | \rho_B}) ] \Delta(q_A J_y)_{B | \rho_B} \\
    &\hspace{10pt}+ \frac{1}{2} \Big[ \beta \sin(\beta(q_A-\rho_A )_{B | \rho_B})(J_x)_{B | \rho_B} 
    + \beta \cos(\beta(q_A - \rho_A )_{B | \rho_B})(J_y)_{B | \rho_B} \\
    &\hspace{10pt}- \beta^2 (\rho_B + \rho_A - q_A)_{B | \rho_B} (O_A^{\rho_A}(J_x))_{B | \rho_B} \Big] (\Delta q_A)^2_{B | \rho_B},
\end{split}
\end{equation}
where we once again use \eqref{eq:DiracObservableClassical}\ for compactness. Generally there would also be terms like $\Delta(q_B J_x)$ and $\Delta(q_B J_y)$, but because of the $B$-gauge condition these are set to zero. They do, however, show up in $A$-gauge.
Combining this result with \eqref{eq:EffectiveJxTransformation}, we find for the covariance \eqref{eq:qubitCovar}
\begin{equation}
\begin{split}\label{eq:EffectiveqBJxTransformation}
    \mathscr{T}_2 \Delta(q_B J_x)_{A | \rho_A} =
    &-\cos(\beta(q_A - \rho_A )_{B | \rho_B}) \Delta(q_A J_x)_{B | \rho_B} +\sin(\beta (q_A - \rho_A)_{B | \rho_B}) \Delta(q_A J_y)_{B | \rho_B} \\
    &+ \frac{1}{2} \left[ \beta \sin(\beta(q_A-\rho_A )_{B | \rho_B})J_x 
    + \beta \cos(\beta(q_A - \rho_A )_{B | \rho_B})J_y \right](\Delta q_A)^2_{B | \rho_B}. 
\end{split}
\end{equation}
Note that, unlike \eqref{eq:EffectiveJxTransformation}, imposing the $A$-gauge conditions $\Delta(q_A y_i)_{A | \rho_A} = 0$\ on the right-hand side of~\eqref{eq:EffectiveqBJxTransformation}, makes it vanish. This operation corresponds to setting $B=A$\ and the result is entirely expected, because $\Delta(q_A J_x)_{A | \rho_A}=0$\ is itself one of the gauge conditions for frame $A$.

The covariance transformation~\eqref{eq:EffectiveqBJxTransformation} is more complicated than its counterpart in the translationally invariant model~\eqref{eq:covarianceTra} in two important ways. First, due to the $\mathfrak{su}(2)$ algebra structure, the relational Dirac observables themselves \eqref{eq:spinDirac} are more complicated, leading to the transformation being explicitly dependent on the classical variables and not just moments. Second, the complete moments expansion of $\left\langle \hat{O}^{\rho_A}_A(\hat{q}_B)\hat{O}^{\rho_A}_A(\hat{q}_C)\right\rangle$\ terminates at second order, meaning that~\eqref{eq:covarianceTra} does not change as one increases the order of truncation. On the other hand, the expansion of  $\left\langle\hat{O}^{\rho_A}_A(\hat{q}_B)\hat{O}^{\rho_A}_A(\hat{J}_x)\right\rangle$\ receives additional terms at all orders due to the power series form of the $\cos$ and $\sin$ functions, so that $\mathscr{T}_M \Delta(q_B J_x)_{A | \rho_A}$, for $M>2$ would have additional higher-order terms not captured in~\eqref{eq:EffectiveqBJxTransformation}. We note that these features of transformation~\eqref{eq:EffectiveqBJxTransformation} are the rule and the simplicity of transformation~\eqref{eq:covarianceTra} is the exception.

\end{document}